\documentclass{article}

\usepackage[utf8]{inputenc}
\usepackage[english]{babel}
\usepackage[x11names]{xcolor}
\usepackage{tabularx}
\usepackage{amsmath}
\usepackage{amsfonts}
\usepackage{amsbsy}
\usepackage{amssymb}
\usepackage{amsthm}
\usepackage{amscd}
\usepackage{enumitem}
\usepackage{float}
\usepackage{hyperref}
\usepackage{siunitx}
\usepackage{comment}
\usepackage{placeins}
\usepackage{multirow}
\usepackage[font=footnotesize]{caption}
\usepackage{subcaption}
\usepackage{graphicx}
\usepackage{longtable}
\usepackage{cancel}
\usepackage{makecell}
\usepackage{lineno}
\usepackage{authblk}

\usepackage[backend=bibtex, giveninits, style=numeric, uniquelist=false, sorting=none, sortcites=true, natbib=true, doi=false, url=false, isbn = false]{biblatex}
\DeclareFieldFormat[article]{title}{#1} %
\DeclareFieldFormat[book]{title}{\itshape #1}
\DeclareFieldFormat[thesis]{title}{#1}
\DeclareFieldFormat[incollection]{title}{\itshape #1}
\DeclareFieldFormat[inproceedings]{title}{\itshape #1}
\DeclareFieldFormat[article]{author}{\color{blue} #1}
\DeclareFieldFormat[article]{pages}{#1.}
\DeclareFieldFormat[incollection]{pages}{#1.} %
\renewbibmacro{in:}{%
	\ifentrytype{article}{}{\printtext{\bibstring{in}\intitlepunct}}}

\DeclareNameAlias{sortname}{last-first}
\DeclareNameAlias{default}{last-first}

\renewbibmacro*{volume+number+eid}{
	\printfield{volume}
	\setunit*{\addnbthinspace}
	\printfield{number}
	\setunit{\addcomma\space}
	\printfield{eid}}
\DeclareFieldFormat[article]{number}{\mkbibparens{#1}}

\renewbibmacro*{publisher+location+date}{
	\printlist{publisher}
	\iflistundef{location}
	{\setunit*{\addcomma\space}}
	{\setunit*{\addcomma\space}}
	\printlist{location}
	\setunit*{\addcomma\space}
	\usebibmacro{date}
	\newunit}

\DeclareFieldFormat[article]{number}{} 

\usepackage{hyperref}
\hypersetup{colorlinks=true,urlcolor=RoyalBlue4,linkcolor=RoyalBlue4,citecolor=RoyalBlue4,naturalnames=true,hypertexnames=false}

\title{Towards active learning: A stopping criterion for the sequential sampling of grain boundary degrees of freedom}

\author[1]{Timo Schmalofski\thanks{timo.schmalofski@icams.ruhr-uni-bochum.de}}
\author[2,3]{Martin Kroll\thanks{martin.kroll@uni-bayreuth.de}}
\author[2]{\\Holger Dette\thanks{holger.dette@ruhr-uni-bochum.de}}
\author[1]{Rebecca Janisch\thanks{rebecca.janisch@icams.ruhr-uni-bochum.de}}
\affil[1]{Interdisciplinary Centre for Advanced Materials Simulation (ICAMS), Ruhr-Universität Bochum}
\affil[2]{Fakult\"at für Mathematik, Ruhr-Universität Bochum}
\affil[3]{Fakult\"at f\"ur Mathematik, Physik und Informatik, Universität Bayreuth}

\date{\today}

\newcommand{\R}{{\mathbb R}}
\renewcommand{\S}{{\mathbb S}}

\newcommand{\kb}{{\mathbf k}}

\newcommand{\Kb}{{\mathbf K}}
\newcommand{\Yb}{{\mathbf Y}}

\renewcommand{\theta}{{\vartheta}}
\renewcommand{\phi}{{\varphi}}

\newcommand{\Ncand}{{N_{\mathrm{cand}}}}
\newcommand{\Ngrid}{{N_{\mathrm{grid}}}}
\newcommand{\Ninit}{{N_{\mathrm{init}}}}
\newcommand{\Nseq}{{N_{\mathrm{seq}}}}
\newcommand{\Ntotal}{{N_{\mathrm{total}}}}
\newcommand{\Nref}{{N_{\mathrm{ref}}}}
\newcommand{\Ncusp}{{N_{\mathrm{cusps}}}}
\newcommand{\Ntheta}{{N_{\theta}}}
\newcommand{\Nphi}{{N_{\phi}}}
\newcommand{\Niterdiv}{{N_{\mathrm{iter, cusps}}}}
\newcommand{\Niterdelta}{N_{\mathrm{iter}, \Delta E}}
\newcommand{\Nstop}{{N_{\mathrm{stop}}}}

\newcommand{\DeltaEref}{\Delta E_{\mathrm{ref}}}
\newcommand{\DeltaEprev}{\Delta E_{\mathrm{prev}}}
\newcommand{\DeltaEstat}{\Delta E_{\mathrm{stat}}}

\newcommand{\thetamax}{{\theta_{\max}}}

\newcommand{\phimin}{{\phi_{\min}}}
\newcommand{\phimax}{{\phi_{\max}}}

\usepackage{listings}
\begin{document}

\maketitle

\begin{abstract}
Many materials processes and properties depend on the \textit{anisotropy}
of the energy of grain boundaries, i.e.~on the fact that this energy is a function of the five geometric degrees of freedom (DOF) of the interface. To access this parameter space in an efficient way and to discover energy cusps in unexplored regions, a method was recently established, which combines atomistic simulations with statistical methods \cite{kroll2022efficient}. This sequential sampling technique is now extended in the spirit of an active learning algorithm by adding a criterion to decide when the sampling has advanced enough to stop. In this instance, two parameters to analyse the sampling results on the fly are introduced: the number of cusps, which correspond to the most interesting and important regions of the energy landscape,  and the maximum change of energy between two sequential iterations. Monitoring these two quantities provides valuable insight into how the subspaces are energetically structured. The combination of both parameters provides the necessary information to evaluate the sampling of the 2D subspaces of grain boundary plane inclinations of even non-periodic, low angle grain boundaries. With a reasonable number of data points in the initial design, only a few \textcolor{black}{appropriately chosen sequential iterations already improve the accuracy of the sampling substantially and unknown cusps can be found within a few additional sequential steps}.
\end{abstract}

\section{Introduction}
Understanding and controlling microstructural evolution in metals and metallic alloys 
is one of the central tasks of materials science and engineering. The dynamics of grain
growth and the evolution of grain shape in metallic microstructures strongly depends on the individual mobility of different grain boundaries (GBs), i.e., on the change of the interface energy with its structural parameters \cite{salama2020role, nino2023influence, conry2022enginering, bhattacharya2021grain}. 
\textcolor{black}{Grain boundaries are commonly divided into low angle grain boundaries (LAGBs), i.e.~with a misorientation angle $\omega$ smaller or equal than $15^\circ$, and high angle grain boundaries (HAGBs) with $\omega > 15^\circ$. LAGBs are of special interest materials science due to their role during dynamic recrystallization and microstructure evolution \cite{Barrett2017Effect, He2016Microstructure, Lin2015EBSD}. Furthermore, they are attractive for segregation \cite{He2020On, Kim2020Influence, Krasnikov2023Effect}, and a high fraction of LAGBs in a material leads to a strengthening effect \cite{Krasnikov2023Effect, Sabirov2013Nanostructured} and causes the material to be more crack resistant \cite{Pope1995Weak, Zhang2000Comparison}. Thus, it is important to know the energies of LAGBs and their inclination dependence.} \\
Nowadays, numerical methods for microstructure modelling and microstructure evolution are available, which explicitly include the variation of interface energy with the geometric degrees of freedom of the grain boundary, once it is known \cite{vakili2020multiphasefield,
	steinbach2011phasefield,moelans2008introduction,lee2000montecarlo,pauza2021computer}. To some extent, this variation can be captured by analytical models based on the Read-Shockley-Wolf (RSW) model \cite{wolf1989read}, see e.g.~\cite{bulatov2014grain,dette2017efficient, chirayutthanasak2022anisotropic,sarochawikasit2021grain}. The improvement of such models, and even more a purely numerical treatment of energy as a function of geometry, rely on comprehensive data bases of grain boundary energies, which can be generated in a systematic fashion via atomistic simulations.
However, being comprehensive is a quite challenging task. On the one hand, the parameter space of GBs is five-dimensional, defined by the misorientation axis and angle, as well as the grain boundary plane inclination. On the other hand, the grain boundary energy does not vary in a \textcolor{black}{monotonic} manner, but exhibits deep cusps at \textcolor{black}{specific} misorientations or boundary plane inclinations.
Thus, a standard high-throughput sampling of the parameter space on a regular grid has a twofold drawback: It is both time consuming and likely to miss the most important features in the energy landscape. To provide an efficient data base, however, these should be included in the sampling. This either requires a sampling strategy based on prior knowledge or at least reasonable assumptions on the topology of the energy landscape  \cite{olmsted2009survey,bulatov2014grain,baird2021five, homer2022examination}, and sometimes even the manual addition of the relevant data  \cite{kim2011identification}.
Based on a symmetry analysis and prior knowledge, Olmsted et al.~\cite{olmsted2009survey} designed a strategy to create an energy data base starting with several 1D subspaces and extending to higher dimensions on from there. Bulatov \cite{bulatov2014grain} used this data to interpolate between the sampled regions by an extended RSW model.
Homer et al.~\cite{homer2015grain} focused on 2D inclination subspaces of coincidence-site lattice based grain boundaries and in a first step reduced the size of the subspace of interest as far as possible by exploiting their point symmetries \cite{patala2013representation,patala2013symmetries}.
Randomly chosen structures from the reduced subspaces were then simulated to further explore it. 
Although impressive progress has been made in these publications, even tackling the complete 5D parameter space \cite{homer2022examination}, none of the mentioned approaches solves the problem of how to find the cusps in the energy landscape automatically.

It is tempting to replace the necessary a priori knowledge by the use of modern machine learning methods, which have become more and more popular and effective in material science \cite{butler2014machine}. Zhang et al.
\cite{zhang2018strategy}, for example, studied how machine learning can be applied accurately to sparse datasets. %
As an example they studied the prediction of the band gap of binary semiconductors.
First machine learning approaches for grain boundary energies have been proposed e.g.~by Restrepo et al.~\cite{restrepo2014artificial}, who successfully trained an artifical neural network to predict GB energies by training it with the data collected in \cite{kim2011identification}.
However, also here information concerning position and energy of the cusps was already part of the training data.
Active learning \cite{settles2012active} provides a promising remedy for this drawback.
In contrast to traditional design of experiments approaches, where the sampling design is fixed beforehand, active learning starts with a comparatively small dataset and then successively proposes where to further explore the parameter space, based on the analysis of the existing data, until a learning goal is reached. 
Such a sequential procedure hopefully results in a better detection of important features than mere high-throughput sampling with a regular sampling design. \textcolor{black}{This would be particularly beneficial for the exploration of the inclination subspace of grain boundary energies of LAGBs, which is large and in which the positions of the cusps can not easily be predicted from symmetry arguments.}

Recently,  Kroll et al.~\cite{kroll2022efficient}  have proposed a method along these lines.
It combines a statistical sampling of the parameter space via a sequential design of experiment approach with a Kriging interpolator to estimate the energy function. Using the jackknife variance, the choice of the next point in the sequential design is a compromise between sampling the region of largest  fluctuations and avoiding a clustering of data points. In this way, the cusps of the energy can be found within a small number of iterations and refined as desired.

To turn this approach into an active learning method, one needs to answer the question: 
when is the sequential sampling good enough to stop the atomistic simulations? The answer sounds simple -- when all relevant cusps of the grain boundary energy  have been found and the predicted energies in unsampled regions are accurate enough. However, to express this answer in measurable quantities and implement it in terms of an automated stopping criterion is not equally obvious, for the following reasons:
\begin{itemize}
	\item There is no way to determine a priori the number of cusps, even in a low-dimensional subspace of the 5D parameter space.
	\item There is no way to calculate the absolute error of the predicted values, since the true energy distribution is unknown.
\end{itemize}
Thus, what is needed is a measure for the convergence of the energy prediction, which is based only on the already calculated data, and a definition of a {sufficient number of cusps} to describe all relevant features of the grain boundary energy variation. In this work, we develop  such a measure, which addresses both aspects and use it to define a stopping criterion for a sequential sampling algorithm. 

In addition to the new  stopping criterion for sequential sampling of grain boundary energies, this work extends the methodology of \cite{kroll2022efficient}  for the 1D subspace of symmetrical tilt grain boundaries (energy as a function of misorientation) to the two-dimensional subspace of energies as function of GB plane inclination. This creates the additional challenges of 
how to choose an initial design for the fundamental zone (FZ) of the 2D subspace, and how to properly interpolate the energy on a suitable grid. The fundamental zone is the minimum area, from which the whole inclination subspace can be constructed by symmetry operations such as rotation and reflections. The concept of fundamental zones itself is described in \cite{homer2015grain}.
The different instances of interpolation in 2D will be explained below.

In the following, the active learning algorithm is explained, starting with a general description of the overall procedure in Section \ref{sec21}. 
The different types of  grids used in various steps of the  algorithm in 2D
are introduced in Section \ref{sec:grid_generation} and further illustrated in \ref{app:grids}.
The stopping criterion itself and the difference between its application to 1D and 2D samplings is elucidated in Section \ref{sec:stopping-criterion}. Roughly speaking it monitors two quantities: the development of the energy profile and the number of cusps.
The results part of this paper in Section \ref{sec:results} starts with a  validation of the stopping criterion by post-processing the data of the 1D STGB subspaces from \cite{kroll2022efficient}. Here it is demonstrated  that the criterion makes the sampling more efficient in the sense that  we can achieve the same accuracy with fewer atomistic simulations
(Section~\ref{sec:1D_Results}). In Section \ref{sec:2D_subspaces} 
we demonstrate the advantages of the new algorithm for sampling  2D subspaces of grain boundary plane inclinations. In particular, we investigate in Section~\ref{sec:Delta_E_stat}  the impact of the choice of the energy threshold (i.e., the desired accuracy)  on the quality  of the prediction and the number of necessary steps to reach it.
Section \ref{sec:contribution} analyses how the two  quantities 
monitored by the stopping criterion evolve throughout the sampling
and it is demonstrated that both criteria are indispensable. 
Finally, the active learning  procedure requires an initial design and the 
influence of  its size towards the quality of the sampling is discussed in Section~\ref{sec:effect}.

\section{Methodology}\label{sec:meth}
\subsection{Basic steps of the procedure}  \label{sec21}
\begin{figure}
	\centering
	\begin{subfigure}{0.9\textwidth}
		\centering
		\includegraphics[width=\textwidth]{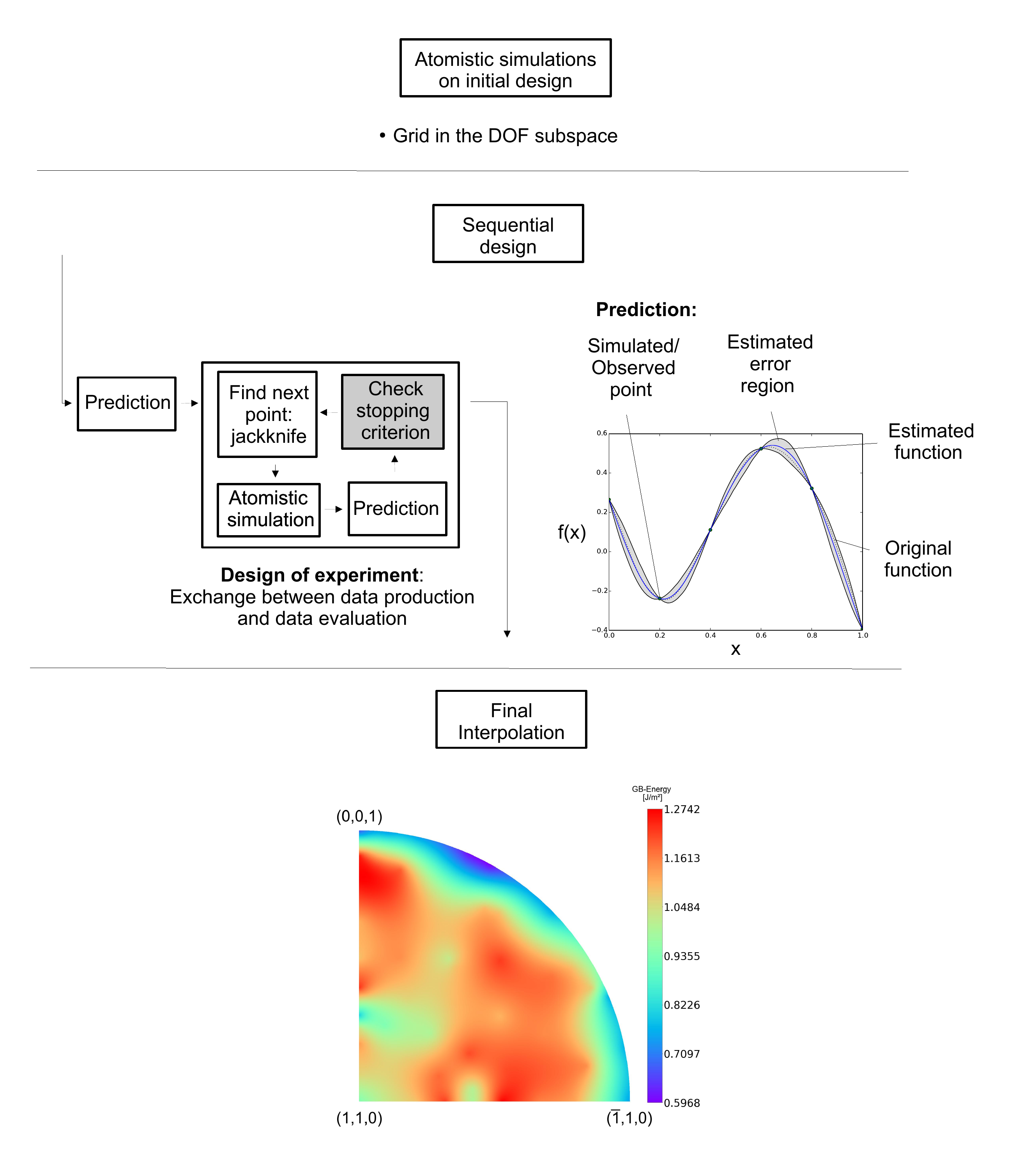}
	\end{subfigure}
	\caption{Flowchart of the overall
		procedure for the sampling of grain boundary energy subspaces. The method consists of three parts: initial design, sequential design and final Kriging interpolation. The stopping criterion (grey box) represents the new part of the algorithm compared to the method proposed in \cite{kroll2022efficient}. The second addition is the extension to 2D energy subspaces of grain boundary inclinations. The example shown for the final interpolation is the fundamental zone of such a subspace for [110]7.5$^{\circ}$ GBs in fcc nickel.} 
	\label{fig:Flowchart}
\end{figure}
The overall sampling approach is schematically shown in Figure \ref{fig:Flowchart}. It consists of three parts, the {initial design}, the {sequential design} step including %
the stopping criterion, and the {final prediction}. The relevant parameters are listed in Table~\ref{table}.
\begin{table}\footnotesize
	\begin{tabular}{ll}
		\multicolumn{2}{l}{\bfseries{Pre-defined parameters}}\\
		\\
		$\Ninit$ & number of initial design points\\
		$\Ncand$ & number of candidate points from which the next point is chosen in each\\
		& sequential step\\	
		$\DeltaEstat$ & 
		threshold to which $\DeltaEprev$ is compared as part of the stopping criterion\\
		$\Niterdelta$ & number of iterations during which $\DeltaEprev$ must stay below the threshold 
		\\ & $\DeltaEstat$ (statistical \textcolor{black}{aspect of the} stopping criterion)\\
		$\Niterdiv$ & number of iterations for which the number of division must not change\\
		&   (topological \textcolor{black}{aspect of the} stopping criterion)\\
		\\
		\multicolumn{2}{l}{\bfseries{Runtime variables}}\\
		\\
		$\Nseq$ & current number of sequential step\\%
		$\Ncusp$ & number of divisions/cusps in the FZ\\
		$\DeltaEprev$ & maximum absolute deviation between Kriging interpolators from two\\
		& consecutive steps\\
		$\DeltaEref$ & in current step, maximum absolute deviation of the Kriging interpolator\\
		& from a reference data set\\
		$\Nstop$ &\makecell[l] {number of iterations required to fullfil the stopping criterion}\\
	\end{tabular}
	\caption{Overview of code-related abbreviations used in the text.}
	\label{table}
\end{table}

In a nutshell, the building blocks of the overall method can be summarised as follows:
\begin{itemize}
	\item[(1)]\label{it:summary:1} The {initial design} 
	defines  $\Ninit$ points 
	in the fundamental zone, for which the corresponding GB energies are calculated from molecular statics, as explained in \ref{sec:atomistics} of the online supplement. These data points can be obtained in a regular high-throughput scheme. 
	\item[(2a)]\label{it:summary:2} The {sequential design} consists of an ongoing sequence of data generation  and data evaluation. Starting with the initial design, the energy distribution in the fundamental zone is predicted by a Kriging estimator (see \ref{app:kriging} for details). Then, in each following iteration the next sampling point is chosen from $\Ncand$ candidate points via a jacknife method (basically a cross-validation method).
	Here we compute 
	for any candidate point the Kriging prediction from the currently available observations and from the reduced sets obtained by deleting exactly one of the current observations at a time.
	The Kriging estimators obtained this way are combined in order to define a jackknife estimate of the uncertainty of the prediction at any of the $\Ncand$ candidate locations, and the next sampling point is then chosen among the maximizers of this estimate.
	Finally, an atomistic simulation is conducted at this new point,
	and an updated Kriging model is fitted to the augmented  data set (previous design + new point).
	We refer to \cite{kroll2022efficient} for more details.
	
	\item[(2b)]
	Next, the validity of the new {\it stopping criterion},  which will be explained in detail in Section \ref{sec:stopping-criterion} below, is checked. If the criterion is satisfied, sequential sampling is terminated, otherwise continued. 
	This means, that, in contrast to \cite{kroll2022efficient}, the number of atomistic simulations, say $\Nseq$, is not fixed  a priori, but determined by the stopping criterion based on the results of the  conducted experiments.
	\item[(3)]\label{it:summary:3} After sequential sampling has stopped, 
	{final prediction} on a dense grid of points is performed via Kriging on the basis of the complete dataset (initial design + sequential design). 
\end{itemize}

In \ref{app:kriging} of the online supplement, we give a brief summary of  Kriging prediction, which is used in steps (2a) and (2b) of the overall procedure
and will be used as a black box method 
in the remaining part of the paper.  We also refer to \cite{rasmussen2006gaussian}, Section \ref{sec:grid_generation}, or \cite{stein1999interpolation} for further details.

The overall procedure extends the one from \cite{kroll2022efficient} mainly in two directions.
\textcolor{black}{
	First, the new methodology is also applicable in the 2D case of grain boundary plane inclinations, in which a prediction of the locations of energy minima simply based on geometric arguments is not possible, but it depends on the atomistic details of the GB plane. The 2D fundamental zone furthermore} requires the definition of suitable sampling grids beyond the 1D case where the definition of regular grids is obvious.
Second, in step (2b) we add the stopping criterion to the overall procedure \textcolor{black}{with the goal} to release us from the task of specifying the number of sequential steps in advance \textcolor{black}{and to turn the statistical sampling into an active learning scheme}.
These two ingredients of the overall procedure will be described explicitly in the following sections.

\subsection{Grid generation}
\label{sec:grid_generation}
All three steps outlined in Section~\ref{sec21} involve Kriging interpolation of the available data which is carried out on a pre-defined grid.
For step~(1) of the algorithm, the initial design, a space-filling $s^2$-grid is used.
This grid is especially suited to achieve a homogeneous distribution of the sampling points and can be motivated from the use of Kriging as the interpolation method of our choice, %
see \ref{sec:s2-grid} of the online supplement for a more detailed discussion.
Such an $s^2$-grid with a larger sample size was also used to define the reference designs for the evaluation of the performance of different predictions. Of course, in a real application these reference designs are not available.
Since the construction of a large $s^2$-grid is rather time consuming, %
a regular equally distant angular grid (see \ref{sec:angular-grid}) is used for the Kriging interpolation in step~(2a) and (2b) of the algorithm, where it is required to evaluate the stopping criterion.
In step~(3), the final interpolation is carried out on a very fine version of this regular equally distant angular grid.
Note that, as the number of points increases, the advantages of the $s^2$-grid become less pronounced, and for a grid with a very high density the regular equally distant angular grids \textcolor{black}{behave similarly}.

Also the candidate points for the jackknife method, which is applied to find the next point during the sequential design step, are supposed to lie on a pre-defined space-filling grid. Here, the space-filling property is desirable to make measurements in all areas of the FZ at least potentially possible. For this purpose, we have taken, mainly for computational reasons, the reduced angular grid (see \ref{sec:reduced-angular-grid}).
In \cite{kroll2022efficient} some heuristics concerning the choice of $\Ncand$ were briefly discussed, but only a fixed value of $\Ncand = 75$ was considered in the simulations.
In this work, after a more detailed investigation of its impact (see \ref{app:Ncand}), $\Ncand$ is increased to 200.

\subsection{Stopping criterion}
\label{sec:stopping-criterion}
\begin{figure}
	\centering
	\includegraphics[width=0.57\textwidth]{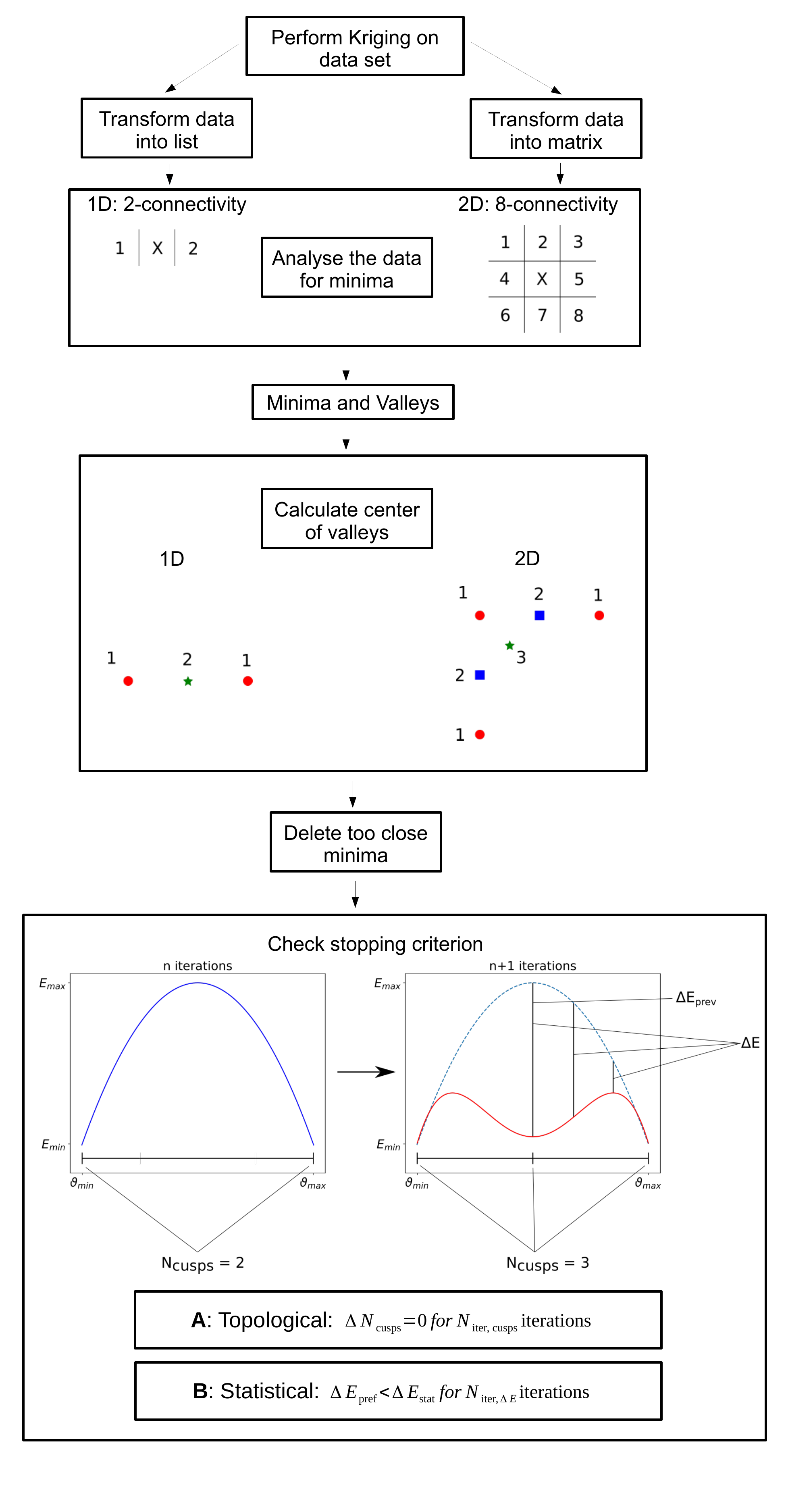}
	\caption{%
		Flowchart illustrating the stopping criterion:
		After Kriging, the interpolated data set is analysed for minima. \textcolor{black}{Well localised} minima are processed directly, while for \textcolor{black}{energy valleys} a central point is determined first. After reducing such minima, which are too close, the topological \textcolor{black}{aspect of the} stopping criterion (expression A) is checked: \textcolor{black}{The number of cusps $\Ncusp$ must not change for $\Niterdiv$ iterations.} For the statistical one, \textcolor{black}{the maximum energy difference between the data points in the current and previous step, $\DeltaEprev$, must be $< \DeltaEstat$ for $\Niterdelta$ iterations (expression B). A pseudo code describing the procedure can be found in \ref{app:pseudo_code}.}}
	\label{fig:matrix}
\end{figure}
In this section we develop a stopping criterion which combines a topological aspect with a statistical one. 
Before it is applied, the available data is interpolated by performing Kriging on a very fine grid with equally distant points. In the 1D subspace of symmetrical tilt grain boundaries, this is a 1D grid of equally distant misorientation angles. In the 2D space that defines the boundary plane normal it is a 2D grid of equally distant polar and azimuthal angles (the regular equally distant angular grid, see Section \ref{sec:angular-grid}.)
To check the topological aspect of the stopping criterion, the cusps in the energy landscape are identified by comparing the energies of neighbouring points. This requires a list of neighbours (defined by the misorientation angle) for the 1D case and a matrix of neighbours (defined by the azimuthal angle in the column and the polar angle in the row) as indicated in Figure \ref{fig:matrix}.

In 1D subspaces, each element of the list is analysed by checking the previous and the following element (2-connectivity).
If both neighbouring elements have a higher energy then the analysed element is classified as a minimum (cusp). If several neighbouring elements have the same energy \textcolor{black}{within $10^{-5}\si{J}/\si{m}^2$} and are surrounded by elements with a higher energy, they form what is called a valley. To define the actual position of the energy minimum, the centre of the valley is calculated, i.e.~the position of the cusp is chosen as the mean value of the misorientation angles of the elements in the valley. 

\textcolor{black}{To identify energy minima in the 2D energy subspaces,  the energy of each point is compared with the energy 
	of its eight neighbours (8-connectivity).
	If all neighbours of the  point  have a strictly higher energy than the point itself, the point is classified as a minimum}.
Extended regions of low energy, the above mentioned valleys, can occur in the 2D case as well and their centres have to be identified. For this purpose a recursive method is applied, which is also illustrated  in Figure \ref{fig:matrix} (calculate centre
of valleys). The middle points between each element of a valley and its nearest neighbours (if still part of the valley) are added to a new list of reduced number of points. This will be repeated with every element in the new list to create a further reduced list. The process repeats until the list only contains one element, which will be the centre of the valley. 

After identification of all minima, the minima which are closer to each other than 
2$^{\circ}$ for 1D and 2D are reduced to the minimum with the lowest energy. %
Here, for 2D subspaces the angular distance is calculated as
\begin{equation}
	\Delta \alpha = \sqrt{\Delta \varphi^2 + \Delta \vartheta^2}.
\end{equation}
\textcolor{black}{The threshold $2^{\circ}$ leads to the fact that no cusps that are closer to each other than $2^{\circ}$ can be distinguished. It can thus be interpreted as the level of resolution, which can be adapted depending on the user's requirements.}

Finally, the stopping criterion is checked, i.e., whether the number of cusps is constant with respect to the previous $\Niterdiv$ steps. 

The statistical aspect of the stopping criterion is defined in a similar way: The change in energy at each point compared to the previous sequential step is calculated. If the maximum difference of the energy towards the previous iterations in the whole subspace is lower than a threshold $\DeltaEstat$ for $\Niterdelta$ iterations \textcolor{black}{in a row}, also the statistical aspect of the stopping criterion is fulfilled. If both subcriteria are met, the overall criterion itself is fulfilled and the sampling stops. A Voronoi tesselation can be applied to the cusps to divide the subspaces into cells. Strictly speaking only the maximum difference in the overall subspace is needed to evaluate energetic aspect, but the calculation of the maximum difference in each Voronoi cell provides a closer look on what is happening in the cells while sampling the subspace.

To summarise, both the overall energy as well as the number of cusps are monitored. Convergence is reached once both, the energy and the number of cusps are stable for a certain number of iterations. 
In this work, $\Niterdelta$ and $\Niterdiv$ are chosen equal to $3$.

\section{Results}\label{sec:results}

\subsection{$1D$ STGB subspaces}
\label{sec:1D_Results}
Recently it was demonstrated in \cite{kroll2022efficient}  that a sequential design is able to identify the cusps in the 1D energy landscape of symmetrical tilt grain boundaries in body-centred cubic iron. 
However, in the cited paper the optimal number of sequential steps was determined {a posteriori} by analysing the maximum error with respect to a reference database. %
In the following, the same data is used to validate the new stopping criterion.

\begin{figure}[H]
	\centering
	\caption*{(a) $\delta  = 0.0$}
	\begin{subfigure}{0.45\textwidth}
		\centering
		\includegraphics[width=\textwidth]{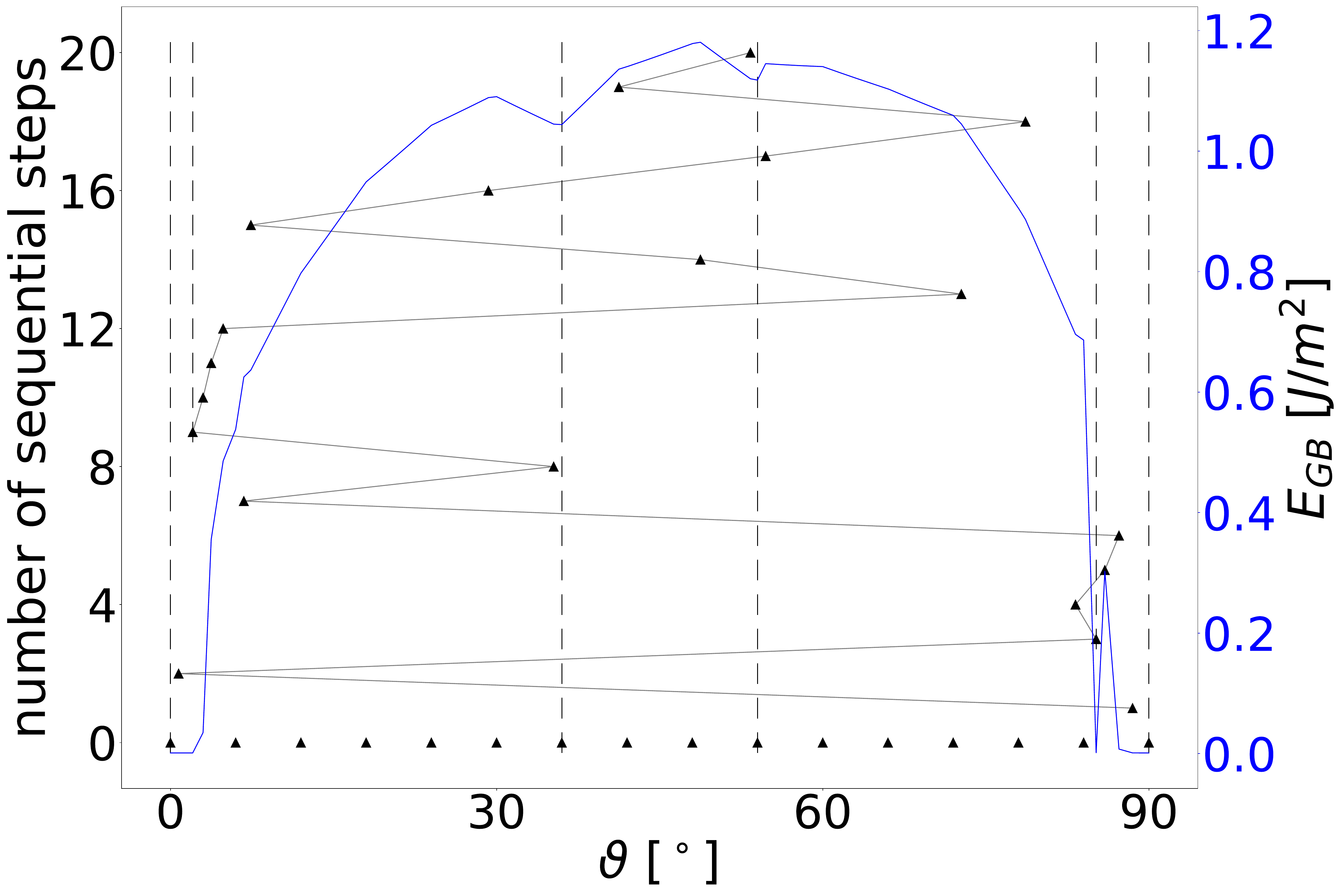}     
	\end{subfigure}
	\begin{subfigure}{0.45\textwidth}
		\centering
		\includegraphics[width=\textwidth]{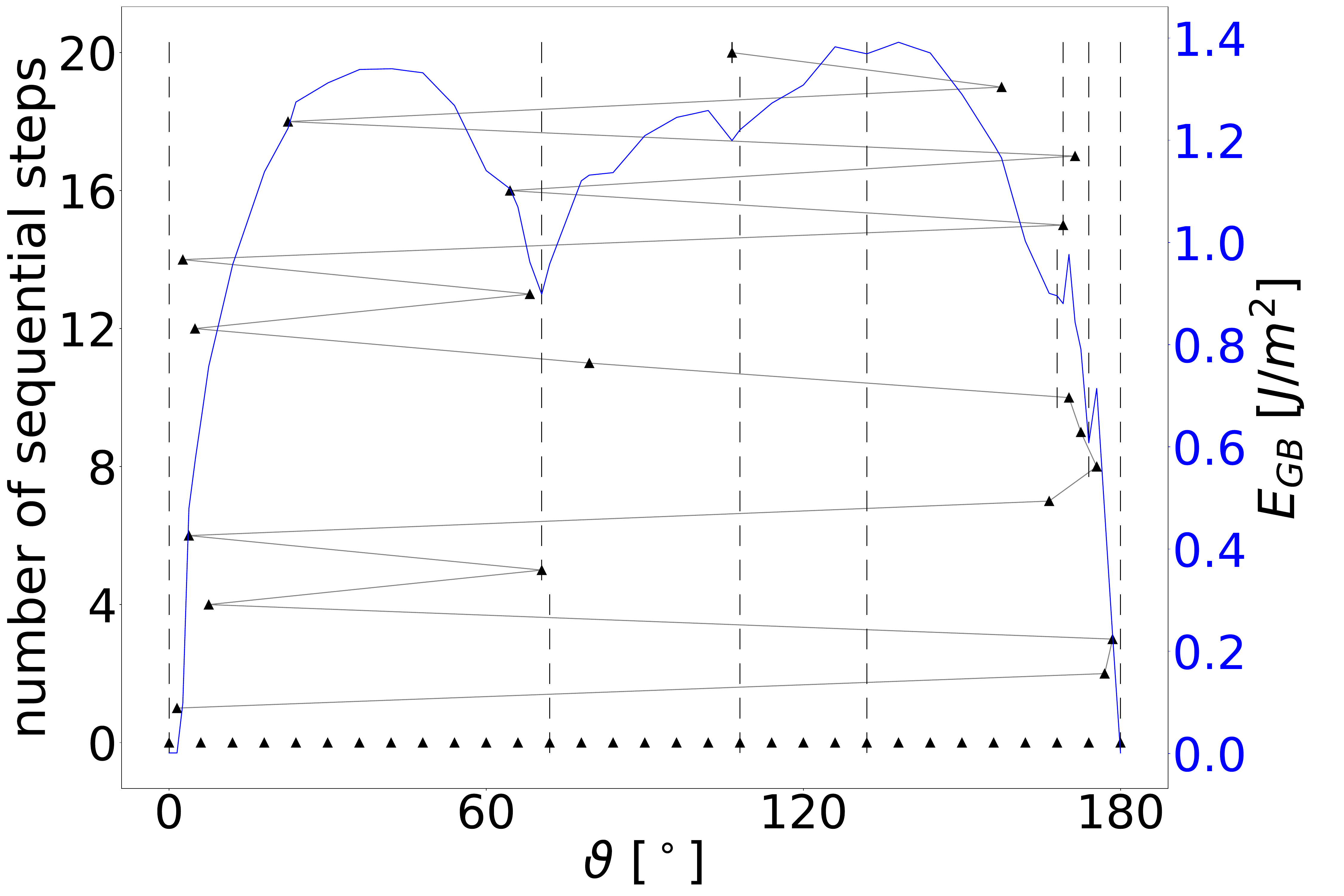} 
	\end{subfigure}
	\caption*{(b) $\delta  = 0.1$}
	\begin{subfigure}{0.45\textwidth}
		\centering
		\includegraphics[width=\textwidth]{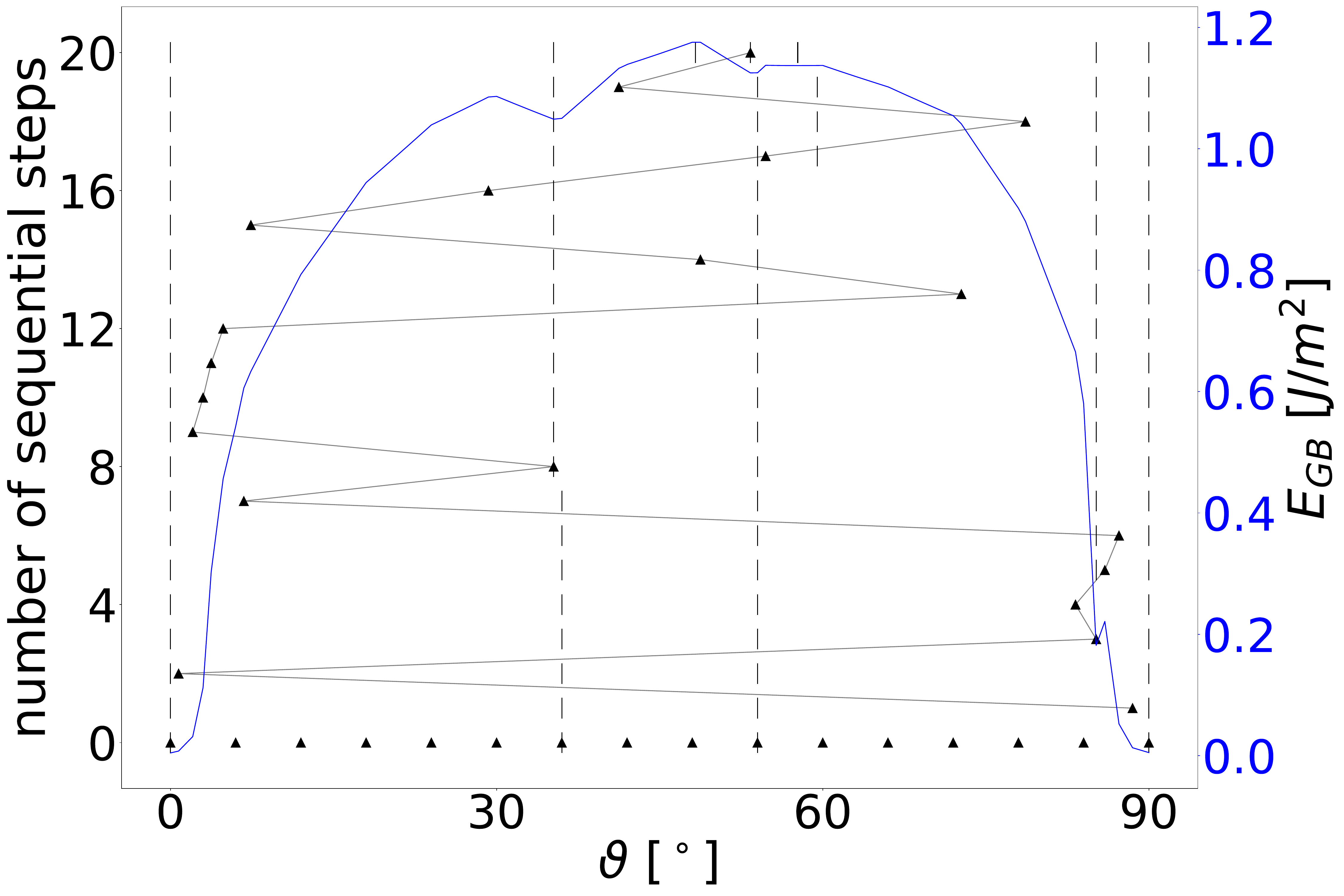}     
	\end{subfigure}
	\begin{subfigure}{0.45\textwidth}
		\centering
		\includegraphics[width=\textwidth]{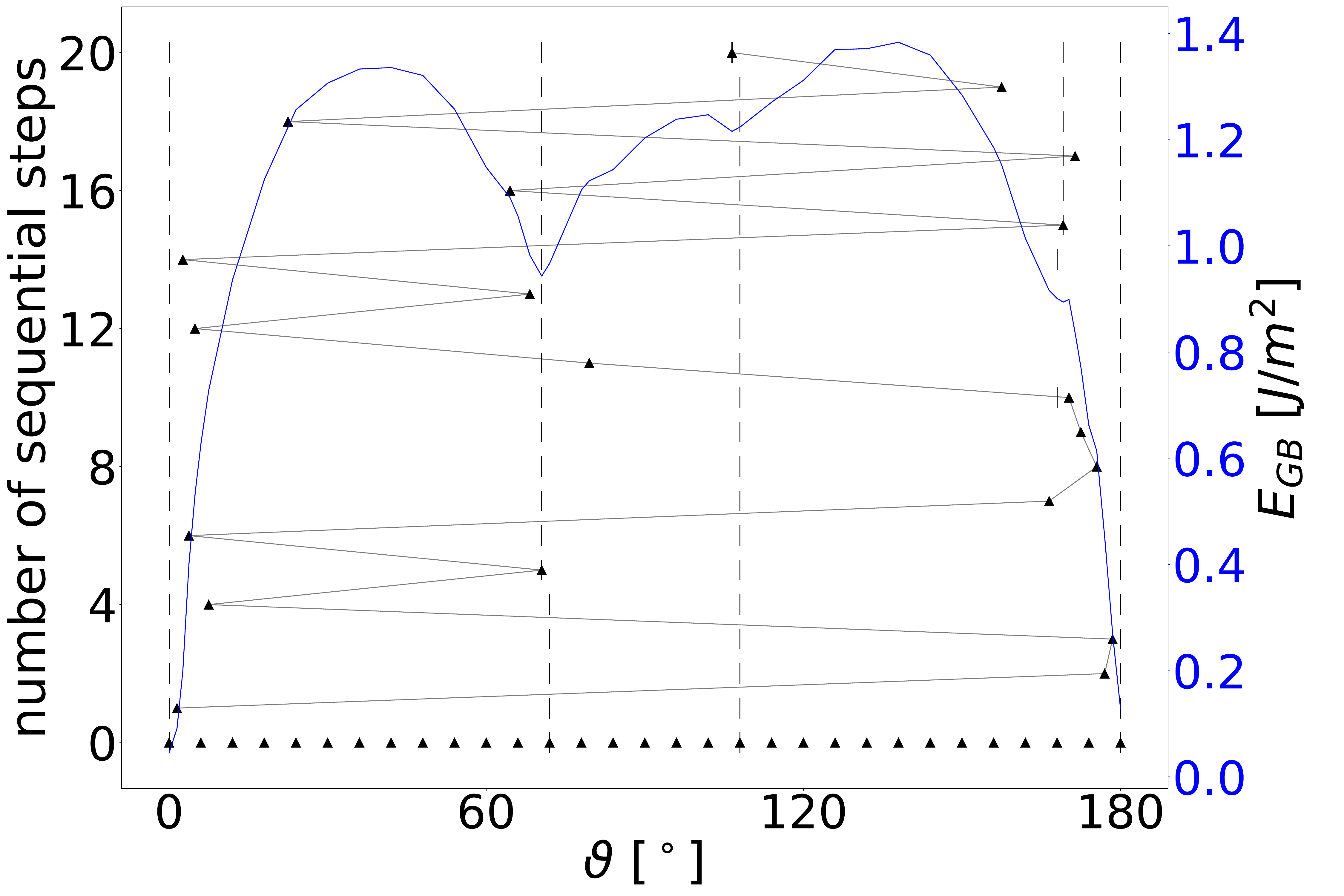} 
	\end{subfigure}
	\caption*{(c) $\delta  = 0.3$}
	\begin{subfigure}{0.45\textwidth}
		\centering
		\includegraphics[width=\textwidth]{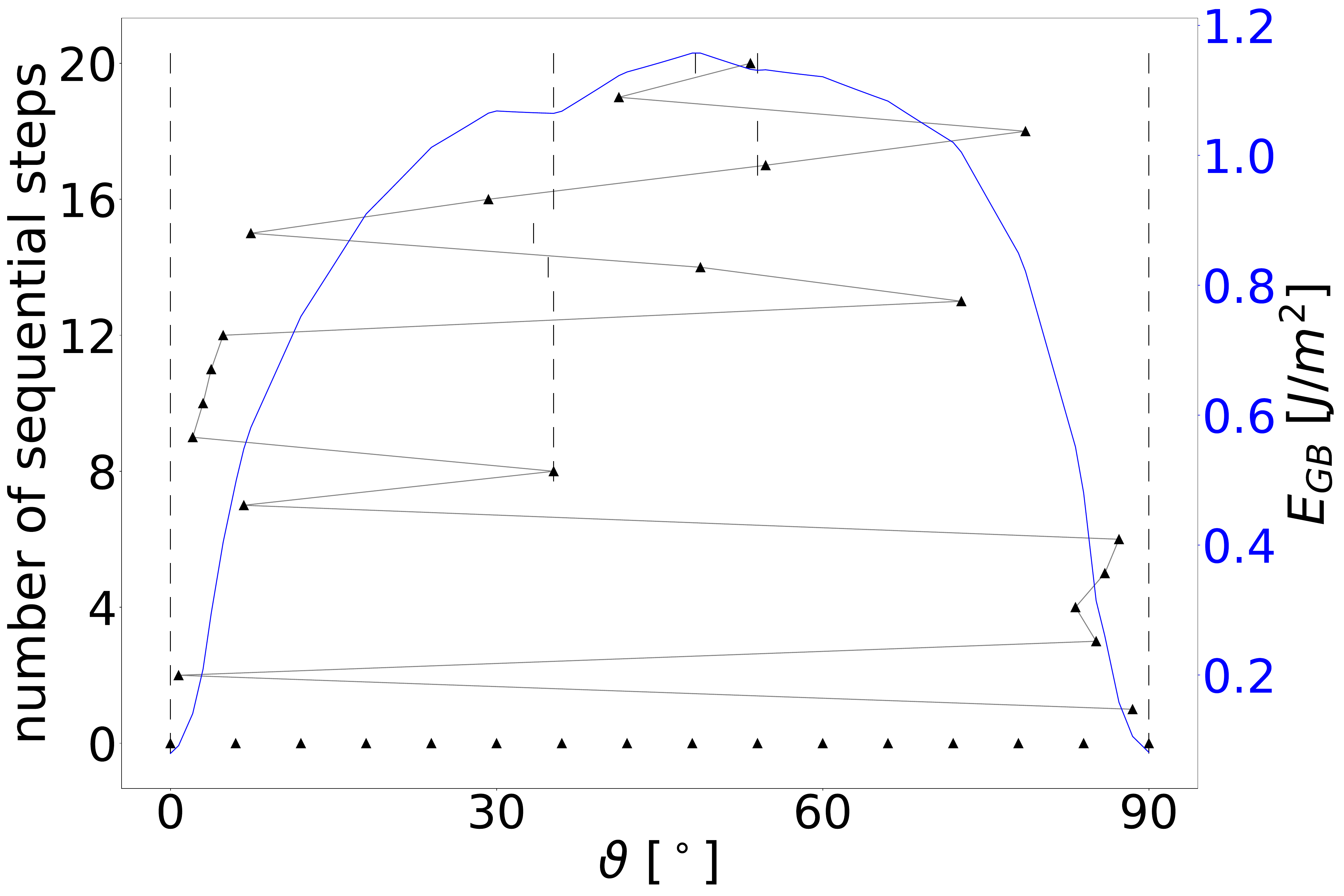}     
	\end{subfigure}
	\begin{subfigure}{0.45\textwidth}
		\centering
		\includegraphics[width=\textwidth]{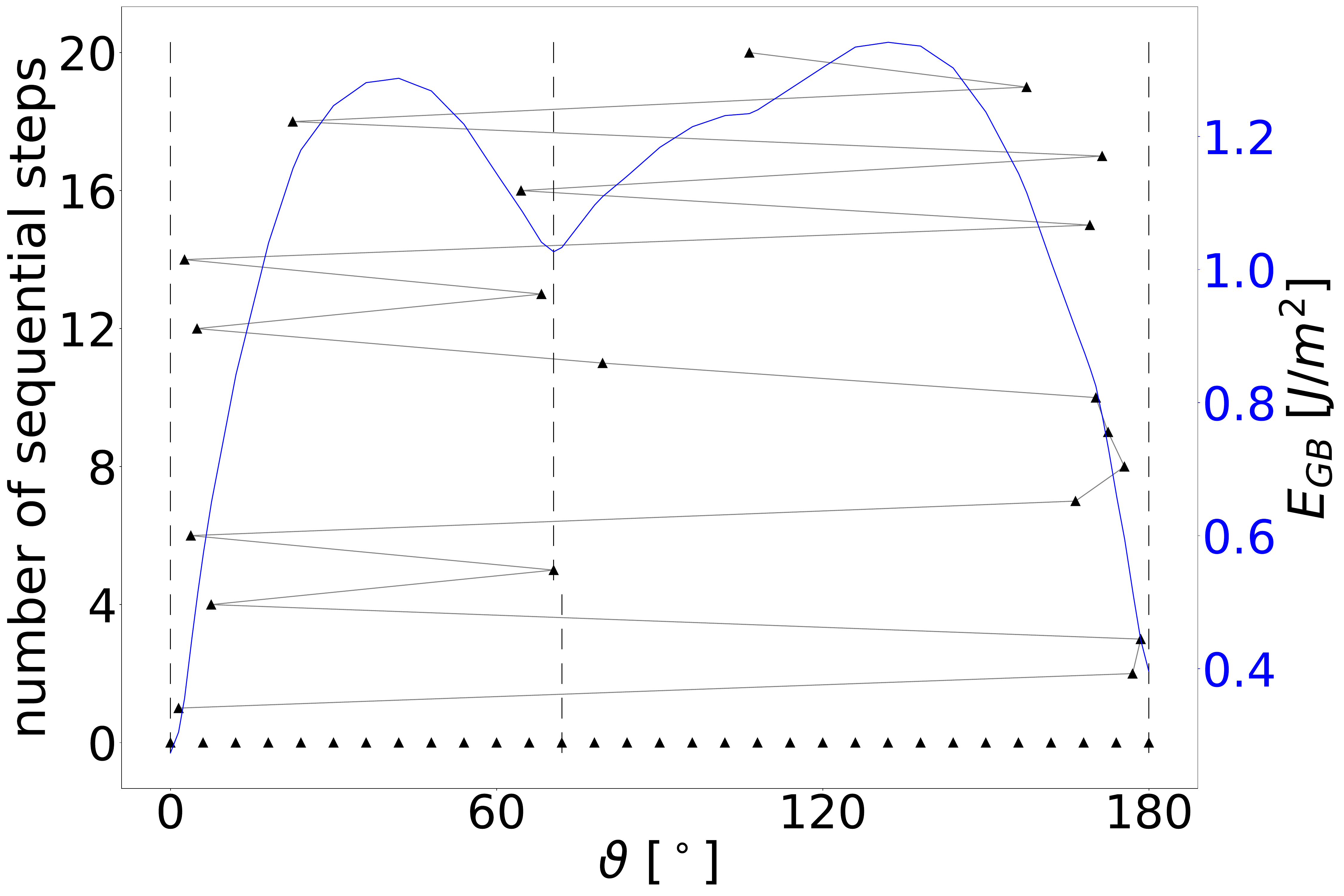} 
	\end{subfigure}
	\caption{Evolution of the sequential design for the $[100]$ 
		(left panel) and $[110]$ STGB subspace (right panel). \textcolor{black}{Kriging is performed in each step with a $\delta$ value of (a) 0.0, (b) 0.1, and (c) 0.3, respectively.} 
		$x$-axis: misorientation angle; left $y$-axis:  current sequential step; right $y$-axis:   energy. The solid blue line is the grain boundary energy as a function of misorientation after the final sampling step. The \textcolor{black}{$\blacktriangle$} on the $x$-axis display the initial design, the other \textcolor{black}{$\blacktriangle$}
		indicate the positions of the new design points calculated by 
		the sequential algorithm. The vertical dashed lines mark the positions of the cusps (they start at the sequential step where a new cusp was discovered).}
	\label{fig:division_boundaries_1D}
\end{figure}

\textcolor{black}{The topological aspect of the stopping criterion is illustrated in Figure~\ref{fig:division_boundaries_1D} for the $[100]$ and $[110]$ STGB subspaces.
	Here, we used the Kriging estimate with an additional parameter $\delta$ (see Appendix \ref{app:kriging}), which determines whether the predicted energy landscape interpolates the simulated data exactly ($\delta =0$) or not ($\delta >0$). As a result, for larger values of $\delta$ the predicted function becomes more smooth. In Figure~\ref{fig:division_boundaries_1D} the parameter  $\delta$ was chosen as (a) $0.0$, (b) $0.1$, and (c) $0.3$, respectively}.
On the $x$-axis of Figure~\ref{fig:division_boundaries_1D}, the locations of the initial design points are displayed.
Going up along the $y$-axis, the evolution of the positions of the cusps 
(vertical dashed lines) and the location of the sequentially chosen design points ($\blacktriangle$) and initial chosen design points ($\blacktriangle$ located at the $x$-axis) can be tracked. For example, for  the [100]-subspace the algorithm chooses 
the $88.5^\circ$ misorientation angle in the first iteration, $0.75^\circ$ in the second, $85.15^\circ$ in the third and so on.
It can be seen that the sequential algorithm allocates a large number of design points in neighbourhoods of the (unknown) cusps. Moreover, it does not select new sampling locations from the same region over several iterations but rather visits neighbourhoods of other cusps. In addition, the plots illustrate that the number of cusps increases with the number of sequential steps performed and finally converges.
For example, for the [110] subspace \textcolor{black}{and $\delta=0$} the algorithm starts with $5$ cusps, and after $8$ and $10$ iterations it detects $6$ and $7$ cusps, respectively. \textcolor{black}{Furthermore, the comparison shows that an increasing value of $\delta$ does not only lead to a smoother energy function but also to a smaller number of detected cusps. For instance, 7 cusps are found after 20 iterations for the $[110]$ STGB subspace if $\delta=0$ is chosen,  but only $5$ 
	and $3$ cusps are found for $\delta = 0.1$  and  $\delta = 0.3$, respectively. Note that in the latter case, the energy function is so smooth that the important cusps at $109.47^\circ$ is hardly visible anymore.
	Therefore, this value can be considered as an upper bound. The difference between the number of cusps for $\delta = 0.0$ and $\delta = 0.1$ shows that also a lower bound is a reasonable choice. In \cite{kroll2022efficient} it is described that for misorientations close to the edges of the fundamental zone the atomic structure can relax to those of the boundary structures, which effects the energy to shrink to $0~\si{mJ}/\si{m}^2$ and results in the spiky shape of the energy curve at the edges. This phenomenon is not present in the estimated function  if $\delta $ is chosen sufficiently large. In summary, using a positive value of $\delta$ improves the prediction, as long as oversmoothing of the energy curve caused by taking $\delta$ too large is avoided.}

\begin{figure}[H]
	\centering
	\begin{subfigure}{0.49\textwidth}
		\centering
		\label{fig:1D:max:error:100}
		\includegraphics[width=\textwidth]{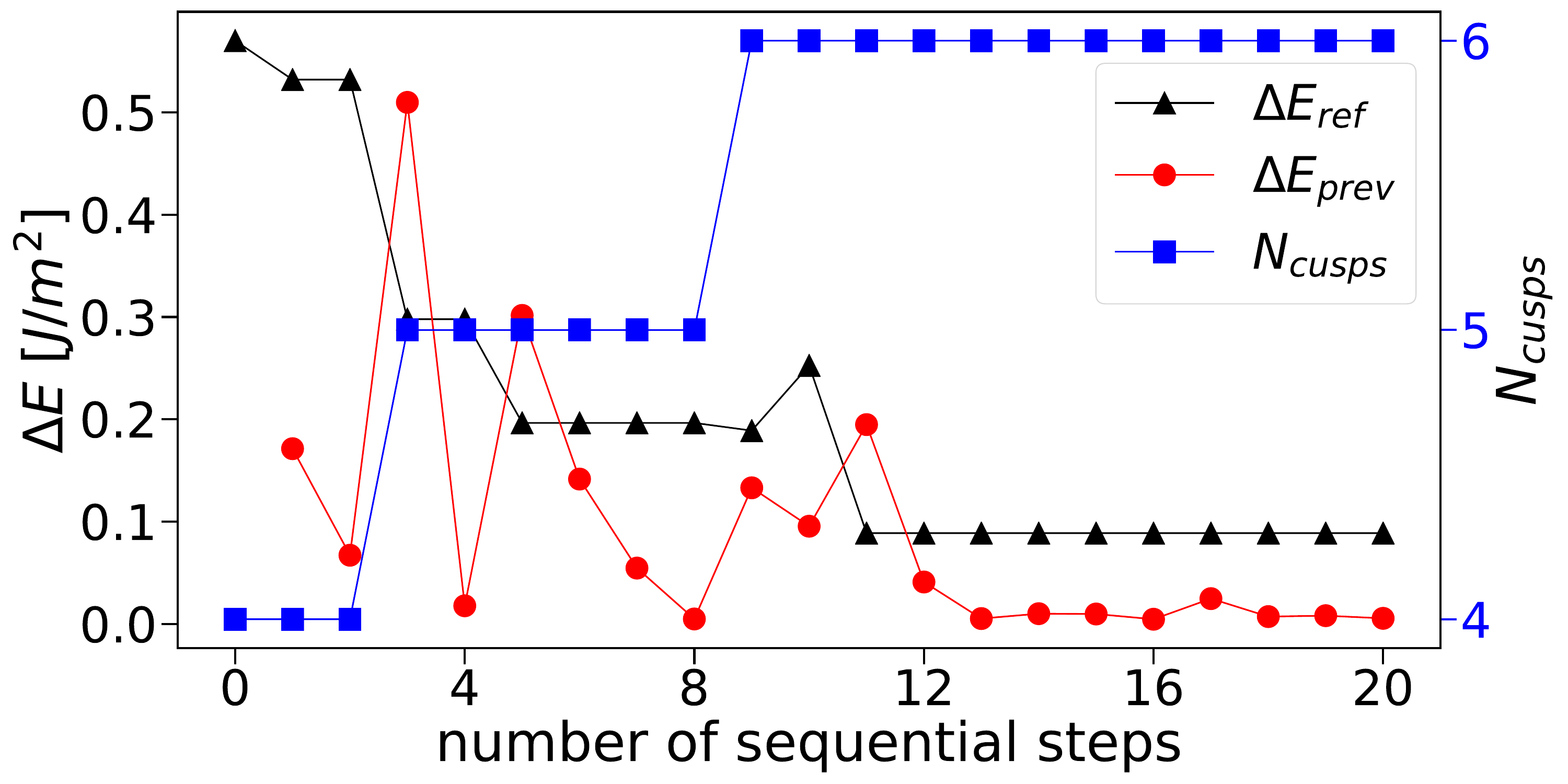}
	\end{subfigure}
	\begin{subfigure}{0.49\textwidth}
		\centering
		\includegraphics[width=\textwidth]{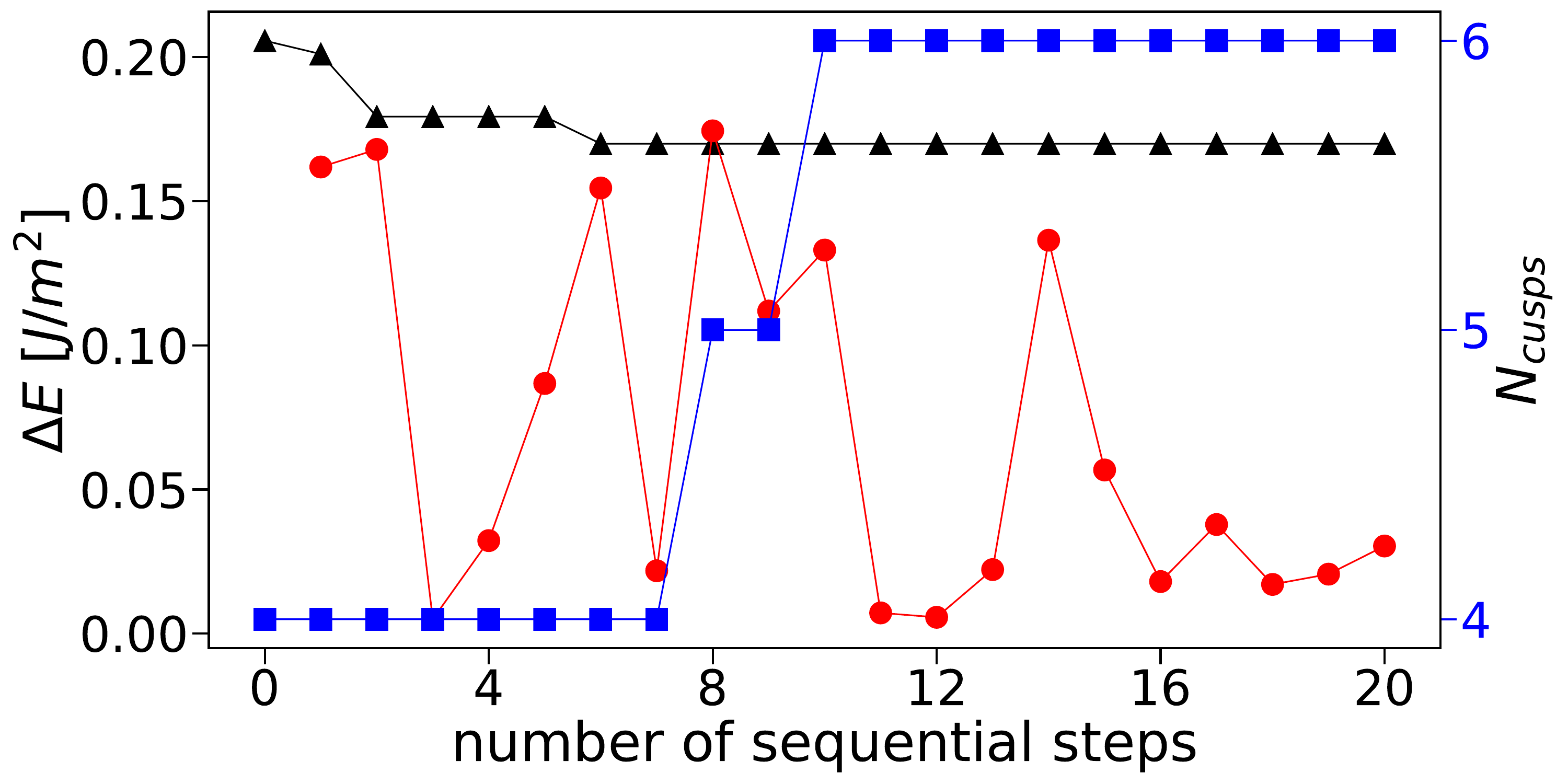}   
	\end{subfigure}
	\caption{
		Evolution of the two contributions to the stopping criterion: maximum absolute error with respect to  the previous sequential step
		(left $y$-axis, \textcolor{black}{\textbullet})
		and the number of cusps, $\Ncusp$ (right $y$-axis, \textcolor{blue}{$\blacksquare$}).
		For the sake of comparison the maximum absolute error  with respect to  a reference database %
		(left $y$-axis, $\blacktriangle$) 
		is also displayed.
		Left panel: $[100]$ subspace with $\Ninit = 16$; Right panel: $[110]$ subspace with $\Ninit = 31$. \textcolor{black}{Analogue plots for alternative error measures in place of the maximum absolute error are included in \ref{app:alt_error}.}}
	\label{fig:STGB_max_error_comparison}
\end{figure}
In \cite{kroll2022efficient}, the quality of sampling was evaluated by the maximum error of the Kriging interpolator with respect to a reference data set. This was done for %
evaluation
purposes.
In a practical application, i.e.,
when sampling a completely unexplored subspace, such a reference data set does of course not exist.
Thus, in this work, the maximum error with respect to the previous iteration, $\DeltaEprev$, is introduced as an alternative error measure, which can be computed from the observed data only and is monitored by the stopping criterion.

Figure~\ref{fig:STGB_max_error_comparison} shows the development of 
this error measure  during the  sequential sampling  (red bullets; the size of the error can be determined from the {left} $y$-axis). Similarly, we display the development of the number of cusps/minima (blue squares; the number of cusps can be determined from {right} $y$-axis). Moreover, the error with respect to the reference database is represented by the black triangles for the sake of a qualitative comparison.
For example, for the $[100]$-subspace we observe from the left part of Figure~\ref{fig:STGB_max_error_comparison} that after $5$ iterations $\DeltaEprev \approx 0.3 \si{J}/\si{m}^2$ and $5$ cusps have been detected.

The comparison of the evolution of $\DeltaEprev$ and $\Ncusp$ with $\DeltaEref$ shows the strength of the new criterion.  While the latter keeps decreasing, $\DeltaEprev$ increases again when a new cusp has been found, and even if $\Ncusp$ remains constant, the sequential design continues until the desired threshold for  $\DeltaEprev$ is reached. For instance, 
in the [110] and the [111] subspaces $\DeltaEref$ remains constant for several iterations (after $6$ iterations for [110] and $2$ iterations for [111]), while $\DeltaEprev$ still varies \textcolor{black}{(indicating that new observations still affect the Kriging estimate)}.
At the same time, Figure~\ref{fig:STGB_max_error_comparison} also illustrates the importance of the topological part of the stopping criterion. \textcolor{black}{In the [110] subspace (right figure) the number of cusps increases in iterations $8$ and $10$, which is accompanied by an increase in $\DeltaEprev$ in both cases.} 
Nevertheless, monitoring only $\DeltaEprev$ is also not sufficient. To see this, consider the [100] subspace and note that after $9$ iterations $\DeltaEprev$ is lower than the threshold $\DeltaEstat = 0.15 \si{J}/\si{m}^2$ for several iterations, but the number of cusps still increases, which means that the stopping criterion is not yet fulfilled here.
\begin{figure}[H]
	\centering
	\includegraphics[width=\textwidth]{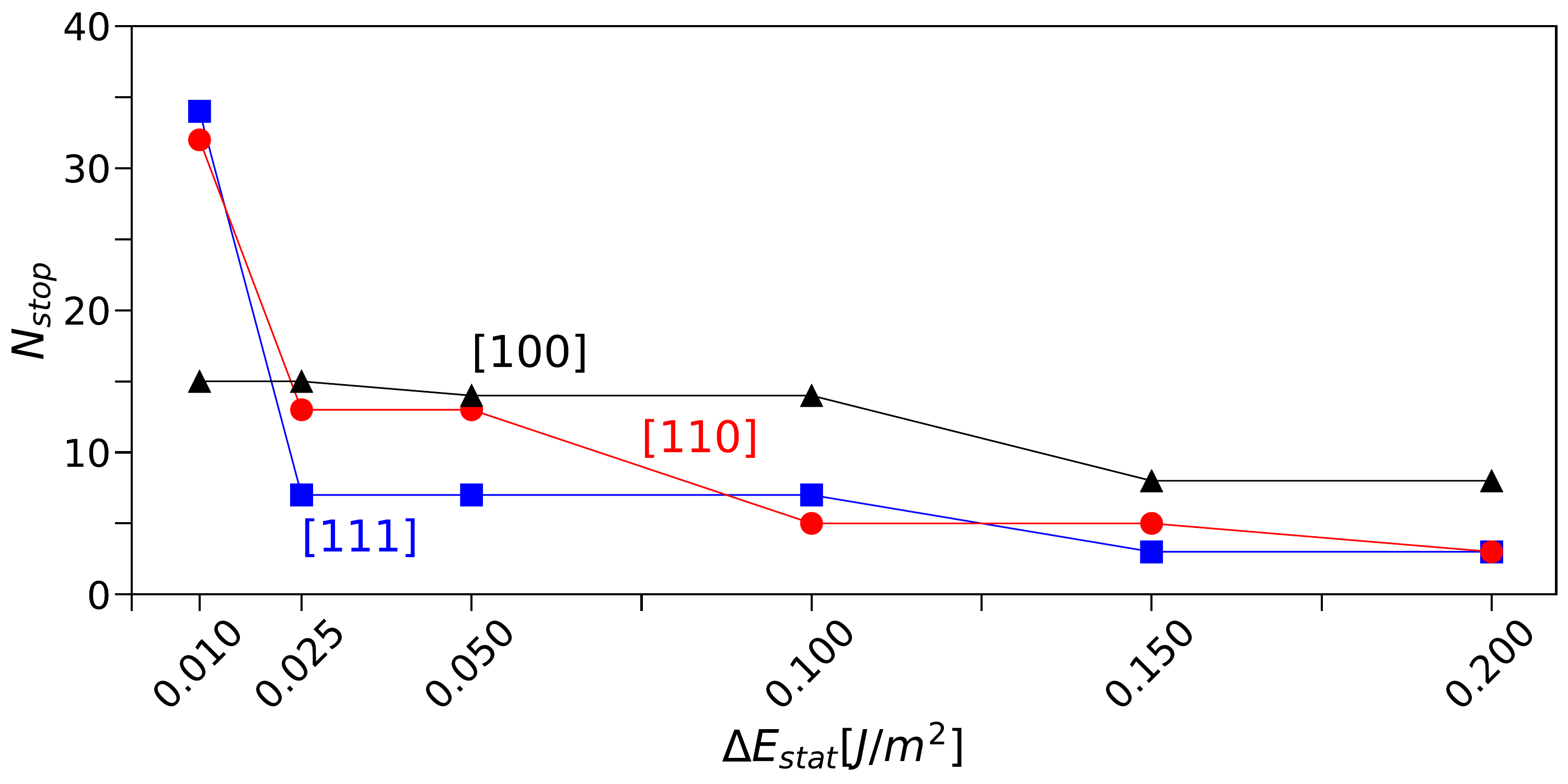}
	\caption{
		Number  of sequential steps ($\Nstop$) required until the algorithm terminates for different values of $\DeltaEstat$ that specify the desired accuracy (measured by $\DeltaEprev$ in each iteration).
		$\blacktriangle$ mark the $[100]$ subspace with $\Ninit = 16$, \textcolor{black}{\textbullet} the $[110]$ subspace with $\Ninit = 31$ and \textcolor{blue}{$\blacksquare$} the $[111]$ subspace with $\Ninit = 21$.}
	\label{fig:STGB_N_stop_3_3}
\end{figure}
In Figure~\ref{fig:STGB_N_stop_3_3} the impact of the required accuracy $\DeltaEstat$ on the stopping criterion is studied for three 1D subspaces (STGB subspaces with a fixed rotation axis of $[100]$, $[110]$ and $[111]$). 
More precisely, for various values of the input parameter $\DeltaEstat$ the figure displays the number of sequential steps, denoted by $\Nstop$, which is required until the algorithm terminates.
Note that the algorithm eventually stops for any choice of $\DeltaEstat$ and for any subspace. For example, for a statistical accuracy of $\DeltaEstat= 0.1 \si{J}/\si{m}^2$ the algorithm stops sequential sampling after $14$, $8$, and $6$ iterations for the $[100]$, $[111]$, and $[110]$ subspaces, respectively.
Clearly, the required number of iterations is a decreasing function of the  desired accuracy $\DeltaEstat$.

The benefit of the stopping criterion becomes clear when the point of termination and resulting accuracy is compared to the empirically chosen number of sequential iterations in \cite{kroll2022efficient}. In that work, the number of iterations was set to $20$, which  corresponds to a sampling where  $\DeltaEstat$ is not more than $0.025 \si{J}/\si{m}^2$ for the [110] and [111] subspaces  and not more than  $0.010 \si{J}/\si{m}^2$ for the [100] subspace.
On the basis of the new stopping criterion the algorithm terminates 
sampling much earlier and still achieves the same precision.
In this regard, the stopping criterion is not only a tool to automatise, but also to optimise the sampling procedure.

\subsection{$2D$ inclination subspaces}
\label{sec:2D_subspaces}
In this section the algorithm with the new stopping criterion is applied to  $2$D inclination subspaces. 
Different samplings of the inclination space of the $\Sigma 3[111]60^\circ$ grain boundaries in bcc Fe, as well as the of the $\Sigma 5[100]36.87^\circ$ and $\Sigma 7[111]38.21^\circ$ boundaries in fcc Al and $[110]7.5^\circ$ boundaries in fcc Ni are considered.

\begin{figure}[H]
	\centering
	\begin{subfigure}{0.25\textwidth}
		\centering
		\caption*{Reference}
	\end{subfigure}
	\begin{subfigure}{0.25\textwidth}
		\centering
		\caption*{Sequential sampling}
	\end{subfigure}
	\begin{subfigure}{0.25\textwidth}
		\centering
		\caption*{High-throughput}
	\end{subfigure}
	
	\caption*{(a) $\Sigma 3$}
	
	\begin{subfigure}{0.25\textwidth}
		\centering
		\includegraphics[width=\textwidth]{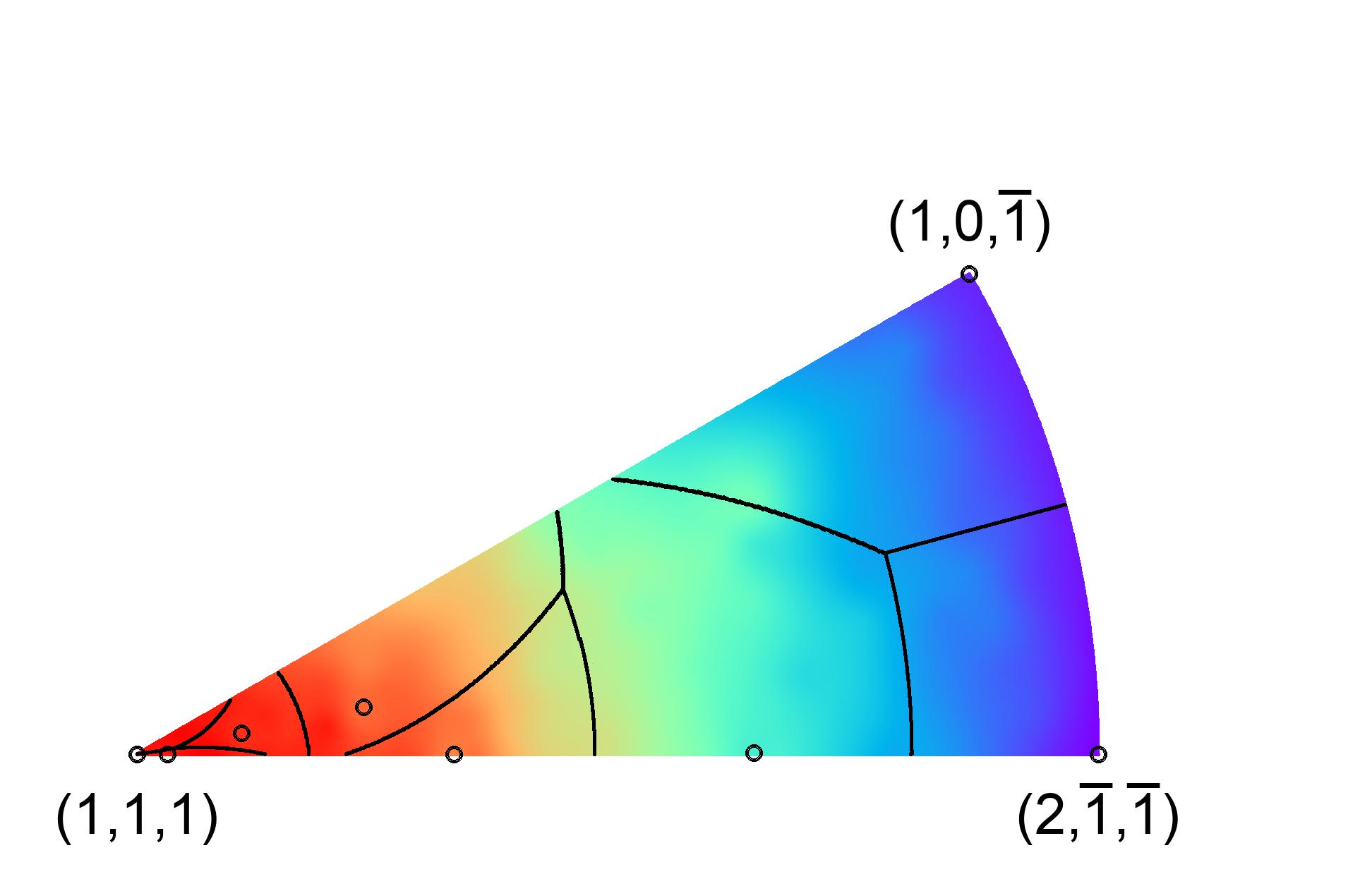} 
	\end{subfigure}
	\begin{subfigure}{0.25\textwidth}
		\centering
		\includegraphics[width=\textwidth]{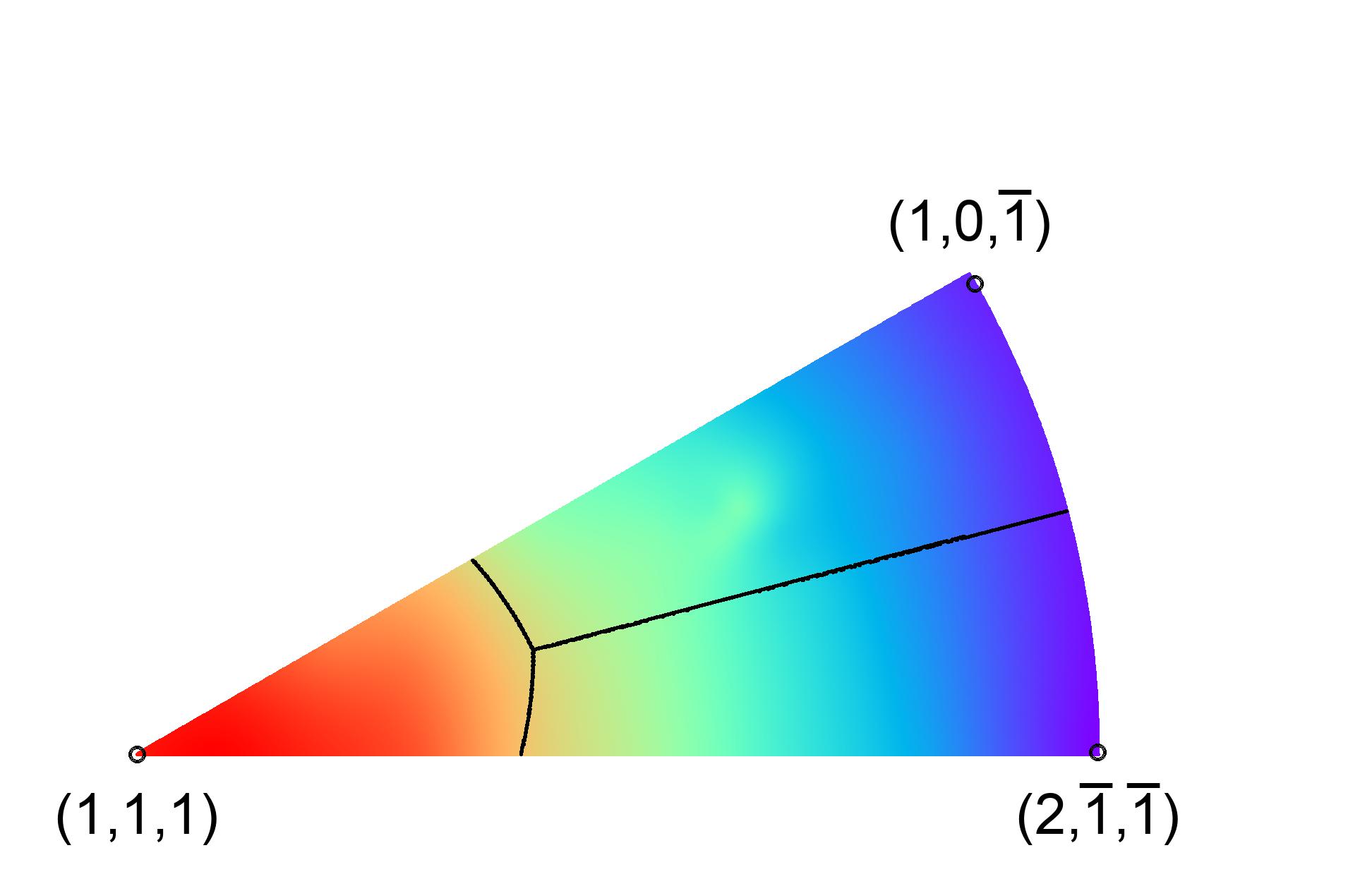}     
	\end{subfigure}
	\begin{subfigure}{0.25\textwidth}
		\centering
		\includegraphics[width=\textwidth]{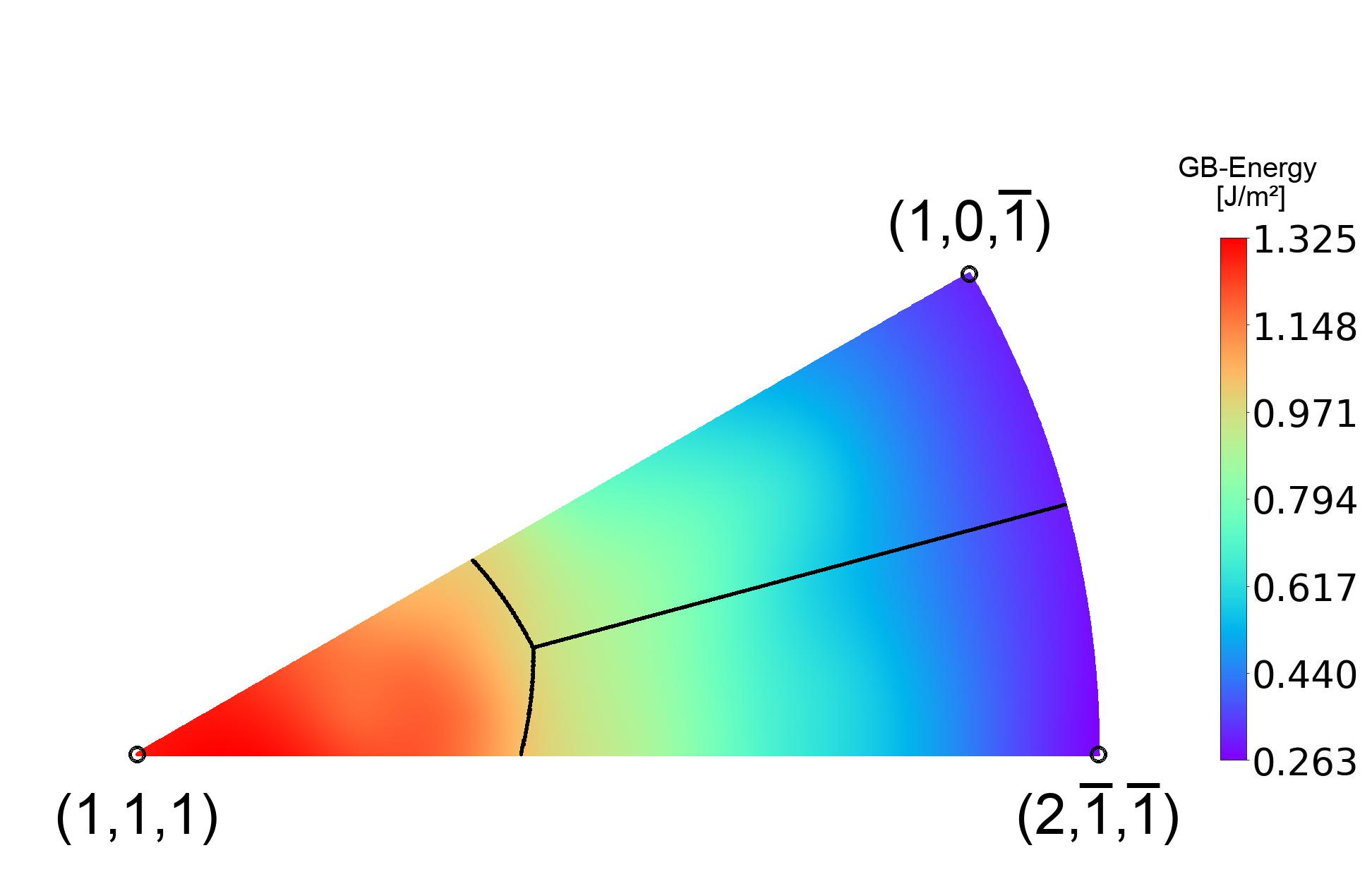} 
	\end{subfigure}
	
	\caption*{(b) $\Sigma 5$}
	
	\begin{subfigure}{0.25\textwidth}
		\centering
		\includegraphics[width=\textwidth]{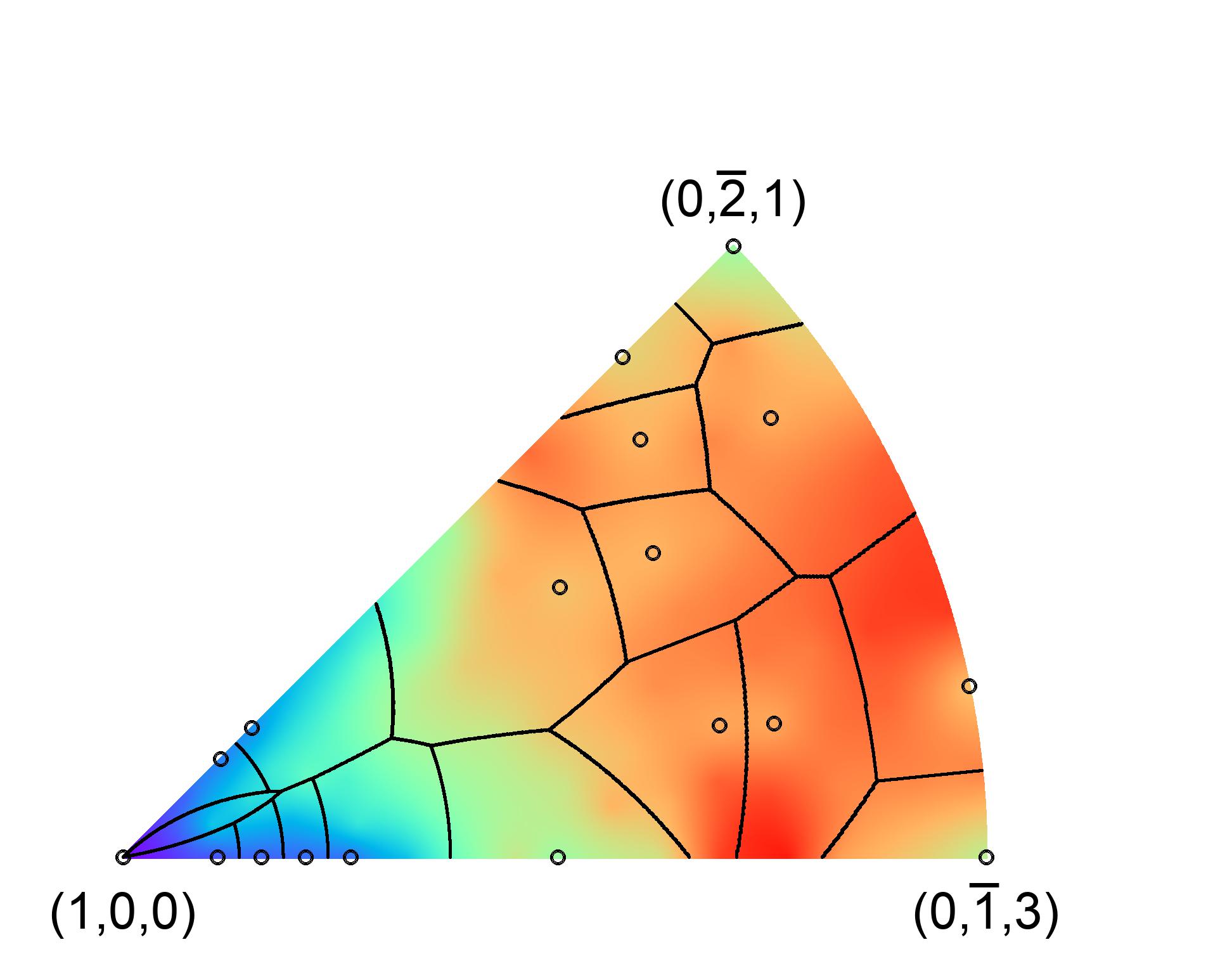} 
	\end{subfigure}
	\begin{subfigure}{0.25\textwidth}
		\centering
		\includegraphics[width=\textwidth]{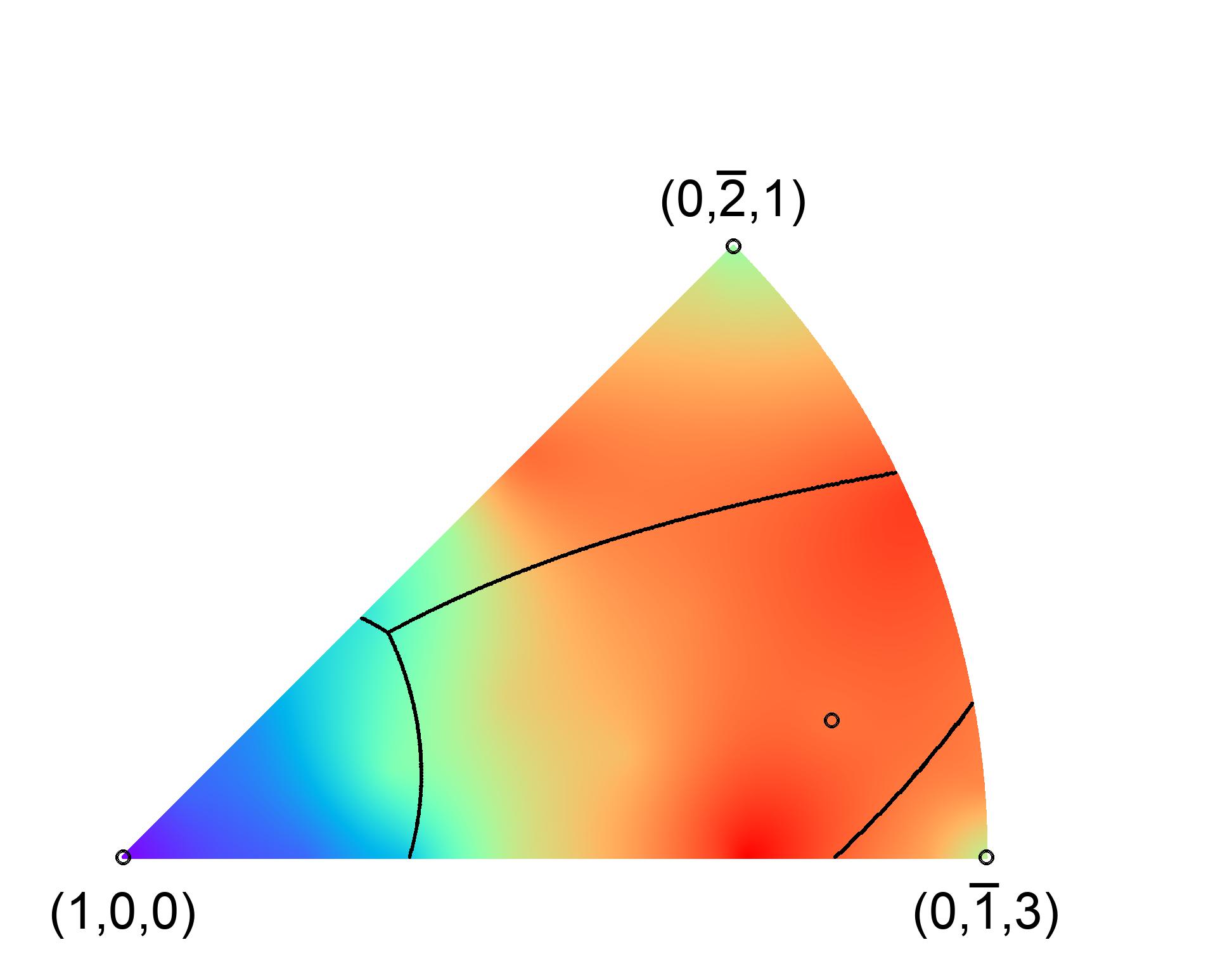}     
	\end{subfigure}
	\begin{subfigure}{0.25\textwidth}
		\centering
		\includegraphics[width=\textwidth]{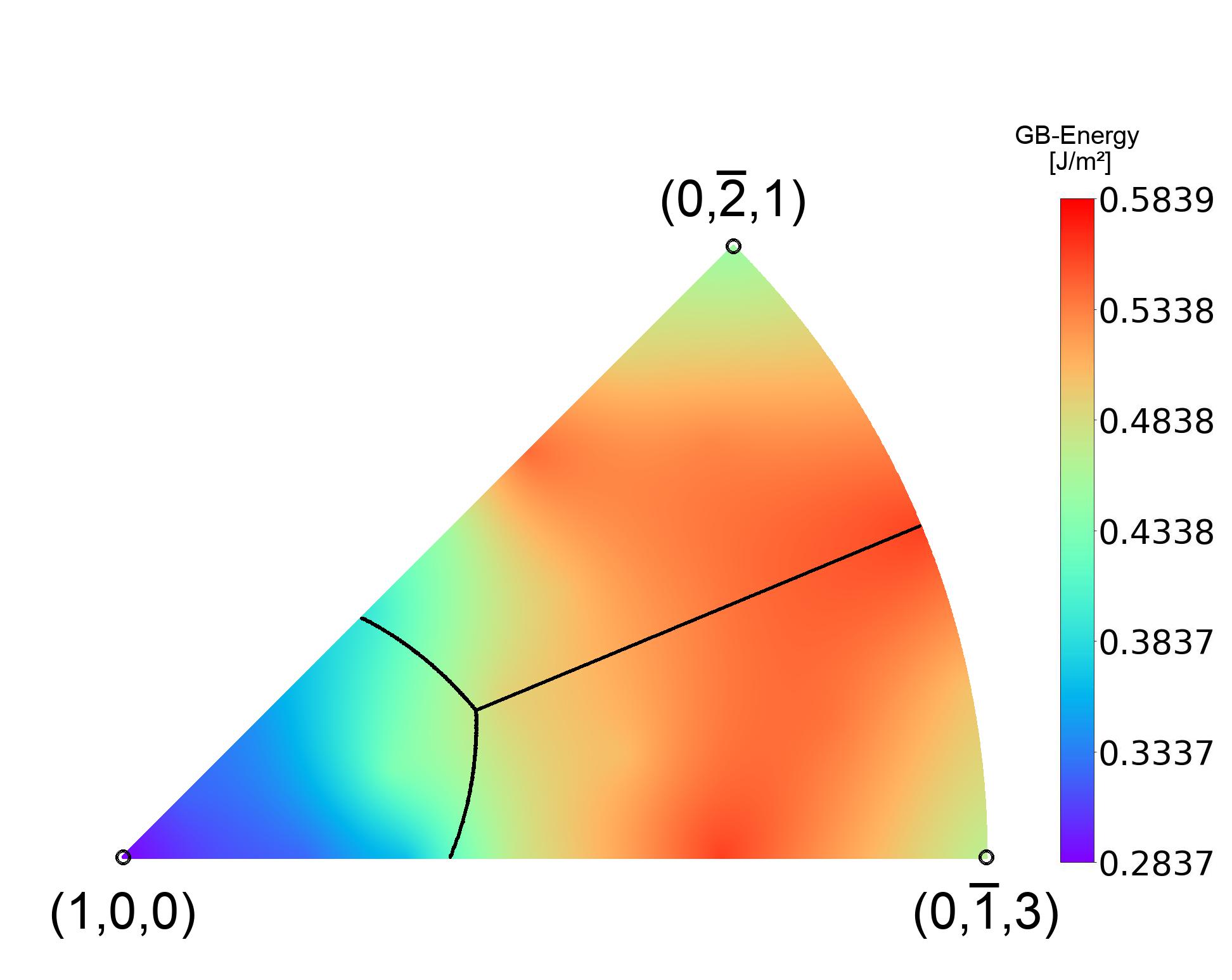} 
	\end{subfigure}
	
	\caption*{(c) $\Sigma 7$}
	
	\begin{subfigure}{0.25\textwidth}
		\centering
		\includegraphics[width=\textwidth]{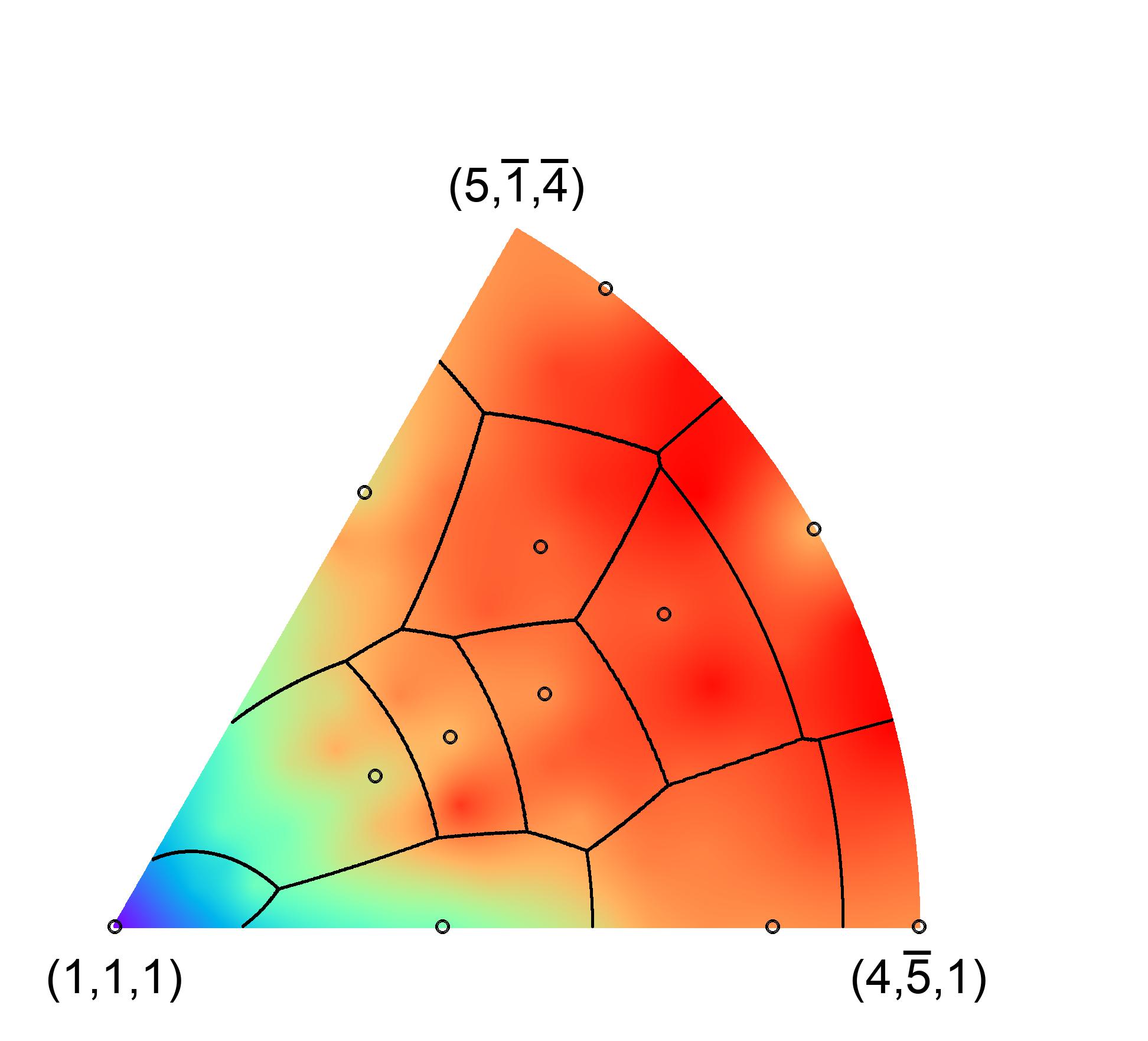} 
	\end{subfigure}
	\begin{subfigure}{0.25\textwidth}
		\centering
		\includegraphics[width=\textwidth]{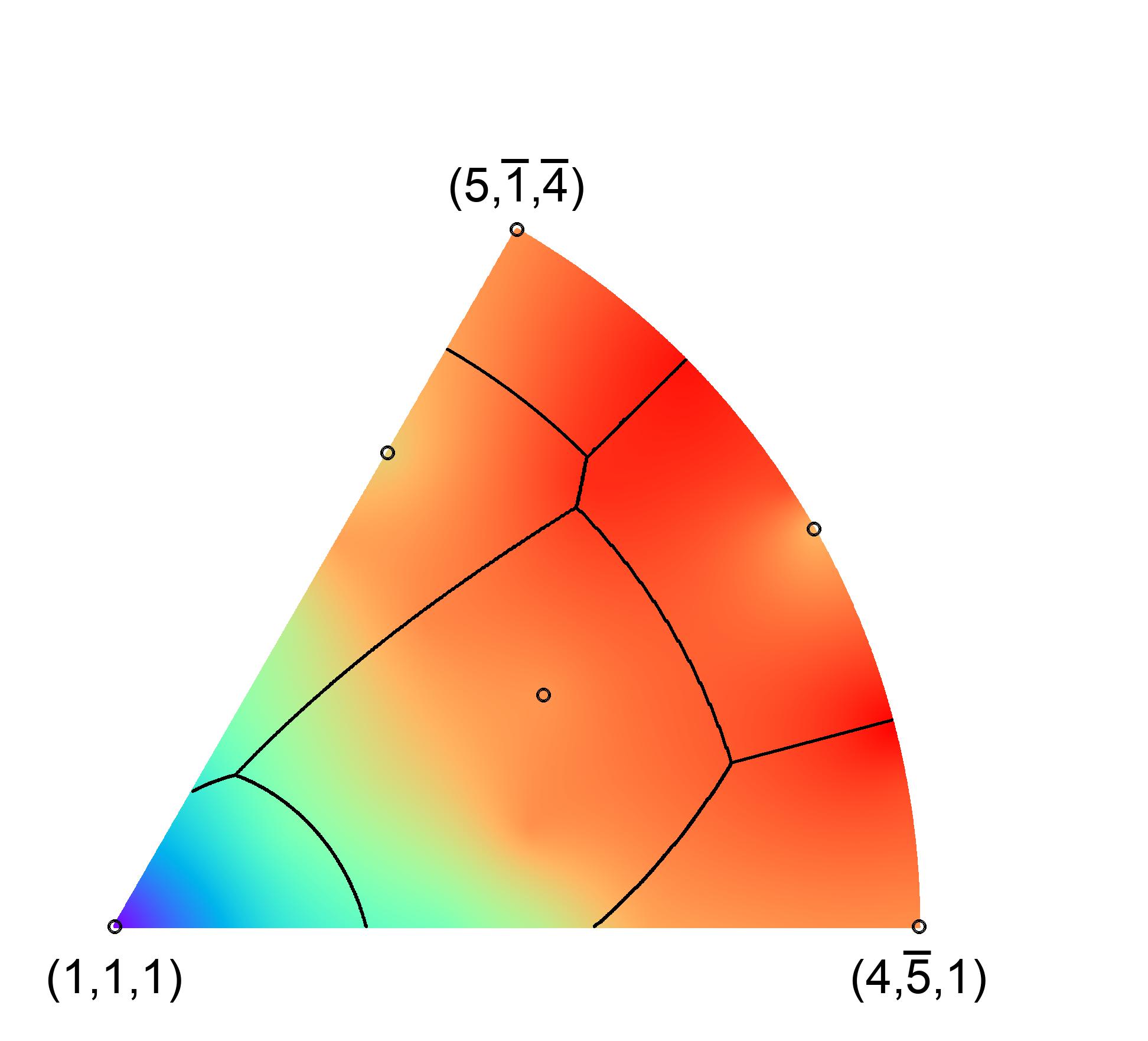}     
	\end{subfigure}
	\begin{subfigure}{0.25\textwidth}
		\centering
		\includegraphics[width=\textwidth]{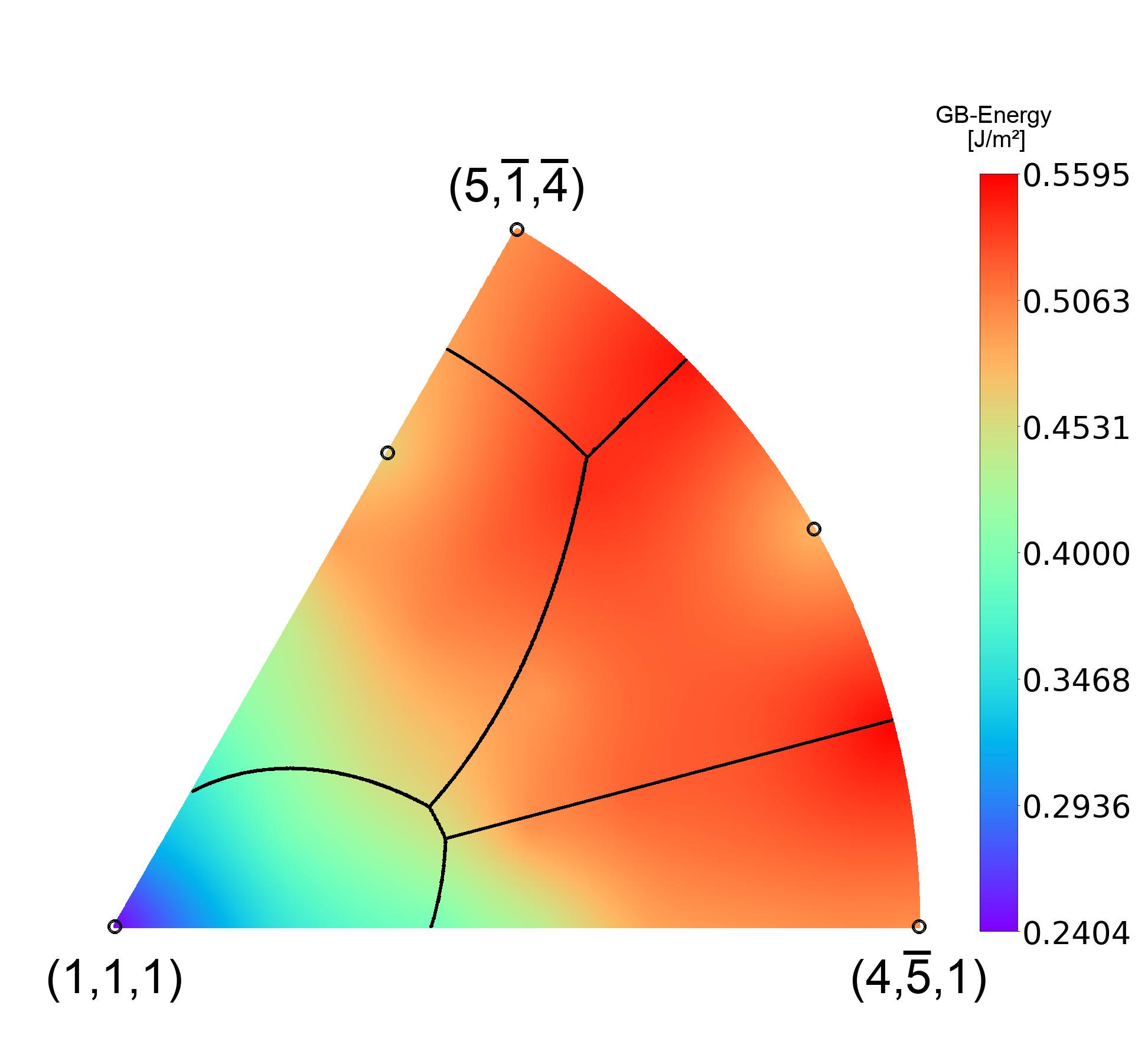} 
	\end{subfigure}\
	
	\caption*{(d)  $[110] \, 7.5^\circ$ }
	
	\begin{subfigure}{0.25\textwidth}
		\centering
		\includegraphics[width=\textwidth]{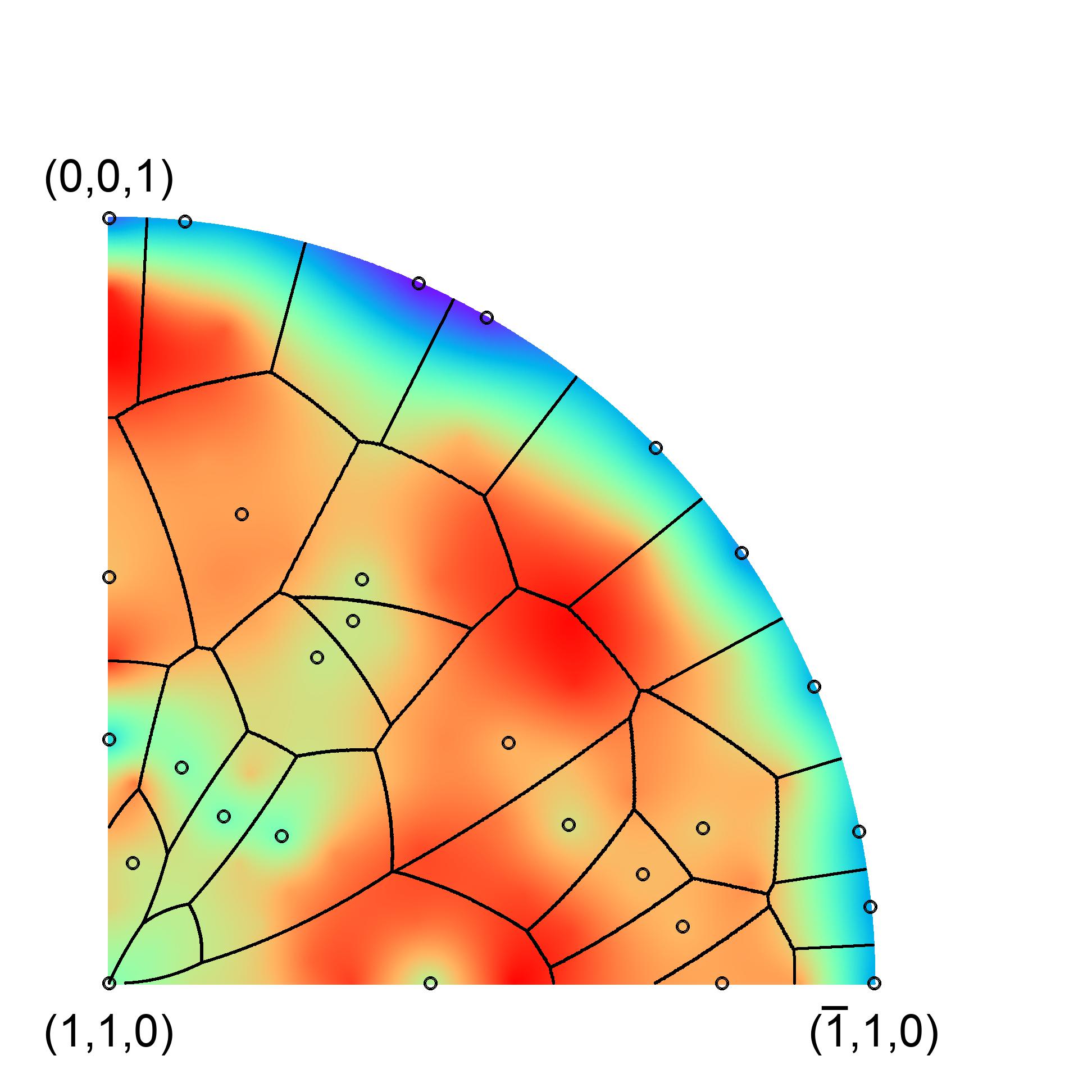} 
	\end{subfigure}
	\begin{subfigure}{0.25\textwidth}
		\centering
		\includegraphics[width=\textwidth]{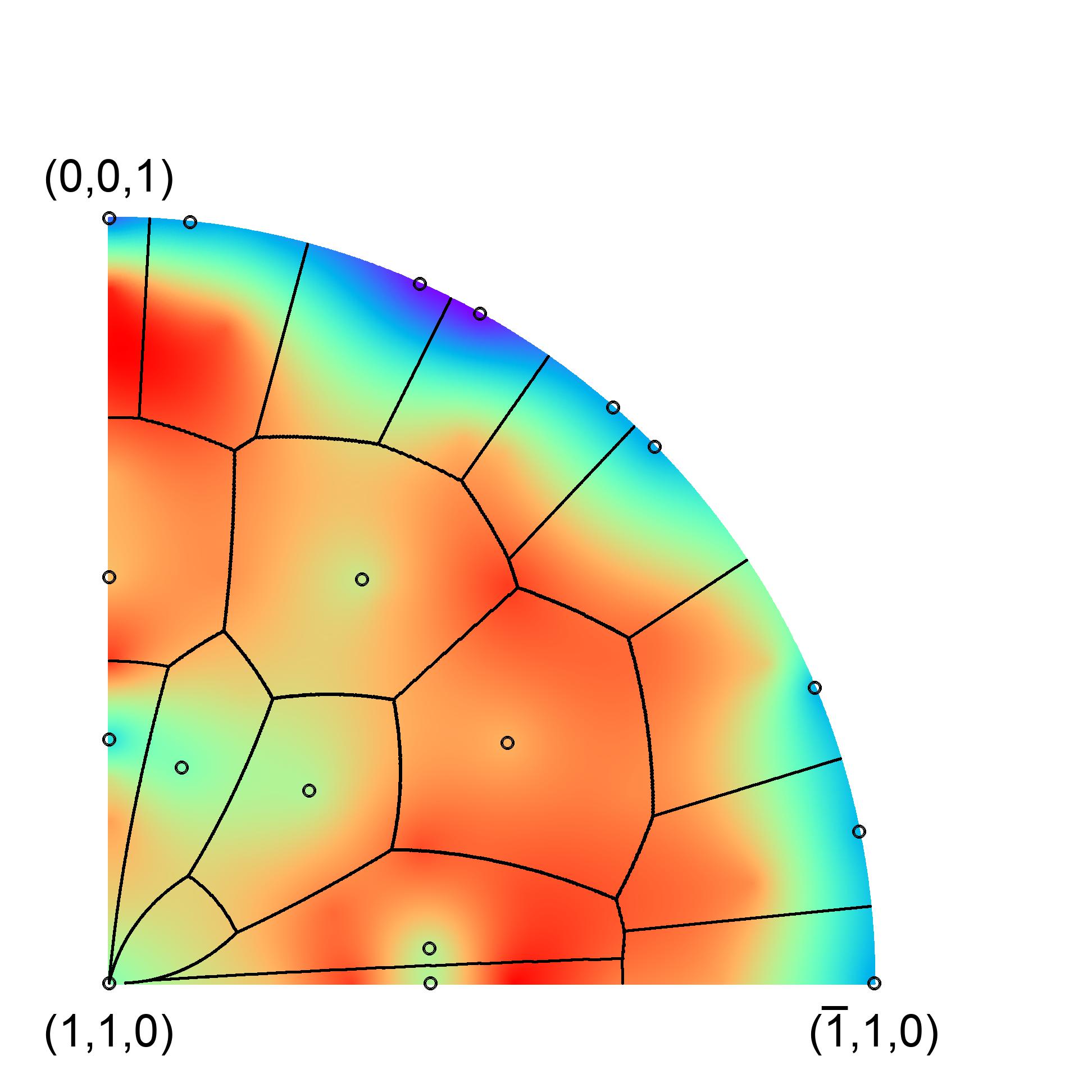} 
	\end{subfigure}
	\begin{subfigure}{0.25\textwidth}
		\centering
		\includegraphics[width=\textwidth]{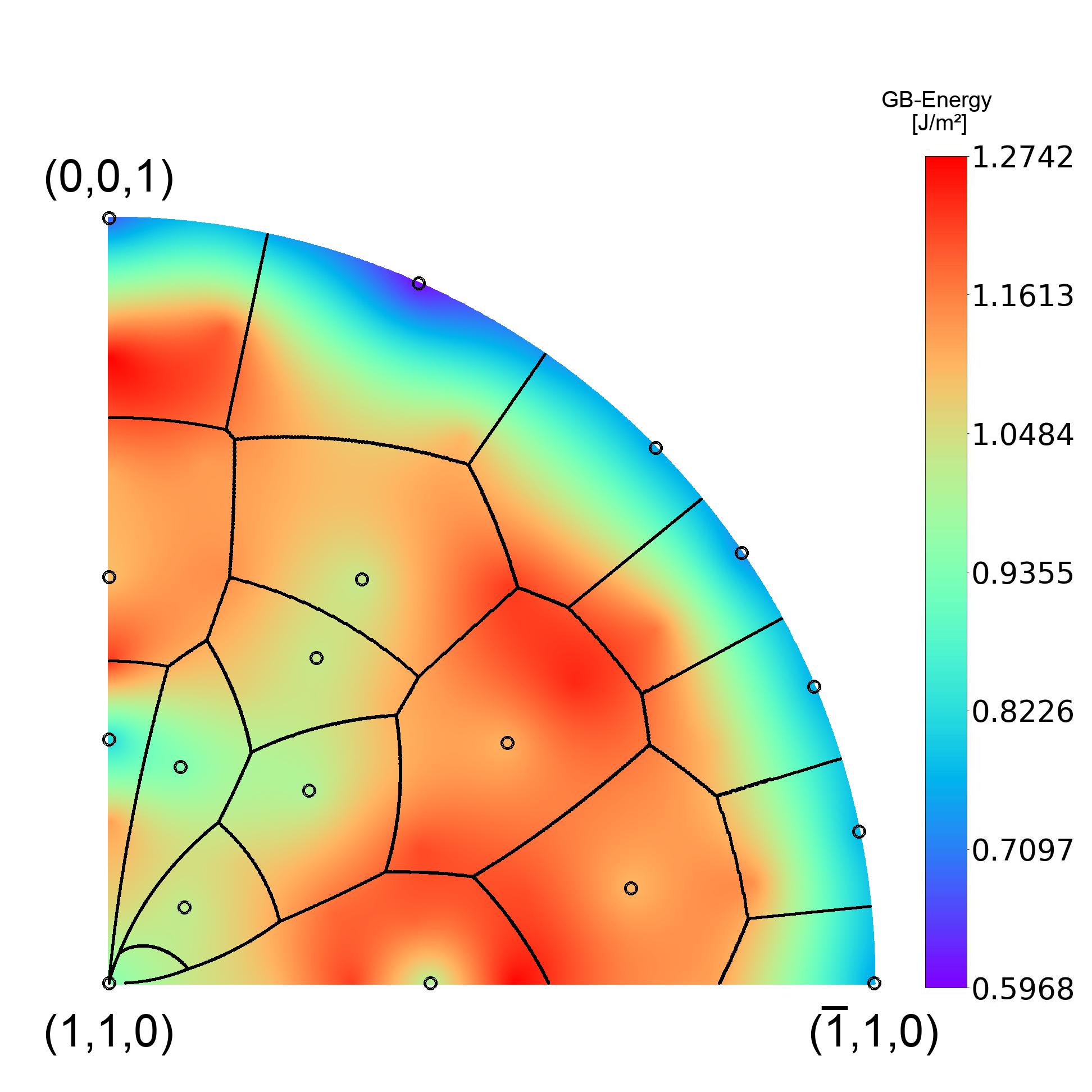}     
	\end{subfigure}
	
	\caption{\textcolor{black}{Predicted energies based on - Left: the reference database. Middle: sequential sampling with $\DeltaEstat=50~\si{mJ}/\si{m}^2$. Right: regular sampling with the same number of points as the sequential sampling.
			The black lines indicate the boundaries of 
			Voronoi cells and the black circles mark the positions of the cusps.
			(a): $\Sigma 3$ subspace; $\Nref = 150$, $\Ninit = 25$ and $\Nseq = 13$. (b): $\Sigma 5$  subspace;  $\Nref = 100$, $\Ninit = 20$ and $\Nseq = 6$. (c): $\Sigma 7$ subspace;  $\Nref = 100$, $\Ninit = 28$ and $\Nseq = 3$. (d): $[110] \, 7.5^\circ$ subspace; $\Nref = 100$, $\Ninit = 40$ and $\Nseq = 14$.} %
	}
	\label{fig:All_Kriging_plots}
\end{figure}

\textcolor{black}{As for the 1D STGB subspaces, a reference database was generated for each subspace to evaluate the quality of the sequential sampling and the stopping criterion. 
	The results of the Kriging interpolation for this database are displayed in the left column of Figure~\ref{fig:All_Kriging_plots}, which shows the energy distribution over the entire fundamental zone.
	The complete space of grain boundary inclinations represents a section of the surface of a sphere, with the normal vector of the GB plane pointing to a point on this sphere. 
	We also display the Voronoi cells around the cusps to visualise the increasing complexity of fundamental zones. The cusps themselves are marked as black circles.
	The individual plots in the left column show that the energy function becomes more complex with increasing value of $\Sigma$, i.e.~with decreasing symmetry (the $\Sigma$ value for the non-periodic $[110] 7.5^\circ$ small angle grain boundaries is infinite).
	The size of the fundamental zone, but also the density of cusps increases from the $\Sigma 3$ to the $[110] 7.5^\circ$ boundary. The variety of potentially possible energy functions makes the development of an efficient sampling procedure challenging and once again motivates the necessity of a reliable stopping criterion.}

\textcolor{black}{In the middle column of Figure~\ref{fig:All_Kriging_plots} we show the prediction based on sequential sampling, in the right column the prediction based on high-throughput sampling with the same number of points as for the sequential sampling.
	We obtain qualitatively very similar energy plots between the sequential sampling and the high-throughput sampling.
	Also the number of cusps detected by both methods is comparable.
	For instance, for the LAGB $[110] 7.5^\circ$ subspace the sequential sampling method finds $18$ cusps, whereas the high-throughput sampling finds $15$ cusps for the same total number of sampling points. In particular, the sequential sampling provides a better description of the tilt line of this subspace (with $9$ cusps) than the regular high-throughput sampling ($7$ cusps). 
	A more detailed illustration of the evolution of the sampling of the $[110]7.5^\circ$ subspace sampling with the number of sequential iterations is shown in Figure~\ref{fig_LAGB_evo} of \ref{app:evolution}}.

In the following, the two contributions to the new criterion are evaluated for the 2D cases. Furthermore the quality of the energy prediction is also governed by two aspects which will be discussed in more detail: the threshold value $\DeltaEstat$ which defines the convergence of the energy
and the number of initial design points $\Ninit$.\\

\begin{figure}
	\centering
	\centering
	\includegraphics[width=\textwidth]{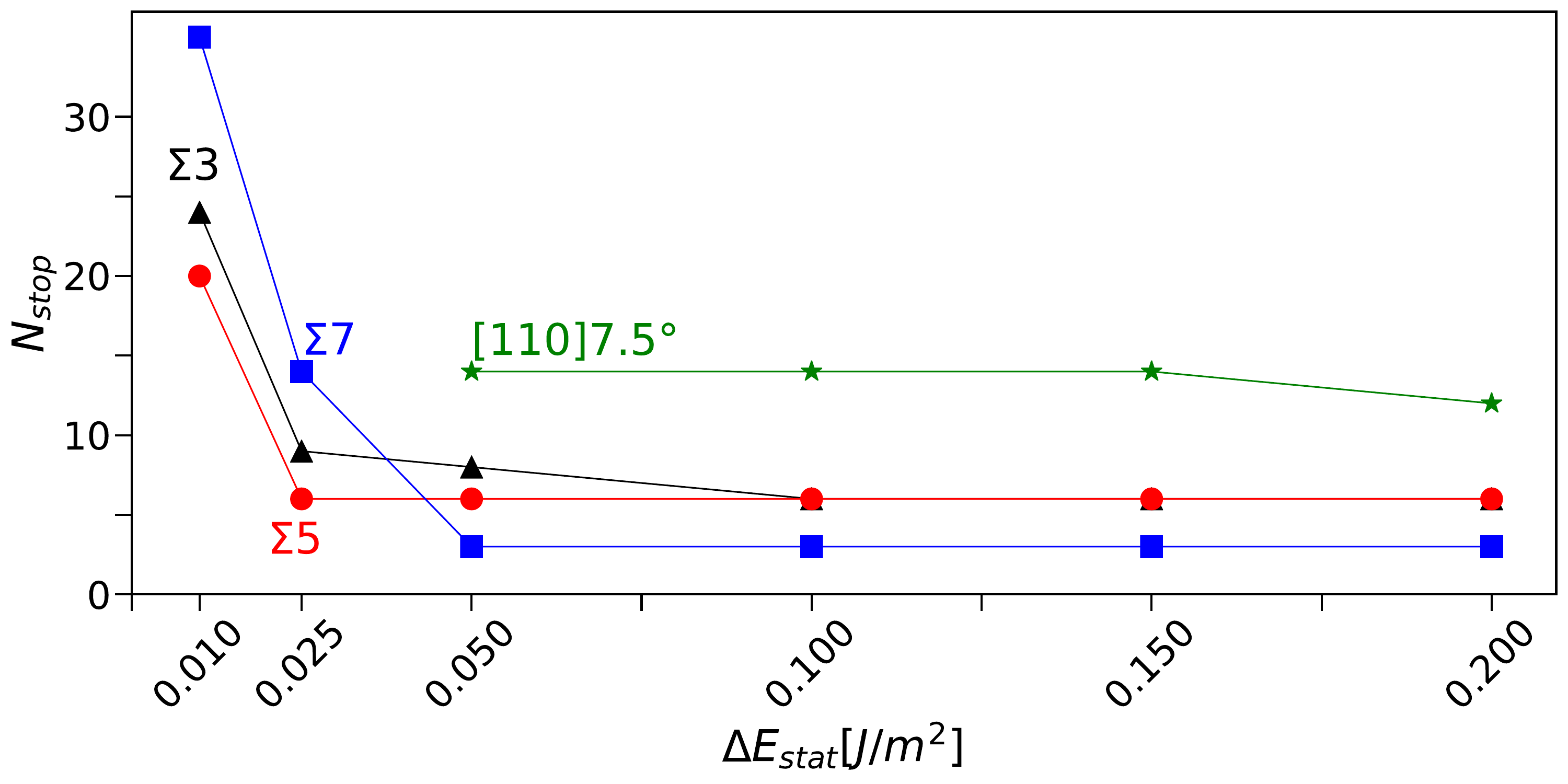}     
	\caption{
		Number $\Nstop$ of sequential steps when the algorithm is terminating.
		$\Sigma 3$ subspace ($\Ninit = 50$): $\blacktriangle$;   $\Sigma 5$ subspace ($\Ninit = 20$): \textcolor{black}{\textbullet};
		$\Sigma 7$ subspace with $\Ninit = 28$: \textcolor{blue}{$\blacksquare$}; $[110]7.5^\circ$ subspace ($\Ninit = 40$): \textcolor{SpringGreen4}{$\star$}.}
	\label{fig:2D_iend}
\end{figure}
\subsection{$\DeltaEstat$ and the speed of convergence}
\label{sec:Delta_E_stat}
The algorithm with the new  stopping criterion is applied to the data of the 2D inclination subspaces and different values for $\DeltaEstat$.
In Figure~\ref{fig:2D_iend}, the number of sequential steps, $\Nstop$, until the algorithm terminates (because the stopping criterion is satisfied) is displayed
as a function of $\DeltaEstat$. Similar to the 1D subspace, $\Nstop$ decreases 
with an increasing $\DeltaEstat$, and, if two neighbouring $\DeltaEstat$ yield the same value  $\Nstop$, the lower value of $\DeltaEstat$ marks the actual accuracy. In other words, a higher  value for  $\DeltaEstat$ does not lead to a gain in speed, because the fluctuations in the energy from one step to the next are small, anyhow. 
The values of  $\Nstop$ for the $\Sigma 3$, $\Sigma 5$ and $\Sigma 7$ subspaces differ only  by $3$, but  for  the $[110]7.5^\circ$ $\Nstop$ is  significantly larger. This is an effect of the complex energy landscape of  the $[110]7.5^\circ$  subspace rather than of the size of the fundamental zone, which becomes apparent when the two contributions to the stopping criterion are analysed. 

\subsection{The impact of $\DeltaEprev$ and $\Ncusp$ on  the stopping criterion}
\label{sec:contribution}

\begin{figure}
	\begin{subfigure}{0.49\textwidth}
		\centering
		\caption*{(a)}
		\includegraphics[width=\textwidth]{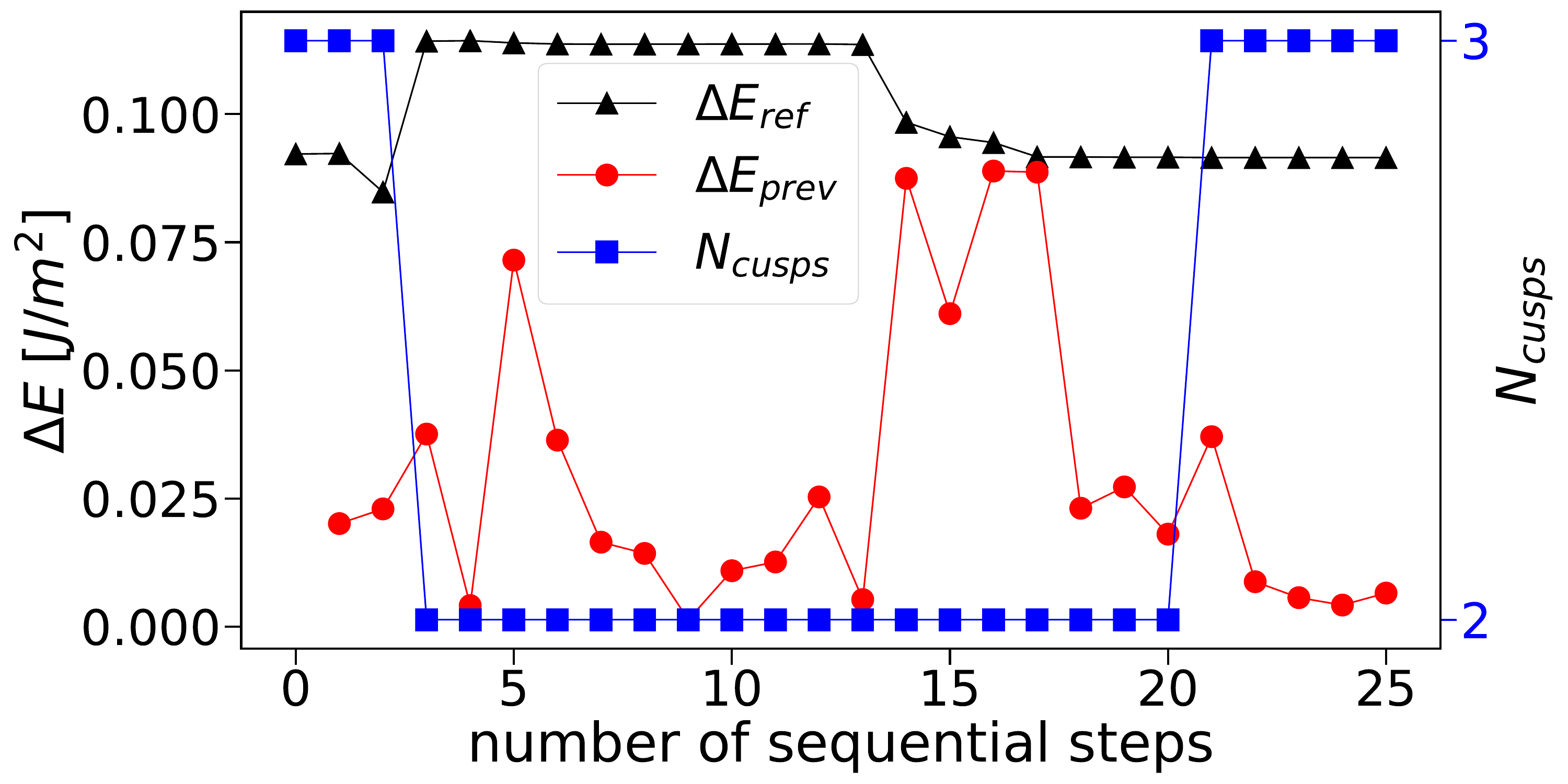}     
	\end{subfigure}
	\begin{subfigure}{0.49\textwidth}
		\centering
		\caption*{(b)}
		\includegraphics[width=\textwidth]{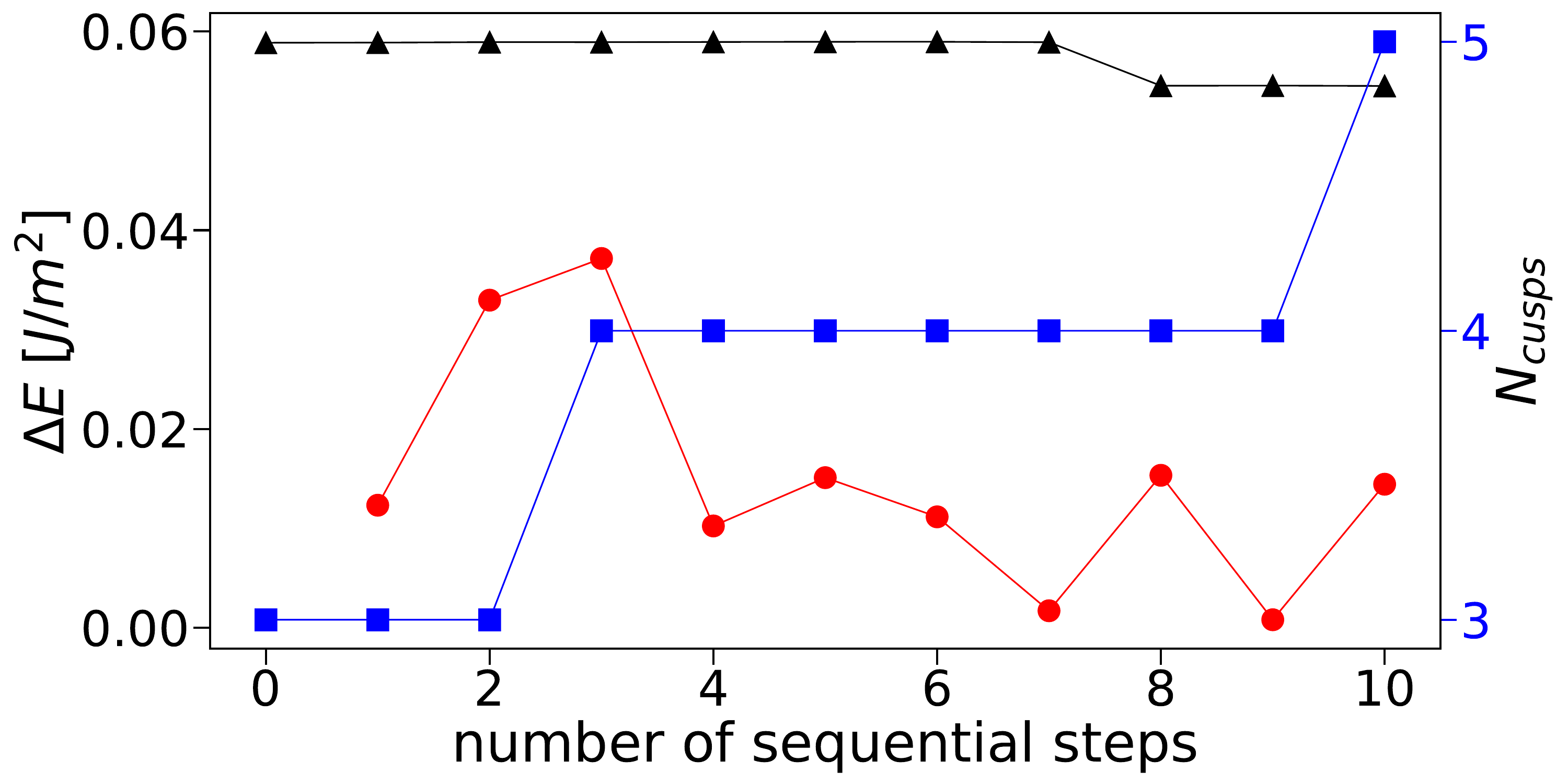}   
	\end{subfigure}
	
	\begin{subfigure}{0.49\textwidth}
		\centering
		\caption*{(c)}
		\includegraphics[width=\textwidth]{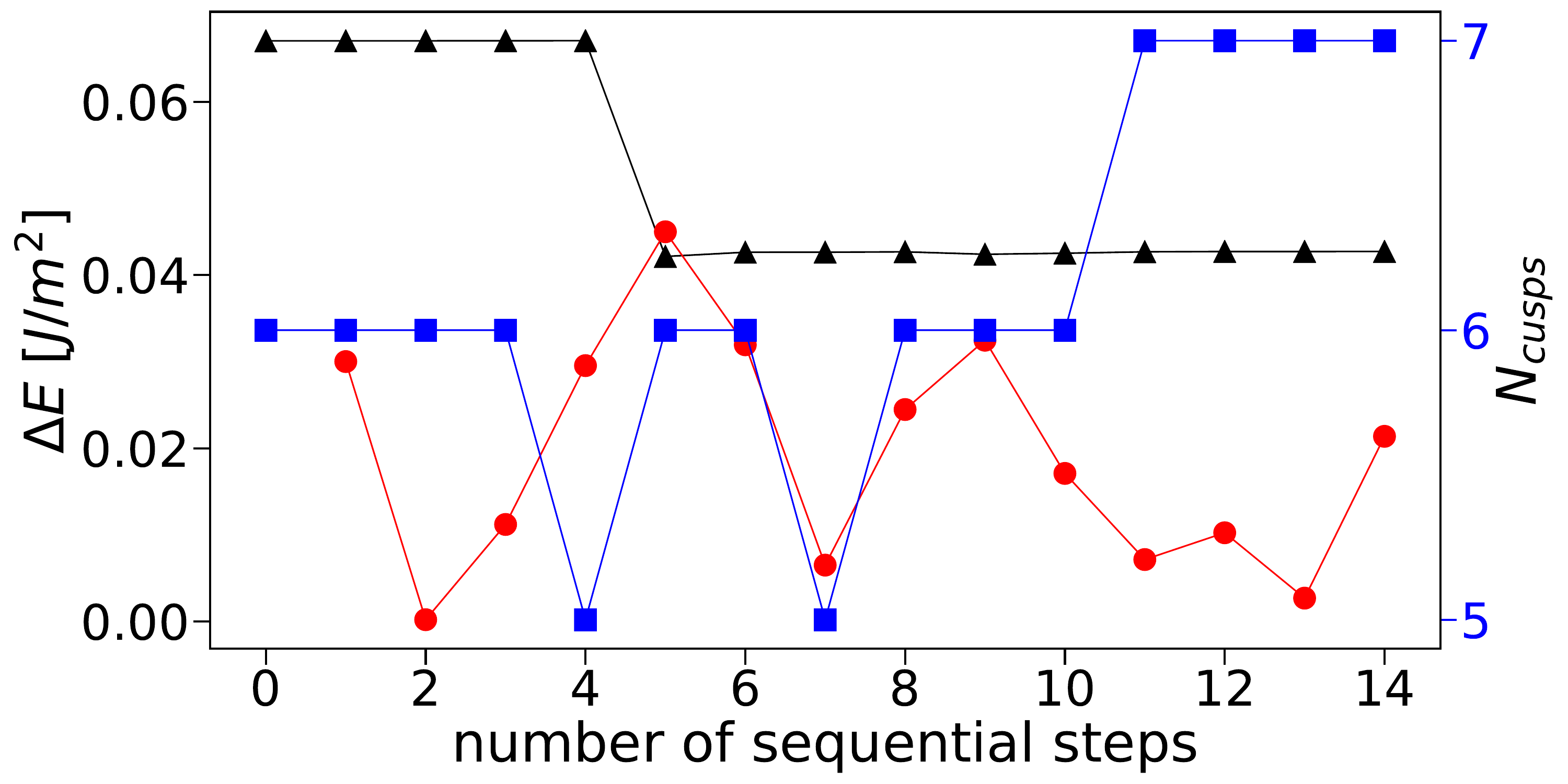}     
	\end{subfigure}
	\begin{subfigure}{0.49\textwidth}
		\caption*{(d)}
		\centering
		\includegraphics[width=\textwidth]{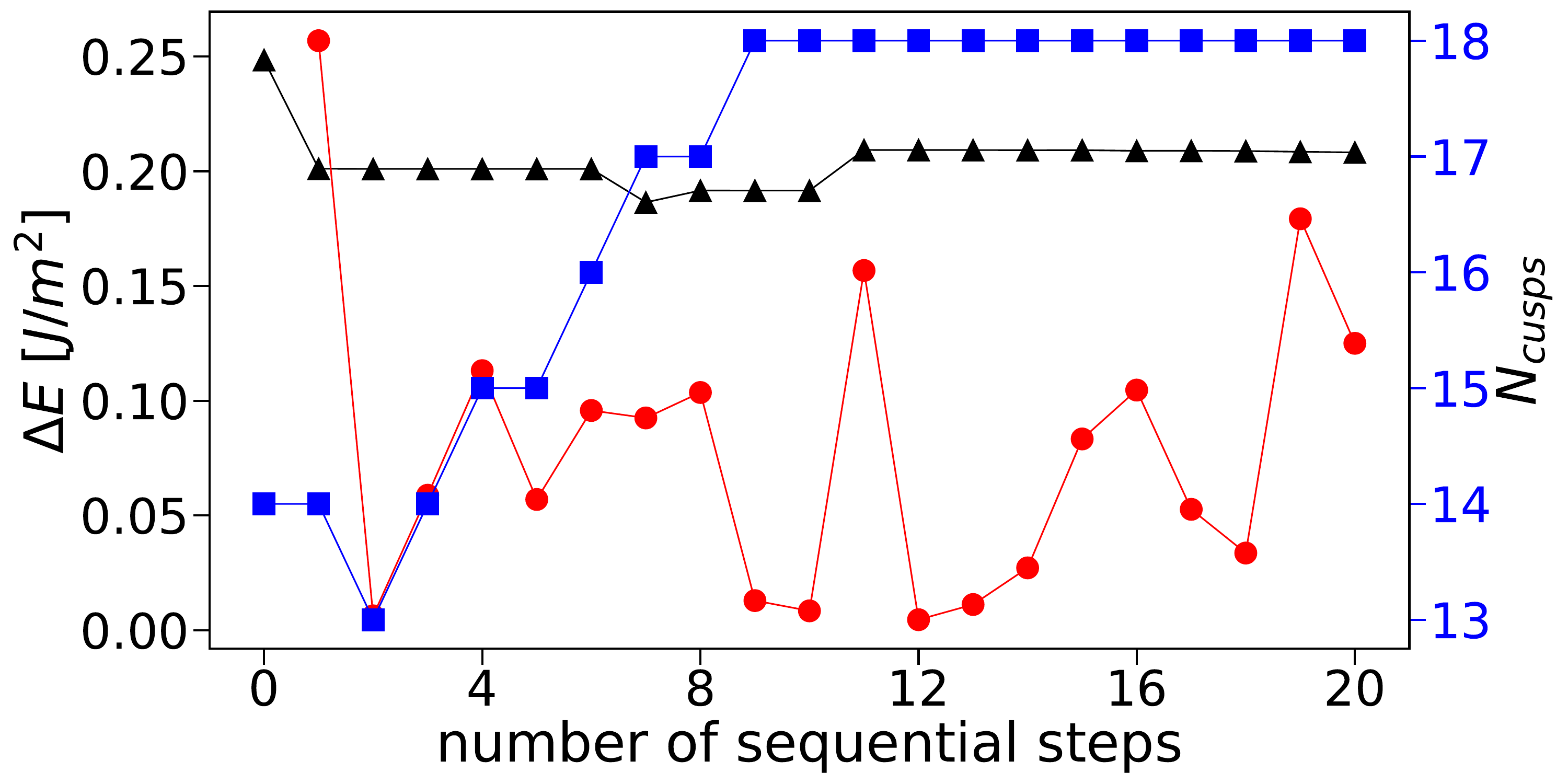} 
	\end{subfigure}
	\caption{Maximum absolute error 
		$\DeltaEref$ with respect to a reference database,  (left $y$-axis, $\blacktriangle$); maximum absolute error $\DeltaEprev$ with respect to the previous sequential step (left $y$-axis, \textcolor{black}{\textbullet}); number of cusps, $\Ncusp$ (right $y$-axis, \textcolor{blue}{$\blacksquare$}).\\
		(a) $\Sigma 3$ subspace with $\Ninit = 50$, (b) $\Sigma 5$ subspace with $\Ninit = 20$, (c) $\Sigma 7$ subspace with $\Ninit = 28$ and (d) $[110]7.5^\circ$ subspace with $\Ninit = 40$.
		\textcolor{black}{Diagrams with alternative error measures in place of the maximum absolute error are provided in \ref{app:alt_error}.}
	}\label{fig:2D_error}
\end{figure}

In Section \ref{sec:1D_Results} it has been argued for 1D subspaces that $\DeltaEprev$ is a reasonable criterion to control the sequential sampling procedure. By monitoring both $\DeltaEprev$ and $\Ncusp$ simultaneously,
a further improvement has been shown. A corresponding comparison in Figure~\ref{fig:2D_error} confirms these findings for 2D subspaces.
\textcolor{black}{The quantity $\DeltaEprev$ already contains sufficient information about the state of the sampling to replace $\DeltaEref$ and the role of $\Ncusp$ becomes even more important now.}
This is particularly visible for the $[110]7.5^\circ$ subspace which is the most complex one among our examples. %
For this case $\Ncusp$ changes several times between the first and the 9th iteration, while $\DeltaEprev$ is already smaller than $0.11 \si{J}/\si{m}^2$ between the second and the 9th iteration. Therefore, without controlling $\Ncusp$  the algorithm  would terminate even though several cusps are still to be discovered in the next iterations. On the other hand $\Ncusp$ does not change after $9$ iterations, while $\DeltaEprev$ goes up to $0.15 \si{J}/\si{m}^2$ after 11 iterations. This shows that both aspects of the stopping criterion are reasonable to optimise the automatised active learning procedure.
\subsection{Interplay of $\Ninit$ and $\DeltaEstat$}
\label{sec:effect}
\begin{figure}
	\centering
	
	\begin{subfigure}{0.46\textwidth}
		\caption{}
		\label{fig:Ncusps_allDeltaEstat}
		\includegraphics[width=0.95\textwidth]{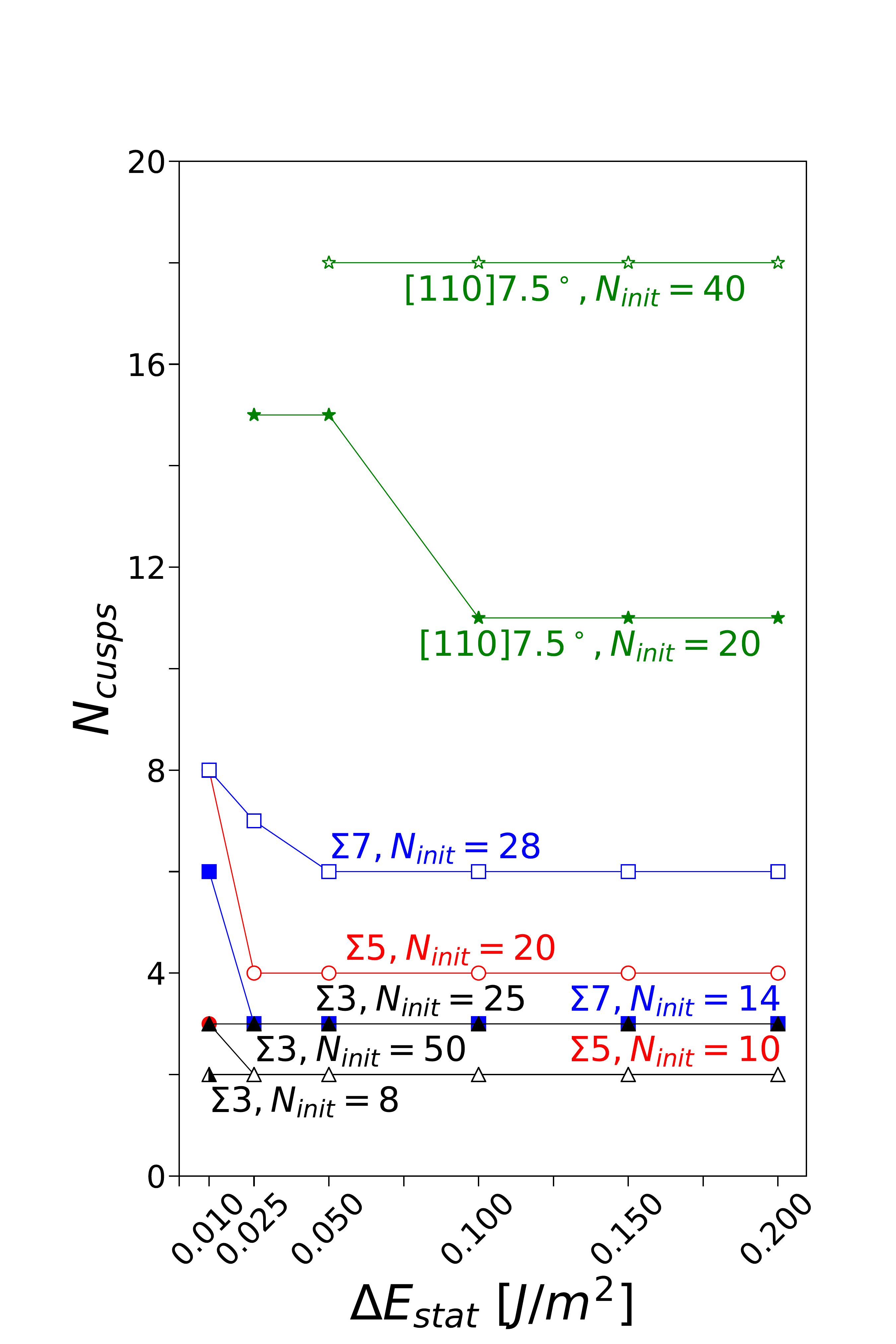}
	\end{subfigure}
	\begin{subfigure}{0.46\textwidth}
		\caption{}
		\label{fig:maxDeltaEref_allDeltaEstat}
		\includegraphics[width=0.95\textwidth]{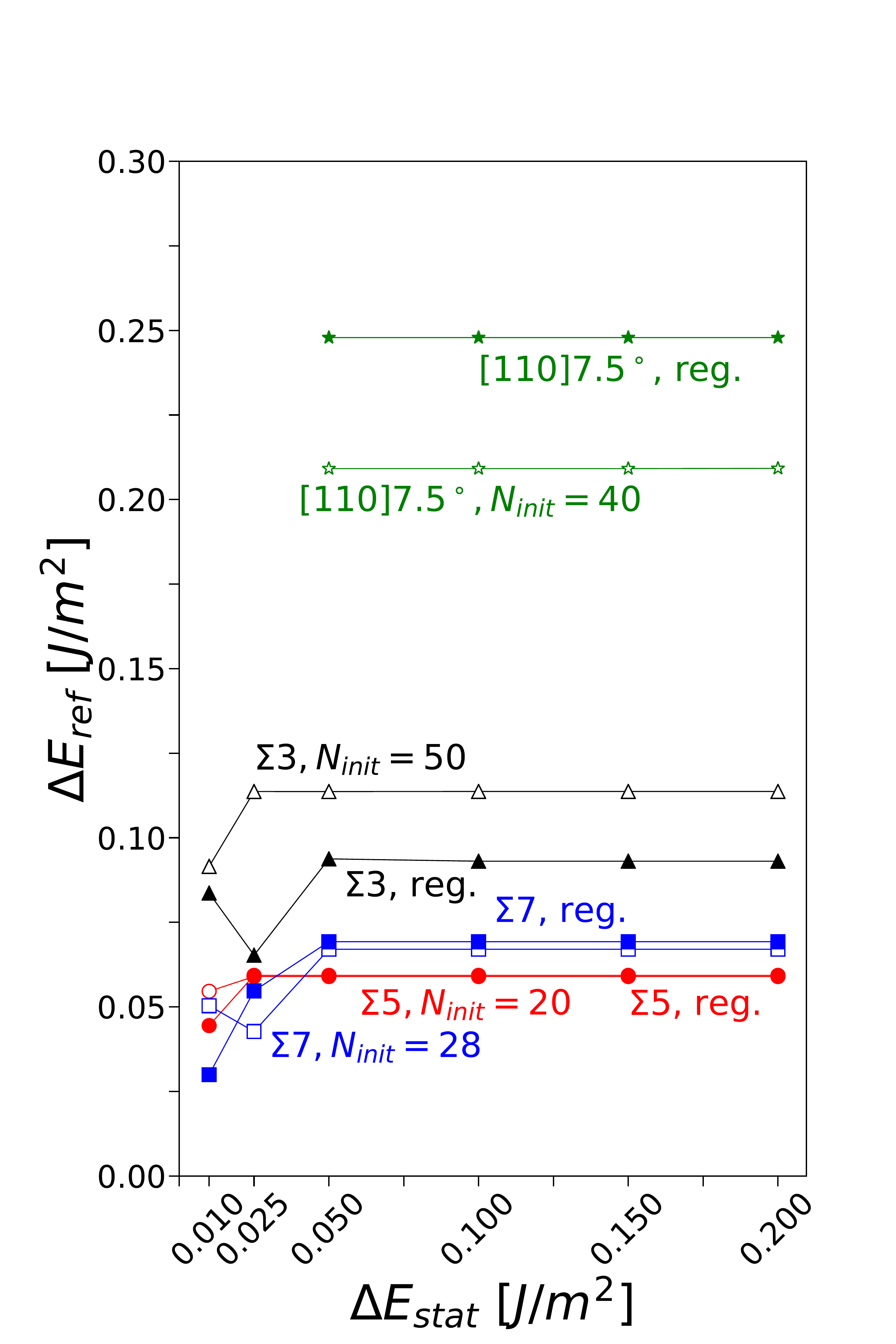}
		\label{fig:Ncusps_allDeltaEstat:b}
	\end{subfigure}
	\begin{subfigure}{0.46\textwidth}
		\caption{}\label{fig:Ncusps_DeltaEstat0_05}
		\includegraphics[width=0.95\textwidth]{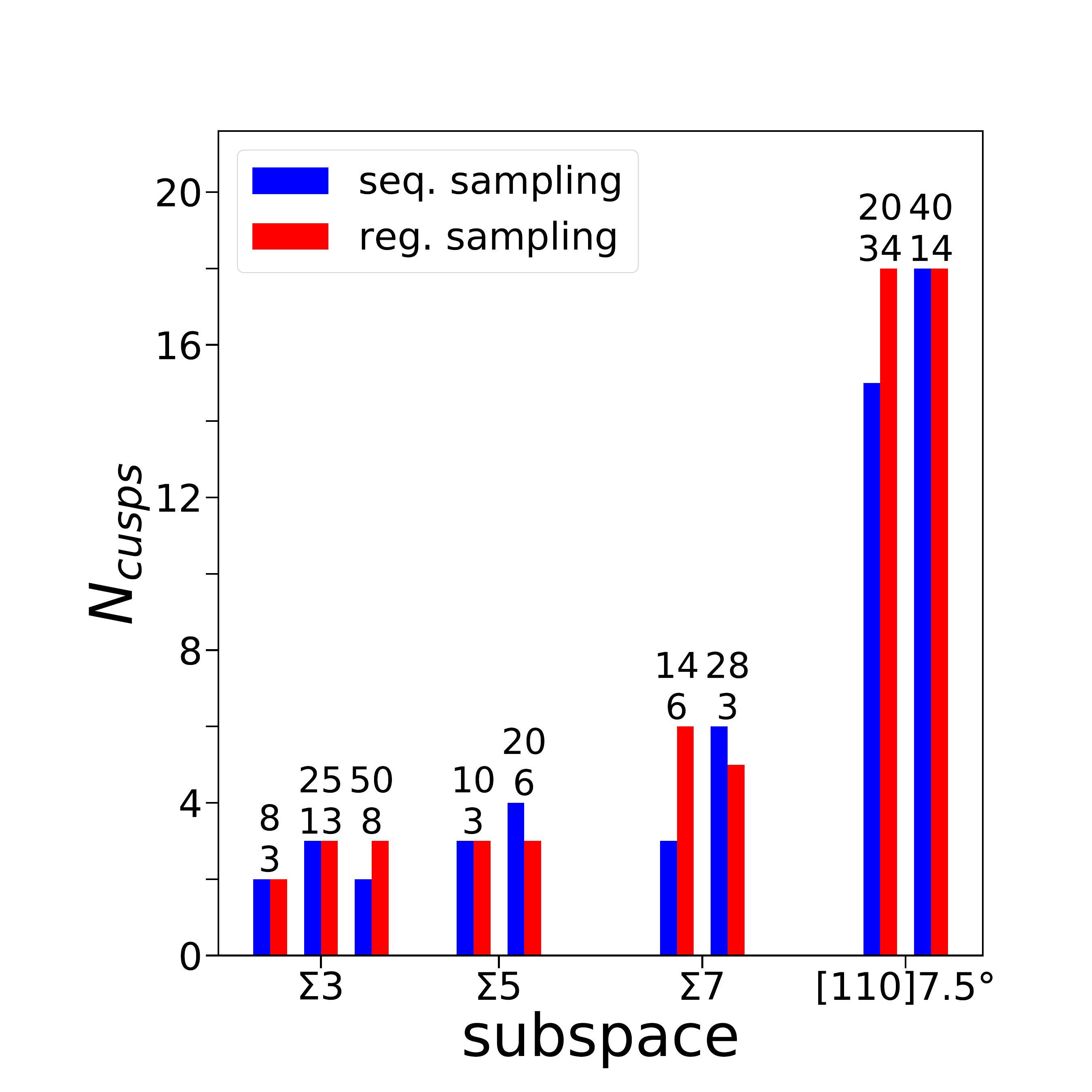}
	\end{subfigure}
	\begin{subfigure}{0.46\textwidth}
		\caption{}\label{fig:maxDeltaEref_DeltaEstat0_05}
		\includegraphics[width=0.95\textwidth]{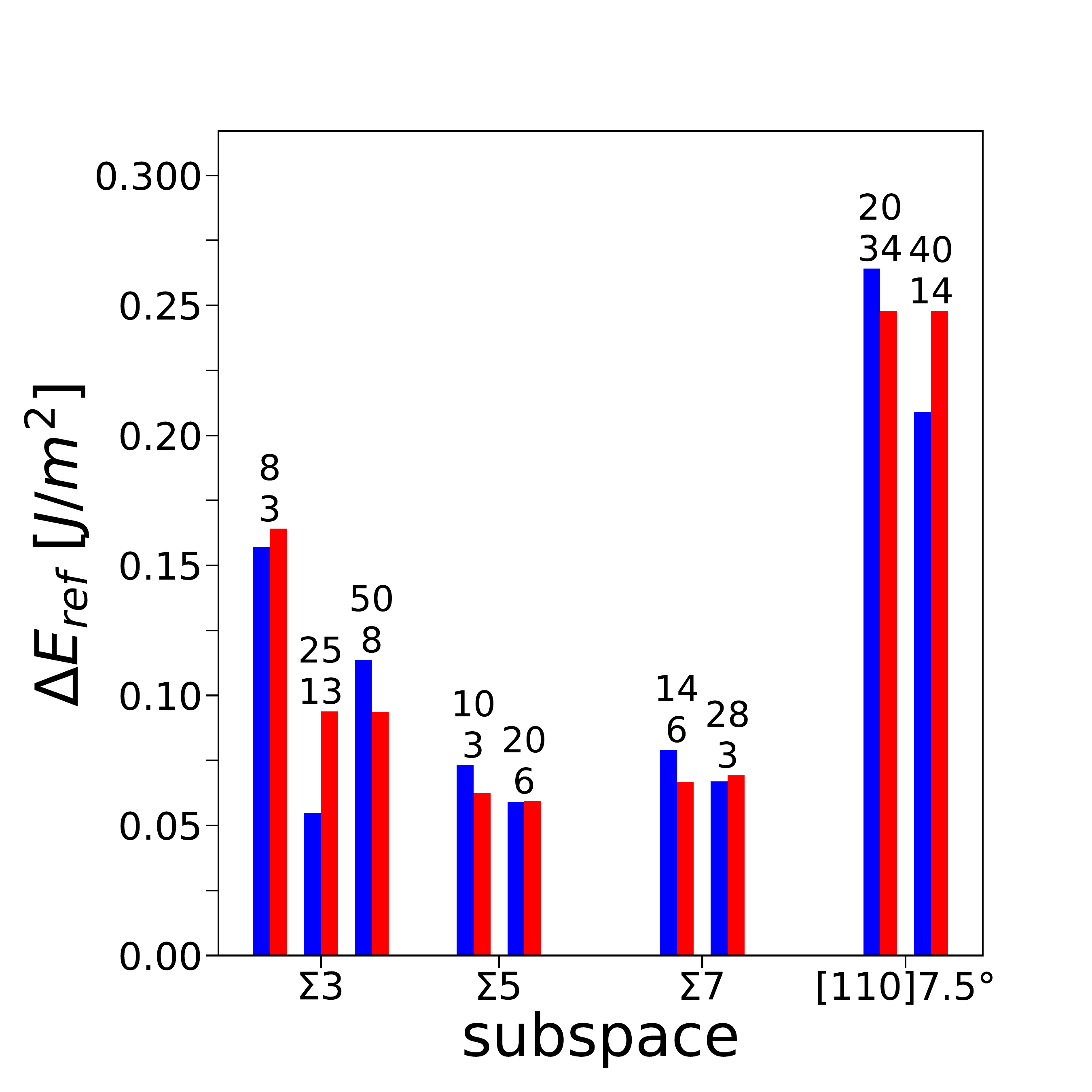}   
	\end{subfigure}
	\caption{Results of sequential and regular sampling (with the same total number of sampling points) for 2D spaces: (a) number 
		of cusps detected by the sequential algorithm
		as a function of $\DeltaEstat$; (b) maximum absolute error with respect to the reference database
		of the sequential algorithm with $\Ninit$ initial design points  as a function of $\DeltaEstat$
		compared to a regular sampling with the same total number of sampling points;
		(c) number of cusps for different initial designs
		for different subspaces
		($\DeltaEstat = 0.05\si{J}/\si{m}^2$);
		(d)  maximum absolute error with respect to the reference database 
		for different initial designs for different subspaces
		($\DeltaEstat = 0.05\si{J}/\si{m}^2$).
	} 
\end{figure}

An aspect which has not been discussed so far, is the influence of the size of the initial design on the quality of the sampling. \textcolor{black}{To do so, we compare the number of cusps as well as the error with respect to the reference database for different sizes of initial and sequential designs, as well as the regular high-throughput sampling. We first look at these two quantities as a function of the desired accuracy, defined by $\DeltaEstat$, in Figures~\ref{fig:Ncusps_allDeltaEstat} and \ref{fig:Ncusps_allDeltaEstat:b}, and then choose the case $\DeltaEstat=0.05\si{J}/\si{m}^2$ for a more detailed analysis of the effect of the initial design in Figures \ref{fig:Ncusps_DeltaEstat0_05} and \ref{fig:maxDeltaEref_DeltaEstat0_05}.}

Figure~\ref{fig:Ncusps_allDeltaEstat} shows the number of cusps which have been found at the end of the sampling as a function of the threshold value $\DeltaEstat$ for different sizes of the initial design.
We observe  that consistently more cusps are found for a larger initial design, with the exception of the $\Sigma 3$ subspace. \textcolor{black}{In this case, also the maximum error w.r.t.~the reference database is larger for sequential than for regular sampling.} This exception can be explained by the rather smooth and \textcolor{black}{steep energy variation from the pure twist grain boundary (the tip of the fundamental zone) to the line of tilt grain boundaries (the right edge of the fundamental zone) for the $\Sigma 3$ misorientation, with only a few cusps. In such a case, the exact Kriging interpolation (i.e., $\delta=0$) between data points within a very small distance can lead to an overfitting of the data resulting in a large error.}
Generally, the number of identified cusps increases with the size and the complexity of the subspace.
In Figure~\ref{fig:maxDeltaEref_allDeltaEstat} we compare the maximum error $\DeltaEref$ with respect to the reference database for a sequential design and a regular design with the same number of atomistic simulations.
\textcolor{black}{For moderate accuracy, $\DeltaEstat\geq0.05 \si{J}/\si{m}^2$, the statistical approach delivers comparable or even better results (in the case of the LAGB), with the exception of the $\Sigma$3 STGB. In this latter case, however, performance can be further improved by choosing a smaller initial design. For $\DeltaEstat < 0.05\si{J}/\si{m}^2$ the regular sampling yields a smaller error $\DeltaEref$ w.r.t.~the reference data. However, as described before, it is in this range of $\DeltaEstat$ that more cusps are identified showing that the tighter convergence criteria are reasonable.}

For the value $\DeltaEstat = 0.05\si{J}/\si{m}^2$, Figures~\ref{fig:Ncusps_DeltaEstat0_05} and \ref{fig:maxDeltaEref_DeltaEstat0_05} show a detailed comparison of the different samplings with the regular one. 
The numbers above the bars refer to the sequential sampling (blue) and represent the number of initial and the number of sequential design points chosen until the stopping criterion is fulfilled. The total number of sample points is the sum of both, i.e., $\Ntotal = \Ninit + \Nseq$. The results for  equivalent regular high-throughput sampling with the same $\Ntotal$ as the sequential sampling are shown as red bars. We observe from the diagrams that energy convergence can also be reached with a small initial design, leading to a small total number of calculations. \textcolor{black}{However, the number of determined cusps is only equal or higher than for the regular sampling (and similarly the error with respect to the reference database is equal or lower) for the larger initial designs} with the above mentioned exception $\Sigma 3$ where sequential sampling with $\Ninit = 25$ identifies the highest number of cusps. 

To choose the optimal $\Ninit$, the size of the different subspaces has to be considered, which goes along with an increase in complexity of the energy function. The maximum polar angle always equals $\vartheta_{\mathrm{FZ}}=90^\circ$. As shown in Figure~\ref{fig:All_Kriging_plots} the maximum azimuthal angle equals $\varphi_{\mathrm{FZ}, \Sigma 3} = 30^\circ$, $\varphi_{\mathrm{FZ}, \Sigma 5} = 45^\circ$, $\varphi_{\mathrm{FZ}, \Sigma 7} = 60^\circ$ and $\varphi_{\mathrm{FZ}, [110]7.5^\circ} = 90^\circ$. The point densities $\rho$ of the initial design can be calculated with the following equation:
\begin{equation*}
	\rho = \frac{\Ninit}{\varphi_{\mathrm{FZ}}}\text{.}
\end{equation*}
In this study we obtained good results for point densities between $0.44$ and $0.66$ per degree.

Figures~\ref{fig:Ncusps_DeltaEstat0_05} and \ref{fig:maxDeltaEref_DeltaEstat0_05} show that the sequential design yields results \textcolor{black}{comparable to} or even better than those of a regular high-throughput sampling in low-symmetry cases, \textcolor{black}{the LAGB, and the $\Sigma$7 STGB}, provided that the threshold for the stopping criterion, $\DeltaEstat$, \textcolor{black}{is reasonably chosen.
	In this work, the grain boundary energies in the fundamental zones vary from $0.30$ ($\Sigma 7$) to $1.10 \si{J}/\si{m}^2$ ($\Sigma 3$) for the low-symmetry cases, and a good result is obtained for $\DeltaEstat = 0.05\si{J}/\si{m}^2$, i.e., for less than 4.5--16.7\% of this variation. This fits with the experience that 
	the experimental determination of grain boundary energies comes with an error of roughly 10\% (see e.g.~\cite{Rohrer2010Comparing}). Thus, 10\% of the minimal energy in a subspace (estimated from the energies on the initial design) seems a reasonable choice for $\DeltaEstat$.}

\section{Discussion and Conclusion}

This work introduces an algorithm for an automated sampling of grain boundary energies in the spirit of an active learning technique. It is based on the sequential sampling strategy introduced in \cite{kroll2022efficient}, which has been developed for 1D STGB subspaces and has now successfully been extended in two directions. On the one hand the new algorithm can be used for 2D applications. On the other hand, and more importantly, the new algorithm is able to decide on the basis of the collected data, when it is reasonable to stop sequential sampling. The major difference to other methods published so far, e.g.~\cite{kim2011identification} and \cite{restrepo2014artificial}, is that no prior knowledge concerning the location of the cusps is needed. The proposed algorithm rather learns the locations of the cusps automatically and terminates when no new cusps are found or major changes in the energy landscape arise over several iterations. %
\textcolor{black}{This is feature will enable future investigations of multidimensional subspaces.}

To arrive at the practical scheme, two quantities to evaluate the quality of the sampling were defined: the maximum deviation of the energy between two following sequential steps and the number of identified minima in the energy landscape. Both quantities can be calculated on the fly, and in combination they are well-suited to evaluate the sampling based on the available data. This allows to define a stopping criterion for automated sequential runs. In our  applications  it  is demonstrated that the total number of sequential steps \textcolor{black}{which is needed to reach a desired accuracy} depends on the variability of the energy landscape.
A larger number of cusps requires more sequential steps. The benefit of monitoring this number becomes particularly clear when looking at low-angle grain boundaries with a rather volatile energy distribution in the fundamental zone and thus a high frequency in the variation of energy versus angles.

It was also shown that the sampling can be further refined  by a careful choice of the size of the initial design. 
\textcolor{black}{In general, the optimal choice of $N_{\mathrm{init}}$ depends sensitively on the 
	topology of the energy landscape. For example, if this is nearly constant only few observations are sufficient whereas for very spiky energy functions the size of the initial design should be larger. 
	A rule of thumb (which gave reasonable results in our studies) is to choose $N_{\mathrm{init}}$ proportional to the size of the FZ, but this choice can be improved if some prior information about the topology of the energy landscape is available.}

With the help of a reasonable metric, the algorithm presented in this paper can  be extended to even higher dimensional subspaces, or the whole parameter space. To some extent, the required data for such an analysis is already available from \cite{baird2021five} and \cite{homer2022examination}, who examined the topology of the 5D parameter space.

To sum up, the  maximum deviation of the energy between two following sequential steps and the number of identified minima in the energy landscape  are useful measures  to describe the quality of the sampling on the fly while giving advanced information about the subspace itself and its complexity. We have developed a  sequential sampling algorithm with a stopping criterion, which is based on a simultaneous monitoring  of these two quantities.  As a result  we obtain a very efficient active learning procedure for the exploration of grain boundary subspaces, %
which opens the door to explore the \textcolor{black}{rather unknown} energy landscape of low angle grain boundaries precisely.

\section*{Code availability}

\textcolor{black}{The code used for this paper is still under development. It is available from the authors upon reasonable request.}

\section*{Acknowledgements}
This research has been supported by the German Research Foundation (DFG), project number 414750139. The authors also gratefully acknowledge the funding of this project by computing time provided by the Paderborn Center for Parallel Computing (PC2).

\printbibliography

\newpage

\pagenumbering{arabic}
\renewcommand*{\thepage}{A\arabic{page}}

\appendix

 \section{Atomistic simulations}
\label{sec:atomistics}
The open-source package LAMMPS \cite{plimpton1995LAMMPS} was used for the construction of the grain boundary structures, their optimisation and the computation of the GB energies via molecular statics. To represent bcc and fcc metals, the embedded atom method type potentials for Fe \cite{mendelev2003development}, Al \cite{zope2003interatomic} and Ni \cite{stoller2016impact} were employed, which are available at \texttt{https://www.ctcms.nist.gov/potentials/}. \textcolor{black}{The basic material properties as they are reproduced by the potentials are shown in \ref{table:atomistic_simulation_parameter_1}.} A spherical grain method introduced in \cite{lee2004computation} and improved, \textcolor{black}{e.g.~}in \cite{Li2019,kroll2022efficient}, was used to model the grain boundaries without periodic boundary conditions. In this approach two spheres are created, one of which is rotated by the misorientation angle around the rotation axis. Subsequently both spheres are cut into half-spheres at the desired grain boundary plane and both half-spheres are combined to construct the grain boundary. To optimise the microscopic degrees of freedom different combinations of trial displacements (parallel and perpendicular to the interface) are applied, and atoms within a certain cut-off radius are deleted. \textcolor{black}{The parameters for the optimisation of the microscopic degrees of freedom for each subspace are displayed in Tables~\ref{table:atomistic_simulation_parameter_2} and \ref{table:atomistic_simulation_parameter_3}}.
\begin{table}
	\begin{center}
		\begin{tabular}{c|c|c|c|c|c|c}
			subspace & 
			$\Delta o_{\min}$ $[\si{\angstrom}]$& 
			$\Delta o_{\max}$ $[\si{\angstrom}]$& 
			$N_{\Delta o}$ &
			$c_{\min}$ $[\si{\angstrom}]$& 
			$c_{\max}$ $[\si{\angstrom}]$& 
			$N_{c}$ \\ 
			\hline
			$\Sigma 3$ & 0 & 1.5 & 2 & 0 & 0.25 & 2\\
			$\Sigma 5$ & 0 & 1.5 & 2 & 0 & 0.25 & 2\\
			$\Sigma 7$ & 0 & 1.5 & 2 & 0 & 0.25 & 2\\
			$[110]7.5^\circ$ & 0 & 1.5 & 2 & 0 & 0.25 & 2\\
		\end{tabular}
		\caption{\textcolor{black}{Minimum and maximum offset values, $\Delta o_{\min}$, $\Delta o_{\max}$ for the displacement perpendicular to the interface and the total number $N_{\Delta o}$ of offset values used in equally distant steps; minimum and maximum cut-off radius, $c_{\min}$,$c_{\max}$ for the deletion of atoms, and the total number of cut off radii $N_c$, used in equally distant steps.}}
		\label{table:atomistic_simulation_parameter_2}
	\end{center}
\end{table}
\begin{table}[H]
	\begin{center}
		\begin{tabular}{c|c|c|c|c|c|c}
			subspace & 
			$\Delta d_{1, \min}$ $[\si{\angstrom}]$& 
			$\Delta d_{1, \max}$ $[\si{\angstrom}]$& 
			$N_{\Delta d_1}$ &
			$\Delta d_{2, \min}$ $[\si{\angstrom}]$& 
			$\Delta d_{2, \max}$ $[\si{\angstrom}]$& 
			$N_{\Delta d_2}$\\
			\hline
			$\Sigma 3$ & 0 & 2.8555 & 6 & 0 & 5 & 6\\
			$\Sigma 5$ & 0 & 5 & 6 & 0 & 5 & 6\\
			$\Sigma 7$ & 0 & 5 & 6 & 0 & 5 & 6\\
			$[110]7.5^\circ$ & 0 & 2.4890 & 6 & 0 & 5 & 6\\
		\end{tabular}
		\caption{\textcolor{black}{Parameters for the two types of displacements parallel to the interface: Minimum and maximum value $\Delta d_{\min}$, $\Delta d_{\max}$, and number $N_{\Delta d}$ of displacements in equally distant steps.}}
		\label{table:atomistic_simulation_parameter_3}
	\end{center}
\end{table}
For all trial structures the interatomic forces are relaxed and the GB energy is calculated from the atoms of an inner sphere (to avoid surface effects to the potential energy of each atom) by using the following equation:
\begin{equation*}
	E_{\mathrm{GB}} = \frac{\sum_{n=1}^{N}E_{\mathrm{pot},n}-N\cdot E_{\mathrm{bulk}} }{\pi r_{\mathrm{i}}^{2}},
\end{equation*}
with $N$ the number of atoms in the inner sphere, $E_{\mathrm{pot},n}$ the energy of the $n$-th atom in the inner sphere, $E_{\mathrm{bulk}}$ the energy per atom in a corresponding bulk structure and $r_{\mathrm{i}}$ the radius of the inner sphere. 
The minimal grain boundary energy which is obtained in this way is taken as the final result. \textcolor{black}{Table \ref{table:atomistic_simulation_parameter_1} shows the used simulation parameters and material properties for each subspace.}

\begin{table}
	\begin{center}
		\begin{tabular}{c|c|c|c|c|c|c}
			subspace & 
			material &
			structure type &
			$a$ $[\si{\angstrom}]$&
			$E_{\mathrm{bulk}} [\si{eV}/atom]$&
			$r_{i}$ &
			$r_{o}$ \\
			\hline
			$\Sigma 3$ & Fe & bcc & 2.86 & -4.12 & 35a & 50a \\
			$\Sigma 5$ & Al & fcc & 4.05 & -3.36 & 40a & 50a \\
			$\Sigma 7$ & Al & fcc & 4.05 & -3.36 & 40a & 50a \\
			$[110]7.5^\circ$ & Ni & fcc & 3.52 & -4.45 & 40a & 50a \\
		\end{tabular}
		\caption{\textcolor{black}{Material properties as reproduced by the empirical potentials, and simulation parameters for each subspace: lattice constant $a$ and energy per atom in a corresponding bulk structure, $E_{\mathrm{bulk}}$; $r_{\mathrm i}$ the radius of the inner sphere and $r_{\mathrm o}$ the radius of the outer sphere of the atomistic model.}}
		\label{table:atomistic_simulation_parameter_1}
	\end{center}
\end{table}

\section{Recap of Kriging}\label{app:kriging}

In the main part of the paper, we have considered the Kriging interpolator as our method of choice for prediction of the GB energy at not observed sampling locations.
This interpolation method is used as a building block both in the sequential step and for final prediction.
Using this method the target GB energy function is predicted by a Gaussian process (GP) model, where one assumes that the target function is the realisation $Y$ of a GP with zero mean and known covariance kernel $K$.
From a given set of actual evaluations, say $\{ (x_1,Y(x_1)),\ldots,(x_N,Y(x_N)))\}$, the aim is to predict the process also at unobserved spots.
This is done by the {best linear unbiased predictor} (BLUP) or {simple Kriging} estimator which is defined as
\begin{equation*}\label{eq:kriging}
	\kb_N(x)^\top \Kb_N^{-1} \Yb_N,
\end{equation*}
where
\begin{itemize}
	\item $x$ is the point of interest where one wants to predict the energy,
	\item $\Kb_N = (K(x_i,x_j))_{i,j=1}^N \in \R^{N \times N}$ is the {Gram matrix},
	\item $\kb_N(x) = (K(x,x_1),\ldots,K(x,x_N))^\top \in \R^{N}$, and
	\item $\Yb = (Y(x_1),\ldots,Y(x_N))^\top$. 
\end{itemize}
\textcolor{black}{The simple Kriging estimator defined this way is guaranteed to interpolate through the simulated data. Alternatively, one can replace the Gram matrix $\Kb_N$ as defined above with $\Kb_N + \delta I_N$ where $I_N$ is the $N \times N$ identity matrix and $\delta > 0$ a small positive constant. In spatial statistics, this practice is referred to as the introduction of a {nugget effect}. Usually, such a nugget effect is used in presence of measurement errors but introducing it also improves numerical stability of the Kriging procedure and yields smoother predicted energy landscapes. Different choices of the parameter $\delta$ are discussed in Section~\ref{sec:1D_Results} of the main text.}

Mostly, the covariance kernel $K$ is assumed to be known only up to some finite dimensional parameter $\theta$, that is, $K=K_\theta$, which is usually estimated from the data using a maximum likelihood approach.
This is commonly called the EBLUP.
For our application we \textcolor{black}{use the maximum likelihood approach for the class of {isotropic}  Mat\'ern covariance kernels} \cite{rasmussen2006gaussian}.
Some of the fundamental zones that serve as the domain of the GP $Y$ in our examples are subdomains of the two dimensional sphere $\S^2 \subset \R^3$. Thus, considering covariance kernels that are defined in terms of the geodesic distance might seem like a natural alternative to such an isotropic kernel, in which the 
value of $K(x,y)$ depends only on the {Euclidean} distance between the points $x$ and $y$.
However, preliminary experiments
have shown that both options yield similar results in our specific application and we have thus restricted ourselves to the more straightforward use of kernels defined in terms of the Euclidean distance.

\section{Grid types for the 2D fundamental zones}\label{app:grids}
The overall algorithm uses three different kinds of grid (see Section \ref{sec:grid_generation}) to sample the FZ of GB plane inclinations, which is defined by the symmetry of the grain boundary. It is described in terms of the azimuthal angle $\varphi$ and the polar angle $\vartheta$. In this work the $s^2$-grid, the regular equally distant angular grid, and the reduced angular grid were used, which will be briefly explained in the following.

\subsection{$s^2$-grid}\label{sec:s2-grid}

\begin{figure}[H]
	\centering
	\begin{subfigure}{0.59\textwidth}
		\centering
		\includegraphics[width=\textwidth]{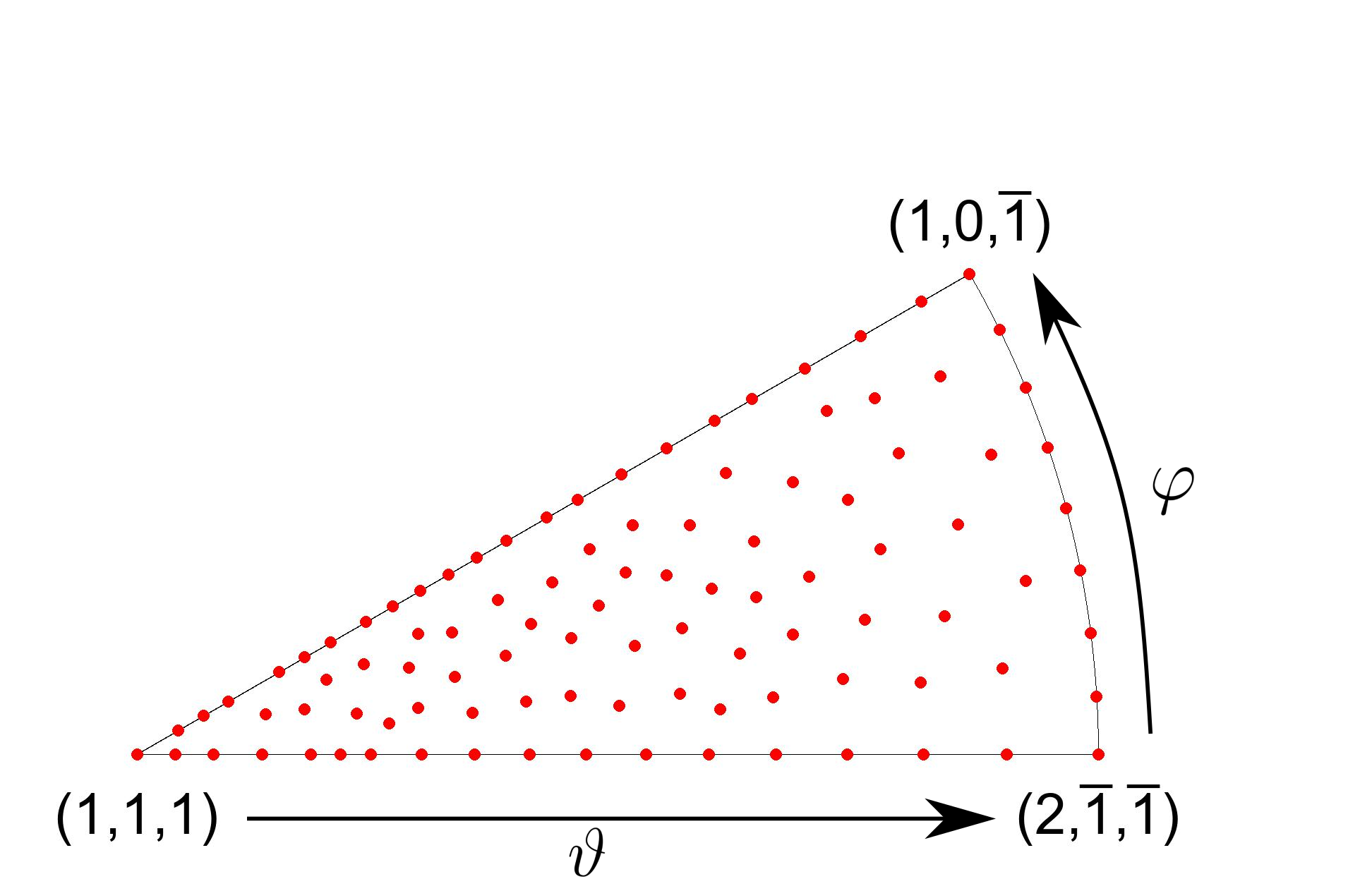}
	\end{subfigure}
	\caption{Exemplary $s^2$-grid with $\Ntotal = 100$.}
	\label{fig:exemplary_s_square_grid}
\end{figure}

The space-filling $s^2$-grid (see Figure~\ref{fig:exemplary_s_square_grid}) is initialised by the three corners of the fundamental zone which is a spherical triangle.
Afterwards, the grid is successively augmented by taking the next design point $x_{N+1}$ as a maximizer of an uncertainty measure which depends on the set of already chosen design points $\{ x_1,\ldots,x_N \}$ and is specific for the Kriging interpolator considered here, see \ref{app:kriging}. Note that, although this grid is also created sequentially, it differs from the sequential design defined in Section~\ref{sec21} in the sense that it is not response adaptive, that is, its construction does not depend on any observed energies.

\subsection{Regular equally distant angular grid}

\label{sec:angular-grid}
\begin{figure}[H]
	\centering
	\begin{subfigure}{0.59\textwidth}
		\centering
		\includegraphics[width=\textwidth]{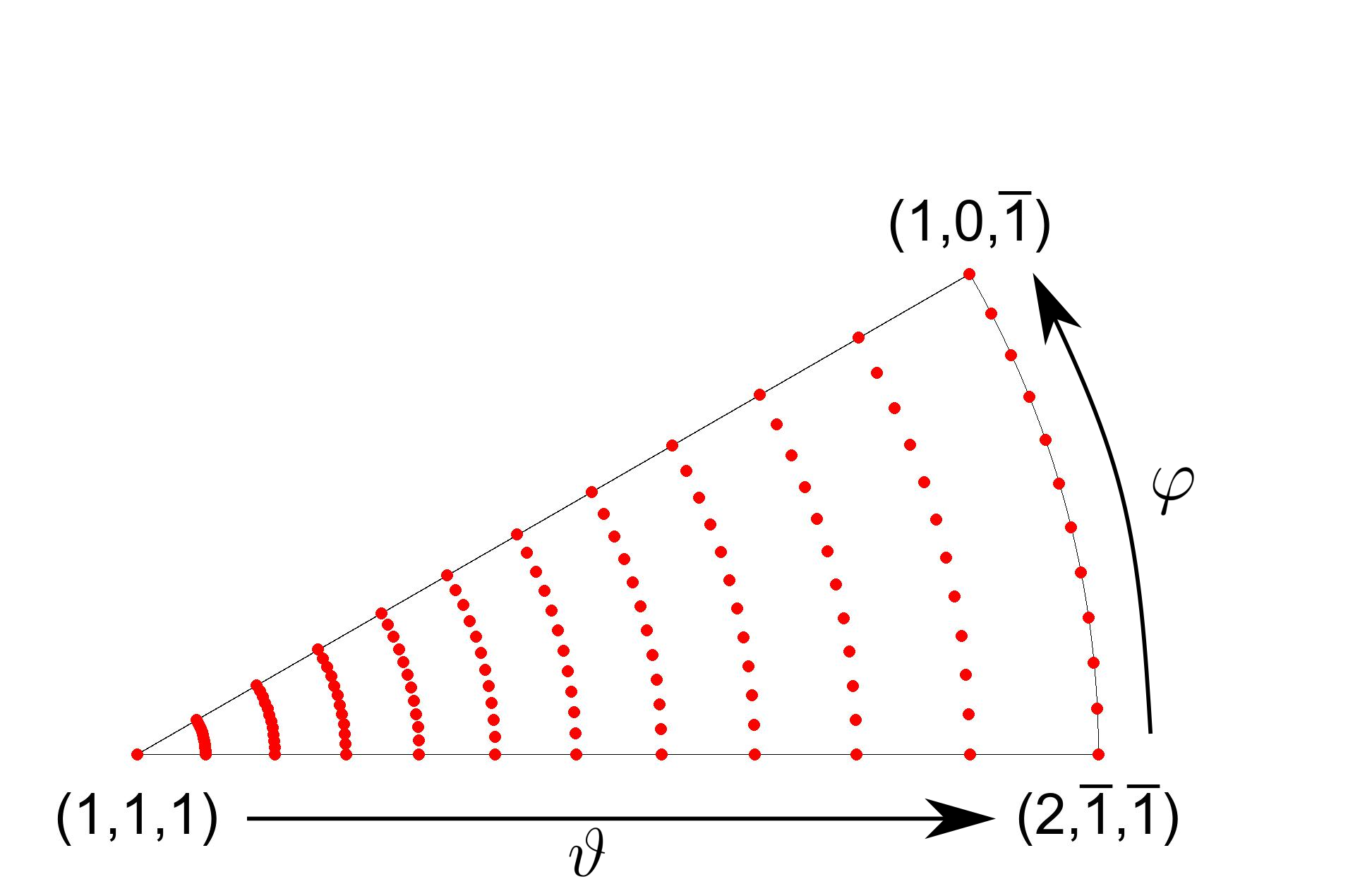}
	\end{subfigure}
	\caption{Exemplary regular equally distant angular grid with $\Ntheta = 12$, $\Nphi = 12$ and $\Ntotal = 144$.}
	\label{fig:examplary_regular_angular_grid}
\end{figure}

A regular equally distant angular grid (see Fig. \ref{fig:examplary_regular_angular_grid}) is defined by a constant azimuthal angle distance $\Delta\varphi$ and a constant polar angle distance $\Delta\vartheta$ between neighbouring points in the grid. Therefore the grid consists of azimuthal lines with a polar angle between $0^\circ$ and $\thetamax$  in $\Delta\vartheta$ steps. The points on each line have a azimuthal angle between $0^\circ$ and  $\phimax$ in $\Delta\varphi$ steps. The number of grid points obtained this way is equal to 
\begin{equation*}
	\Ngrid = \Nphi \cdot \Ntheta,
\end{equation*}
where 
$N_{\varphi}$ the number of azimuthal lines and $N_{\vartheta}$ the number of polar lines.

\subsection{Reduced angular grid}
\label{sec:reduced-angular-grid}

\begin{figure}[H]
	\centering
	\begin{subfigure}{0.59\textwidth}
		\centering
		\includegraphics[width=\textwidth]{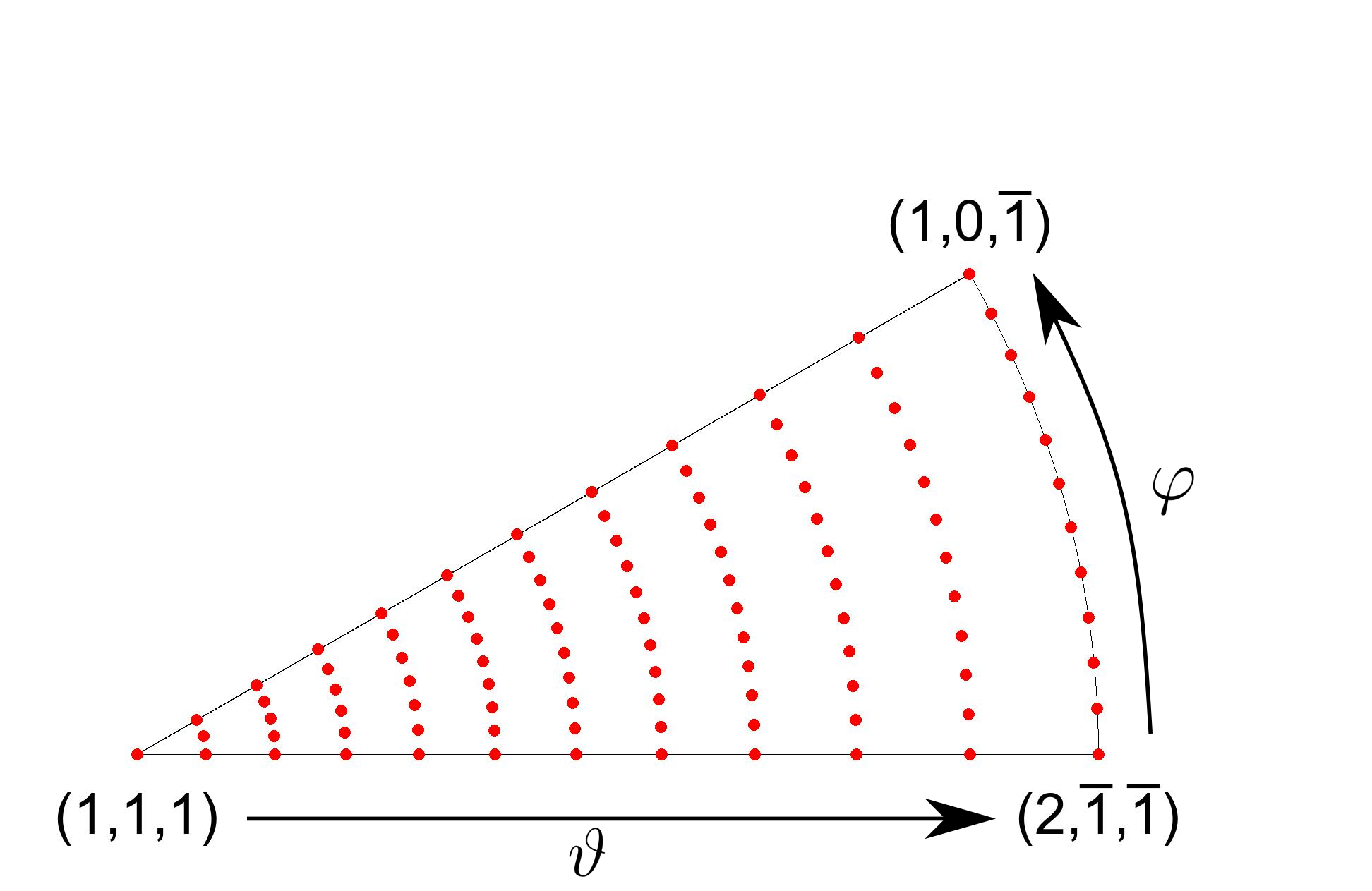}
	\end{subfigure}
	\caption{Exemplary reduced angular grid with $\Ntheta = 12$, $\Nphi = 12$ and $\Ntotal = 100$.}
	\label{fig:exemplary_reduced_angular_grid}
\end{figure}

The reduced angular grid (Fig. \ref{fig:exemplary_reduced_angular_grid}) is also generated in dependence of the two parameters $N_\varphi$ and $N_\vartheta$.
Again, $N_\vartheta$ is the number of different latitude values in the final grid which is given as an equidistant grid on $[\phimin,\phimax]$.
In contrast to the regular grid, the number of grid points on every latitude is chosen roughly proportional to the length of the considered line segment parallel to the equator.
Hence the number of grid points with the same latitude coordinate is maximal at the equator and decreases when moving towards the pole ($\theta=0^{\circ})$.

\section{Effect of number of candidate points}\label{app:Ncand}
In this section the effect of candidate points $\Ncand$ on the performance of the algorithm is studied by comparing the results for the $\Sigma 3[111]60^\circ$ inclination subspace with two different choices $\Ncand$ of candidate points from which the next sampling point is chosen by the jackknife criterion.
Originally, without using the stopping criterion, this subspace was sampled with $\Ninit = 25$, $\Nseq = 50$ and $\Ninit = 50$, $\Nseq = 25$, and both samplings were performed with $\Ncand = 75$ and $\Ncand = 200$, respectively. Note, that the $\Sigma 3[111]60^\circ$ STGB has the highest symmetry and the least complex energy distribution in the fundamental zone among the examples considered in the main part of the paper.
Therefore, the $\Ncand$ value which is appropriate for the sampling of this zone should be considered the minimum value for the other subspaces.

\begin{figure}
	\centering
	\begin{subfigure}{\textwidth}
		\centering
		\includegraphics[width=\textwidth]{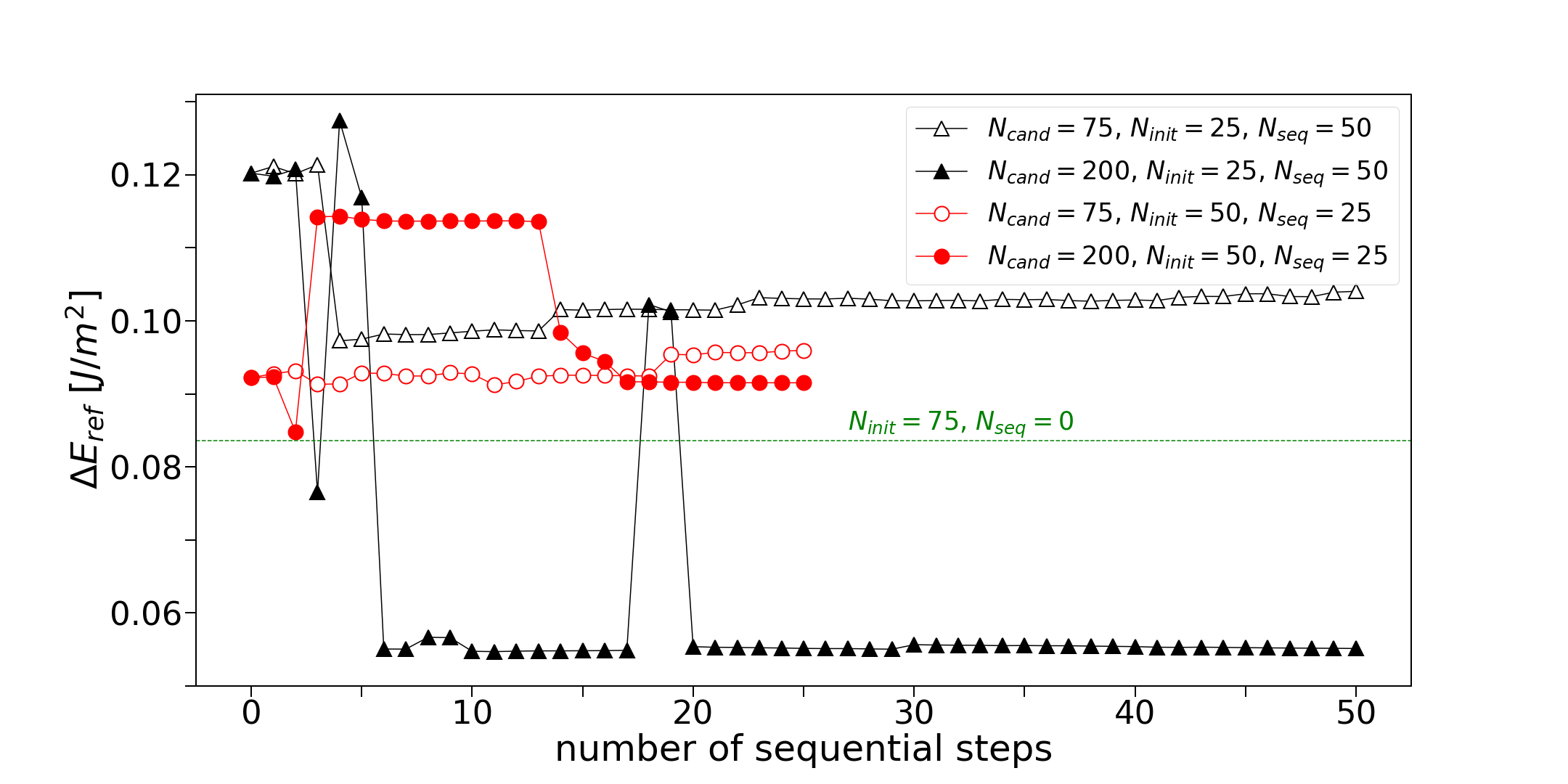}     
	\end{subfigure}
	\caption{Maximum error with respect to a reference database of the energy in the inclination subspace of the $\Sigma 3[111]60^\circ$ grain boundaries, evaluated for different designs and different 
		numbers of candidate points ($\Ncand$). The green dotted line represents the error of a regular high throughput sampling of 75 points with respect to the database.}
	\label{fig:ncand}
\end{figure}

The maximum error with respect to  the reference database for the $4$ scenarios is shown  in Figure~\ref{fig:ncand}. 
As a benchmark, we also consider the error of a regular sampling of $75$ points (green dotted horizontal line).  
The figure shows that on the long run the sampling with $\Ncand = 200$ outperforms the sampling with $\Ncand = 75$ in both cases ($\Ninit = 25$ and $\Ninit = 50$). 
Note that it can be misleading to compare the error at isolated sequential steps because the discovery of a new cusp might cause a, hopefully short-term, increase in the maximum absolute error.
Of course, the discovery of new cusps can take place at different sequential steps for the different designs and $\Ncand$ values.

The advantage of a higher number of candidate points becomes particularly apparent for the  scenario with $\Ninit = 25$. Here, for the choice  $\Ncand = 200$, 
sampling with $\Ninit = 25$ and only $7$ additional points even outperforms the regular high throughput sampling with $75$ points. Accordingly, only $\Ncand = 200$ is considered for the algorithm in the main part of the paper.

It is also observed from  Figure~\ref{fig:ncand} that a higher value of $\Ninit$ does not necessarily improve the quality of the sampling which again demonstrates the potential superiority of the sequential procedure. To explore this further, for the $\Sigma 3$ case the choice $\Ninit = 8$ is  also discussed in the main part of the paper.

\section{Pseudo code of the stopping criterion}\label{app:pseudo_code}

\textcolor{black}{The following pseudo code describes the procedure of checking the subcriteria of the stopping criterion:}
\begin{lstlisting}
	# initialization with initial design
	
	- perform Kriging and reduce to set of relevant minima
	- $N_{\mathrm{cusps,old}}$ := number of relevant minima
	
	# sequential stage
	
	$N_{\mathrm{iter, cusps}} := 3$ # number of successive stages the topolocical subcriterion 
	must be met
	$N_{\mathrm{current, cusps}} := 0$ # current number of stages the topolocical subcriterion 
	is met
	
	$N_{\mathrm{iter, \Delta E}} := 3$ # number of successive stages the statistical subcriterion
	must be met
	$N_{\mathrm{current, \Delta E}} := 0$ # current number of stages the statistical subcriterion 
	is met
	
	REPEAT
	- add new point to dataset
	- perform Kriging and reduce to set of relevant minima
	- $N_{\mathrm{cusps,new}}$ := number of relevant minima
	- $\Delta N_{\mathrm{cusps}} := \lvert N_{\mathrm{cusps,new}} - N_{\mathrm{cusps,old}} \rvert$
	- $N_{\mathrm{cusps,old}} := N_{\mathrm{cusps,new}}$
	- compute $\Delta E_{\mathrm{prev}}$
	
	IF $\Delta N_{\mathrm{cusps}} = 0$: # topolocical subcriterion is met
	$N_{\mathrm{current, cusps}} := N_{\mathrm{current, cusps}} + 1$ 
	ELSE: # topolocical subcriterion was not met
	$N_{\mathrm{current, cusps}} := 0$
	$N_{\mathrm{current, \Delta E}} := 0$
	
	IF $\Delta E_{\mathrm{prev}} < \Delta E_{\mathrm{stat}}$: # statistical subcriterion is met
	$N_{\mathrm{current, \Delta E}} := N_{\mathrm{current, \Delta E}} + 1$ 
	ELSE: # statistical criterion was not met
	$N_{\mathrm{current, cusps}} := 0$
	$N_{\mathrm{current, \Delta E}} := 0$
	
	UNTIL # algorithm stops when 
	the topolocial subcriterion was met for $N_{\mathrm{iter, cusps}}$ times 
	and the statistical criterion for $N_{\mathrm{iter, \Delta E}}$ times
	$N_{\mathrm{current, cusps}} = N_{\mathrm{iter, cusps}}$
	$N_{\mathrm{current, \Delta E}} = N_{\mathrm{iter, \Delta E}}$
\end{lstlisting}

\section{Alternative error measurements (RMSE and MAE)}\label{app:alt_error}
\begin{figure}[H]
	\centering
	\begin{subfigure}{0.49\textwidth}
		\centering
		\label{fig:1D:max:error:100}
		\includegraphics[width=\textwidth]{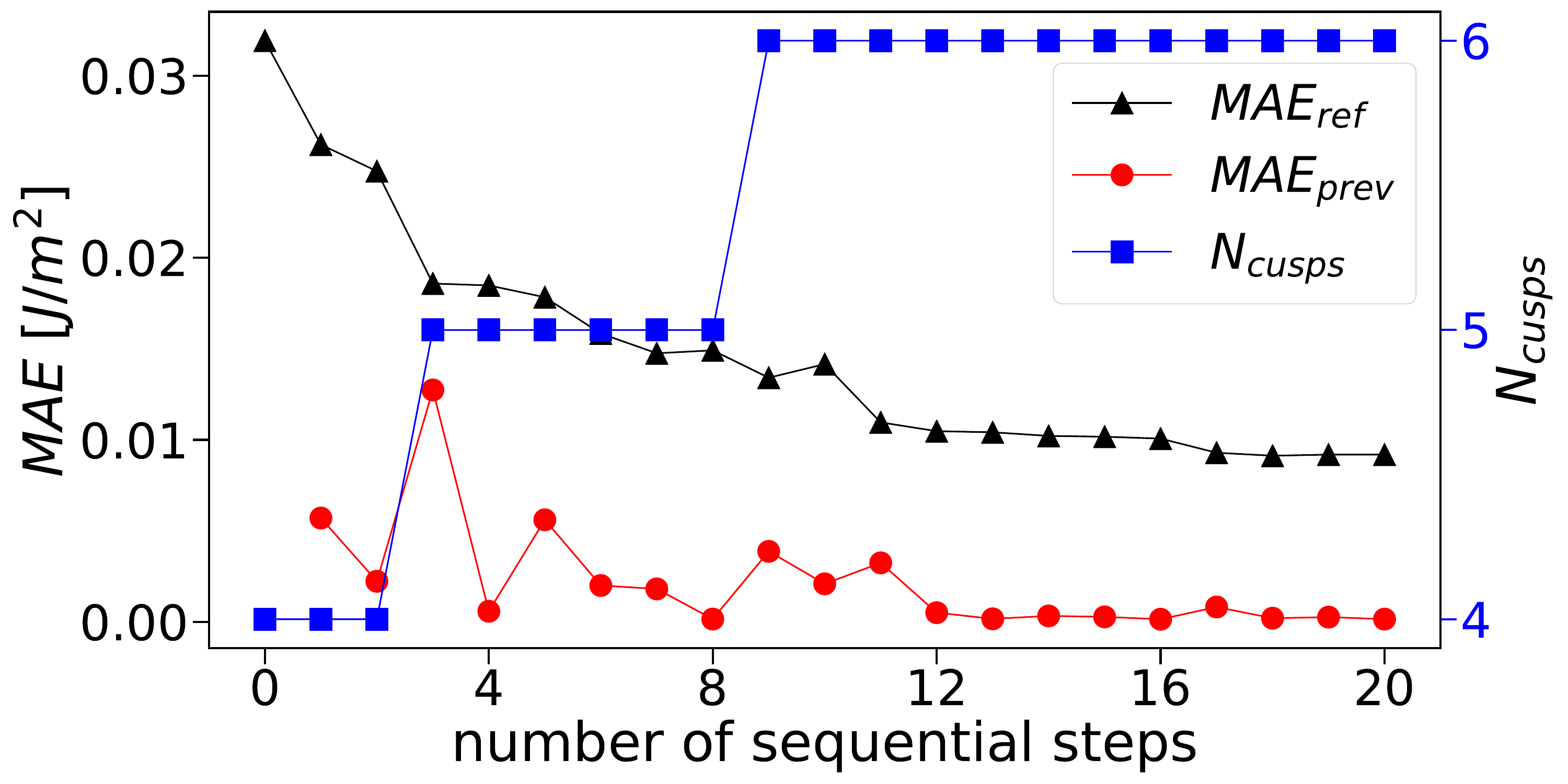}
	\end{subfigure}
	\begin{subfigure}{0.49\textwidth}
		\centering
		\includegraphics[width=\textwidth]{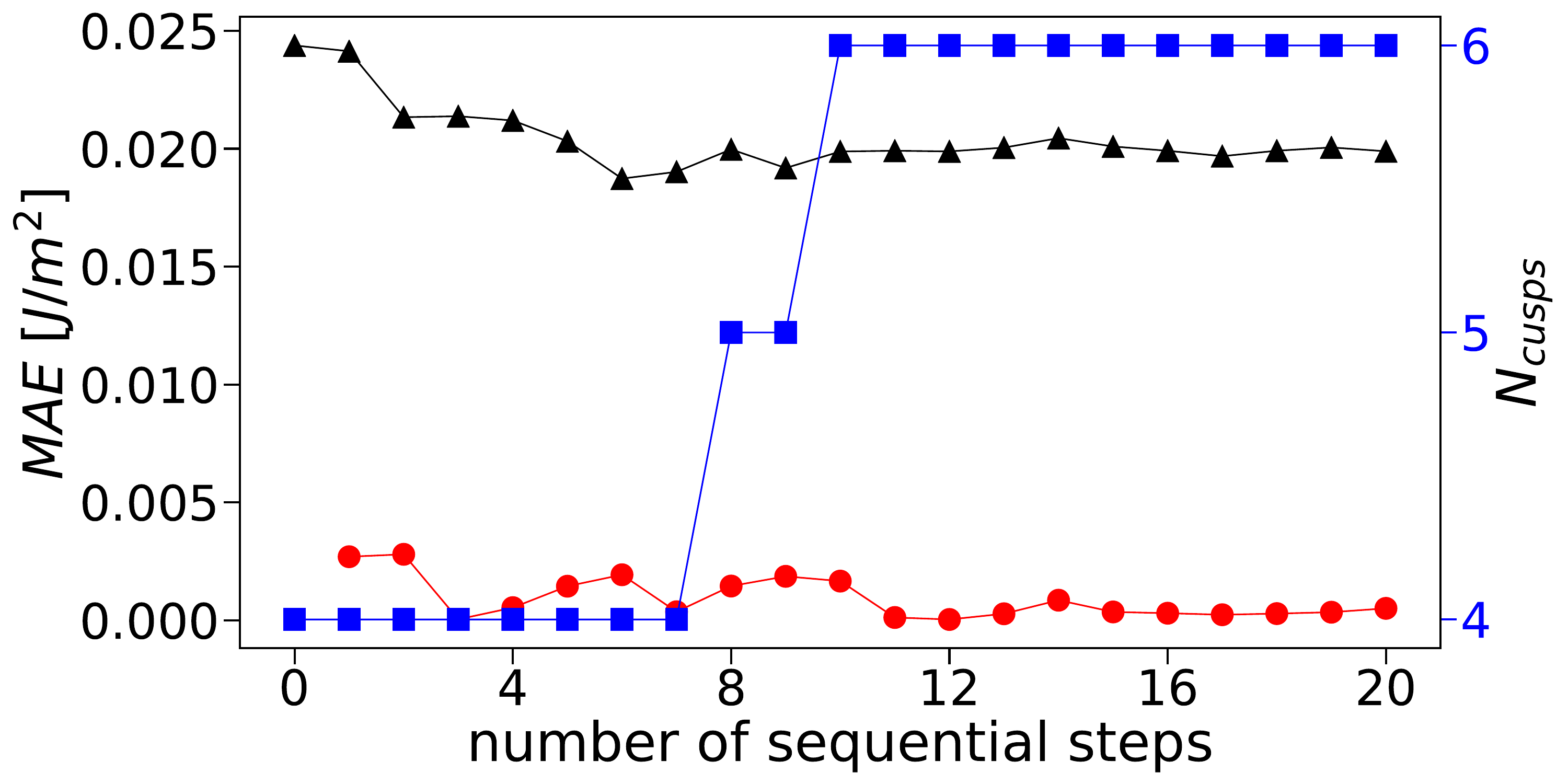}
	\end{subfigure}
	\caption{\textcolor{black}{Mean absolute error
			$\mathrm{MAE}_{\mathrm{ref}}$ with respect to a reference database,  (left $y$-axis, $\blacktriangle$); mean absolute error
			$\mathrm{MAE}_{\mathrm{prev}}$ with respect to the previous sequential step (left $y$-axis, \textcolor{black}{\textbullet}); number of cusps, $\Ncusp$ (right $y$-axis, \textcolor{blue}{$\blacksquare$}).\\
			Left panel: $[100]$ subspace with $\Ninit = 16$; Right panel: $[110]$ subspace with $\Ninit = 31$.}}
	\label{fig:STGB_mae_comparison}
\end{figure}

\begin{figure}[H]
	\centering
	\begin{subfigure}{0.49\textwidth}
		\centering
		\label{fig:1D:max:error:100}
		\includegraphics[width=\textwidth]{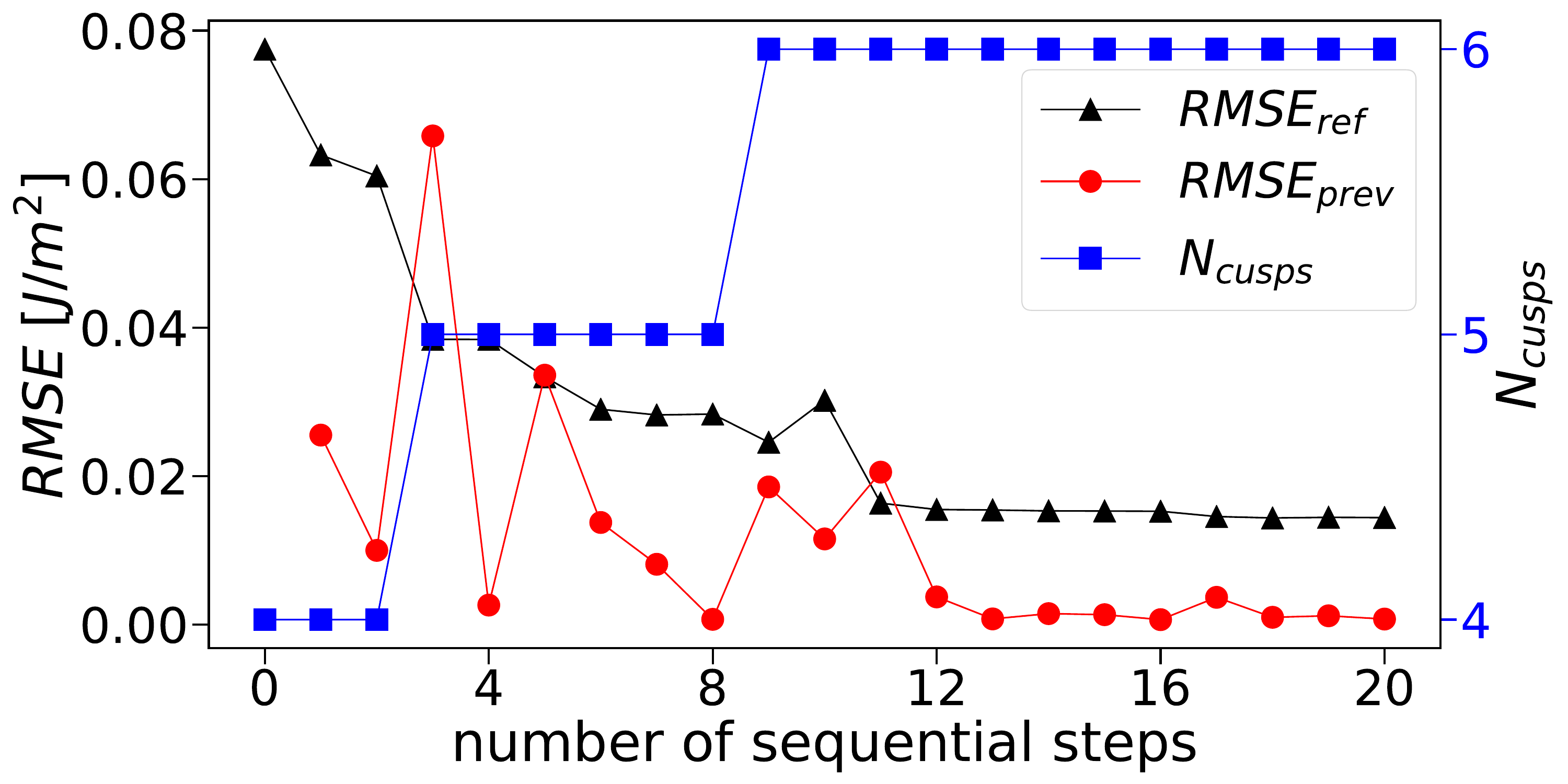}
	\end{subfigure}
	\begin{subfigure}{0.49\textwidth}
		\centering
		\includegraphics[width=\textwidth]{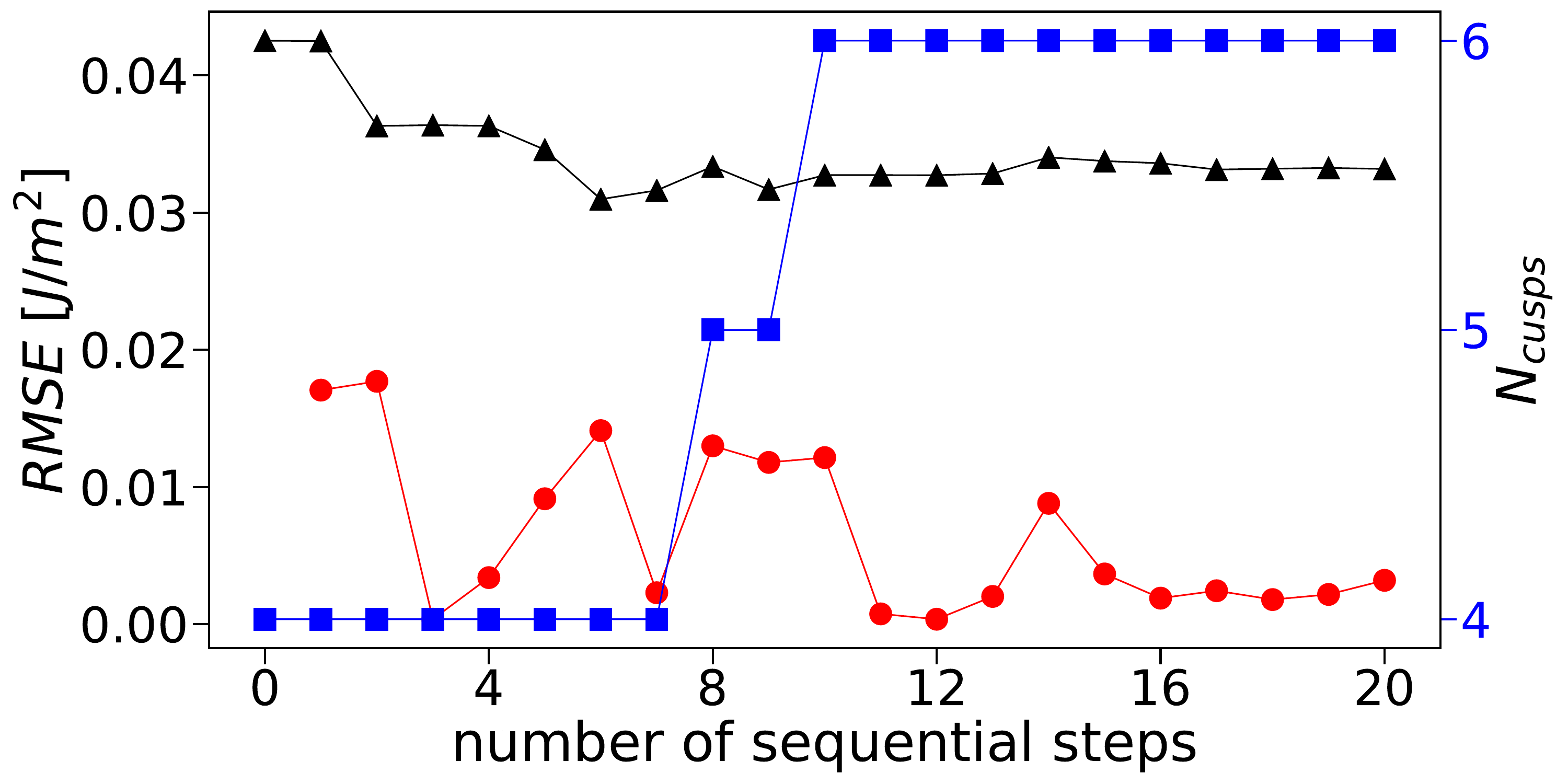}
	\end{subfigure}
	\caption{\textcolor{black}{Rooted mean squared error
			$\mathrm{RMSE}_{\mathrm{ref}}$ with respect to a reference database,  (left $y$-axis, $\blacktriangle$); rooted mean squared error
			$\mathrm{RMSE}_{\mathrm{prev}}$ with respect to the previous sequential step (left $y$-axis, \textcolor{black}{\textbullet}); number of cusps, $\Ncusp$ (right $y$-axis, \textcolor{blue}{$\blacksquare$}).\\
			Left panel: $[100]$ subspace with $\Ninit = 16$; Right panel: $[110]$ subspace with $\Ninit = 31$.}}
	\label{fig:STGB_rmse_comparison}
\end{figure}

\begin{figure}[H]
	\begin{subfigure}{0.49\textwidth}
		\centering
		\caption*{(a)}
		\includegraphics[width=\textwidth]{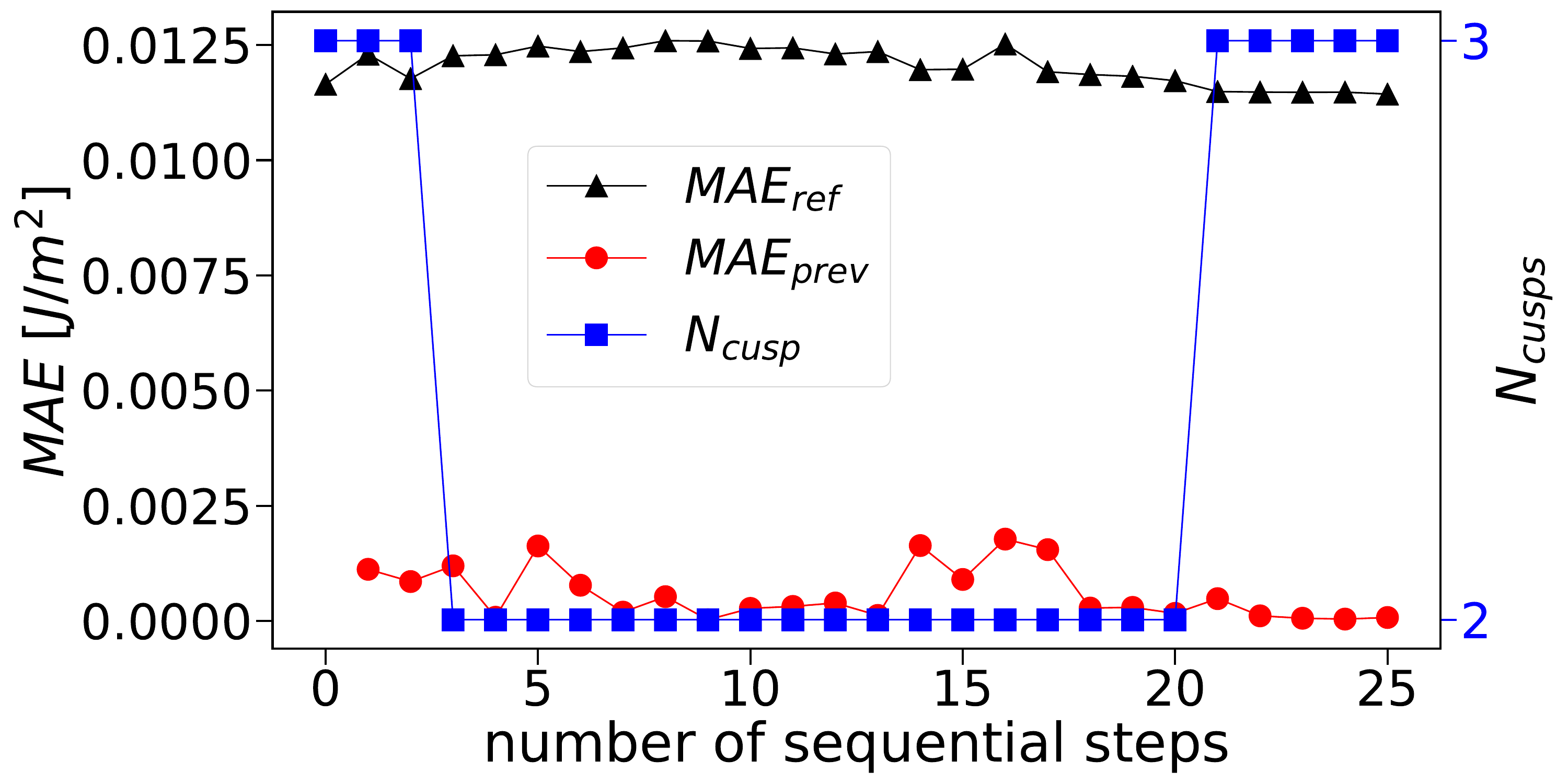}     
	\end{subfigure}
	\begin{subfigure}{0.49\textwidth}
		\centering
		\caption*{(b)}
		\includegraphics[width=\textwidth]{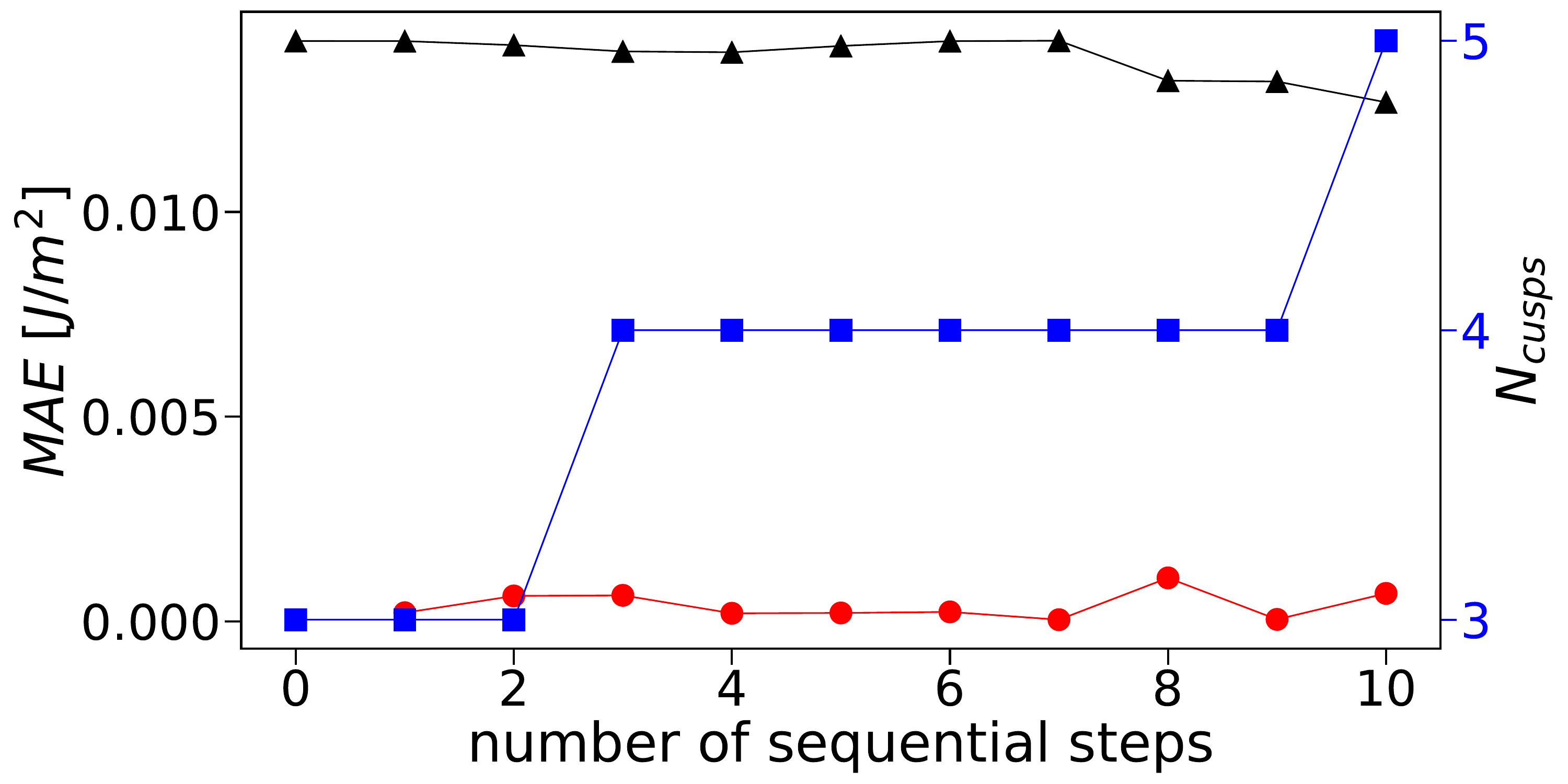}   
	\end{subfigure}
	
	\begin{subfigure}{0.49\textwidth}
		\centering
		\caption*{(c)}
		\includegraphics[width=\textwidth]{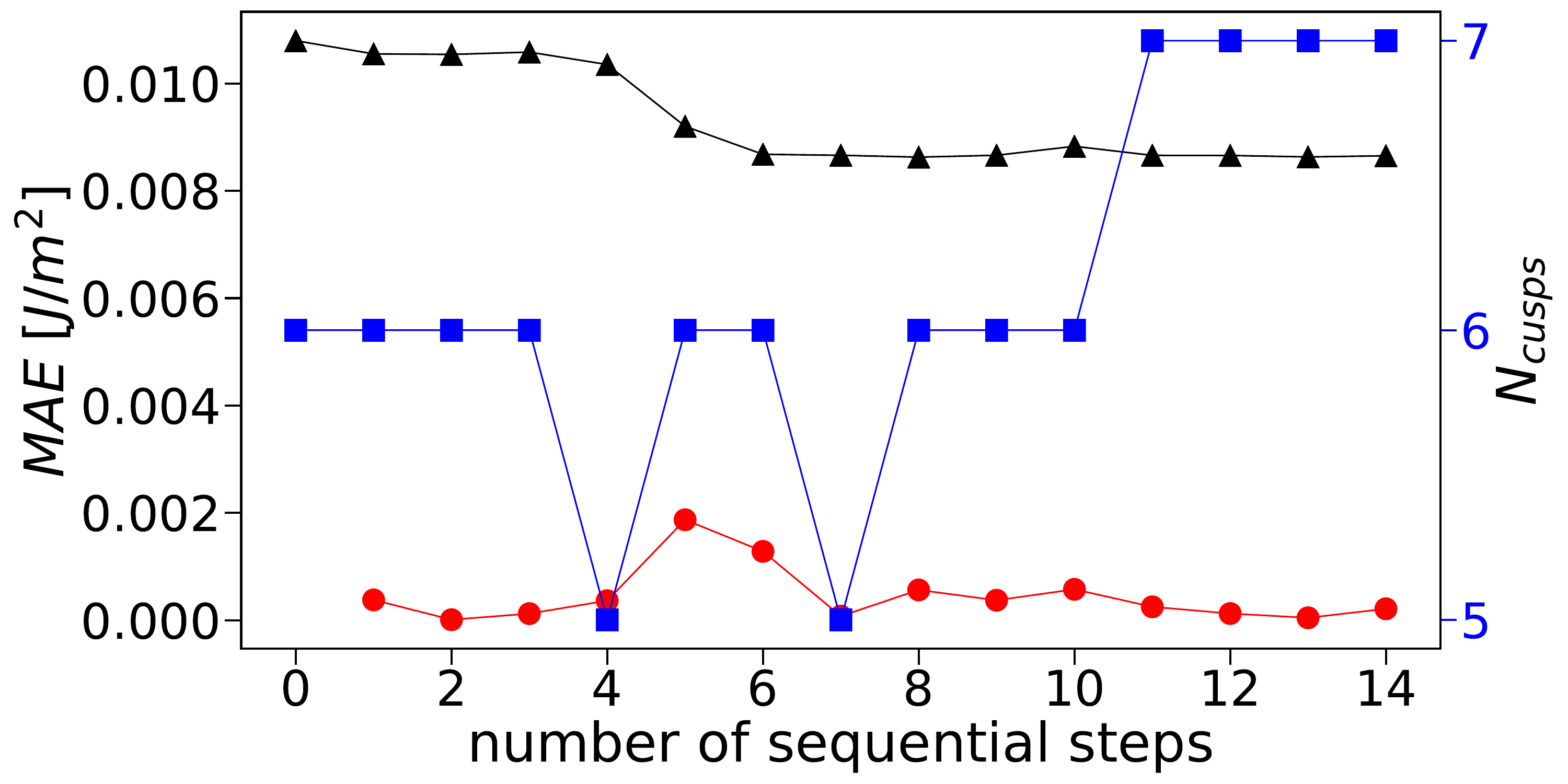}     
	\end{subfigure}
	\begin{subfigure}{0.49\textwidth}
		\centering
		\caption*{(d)}
		\includegraphics[width=\textwidth]{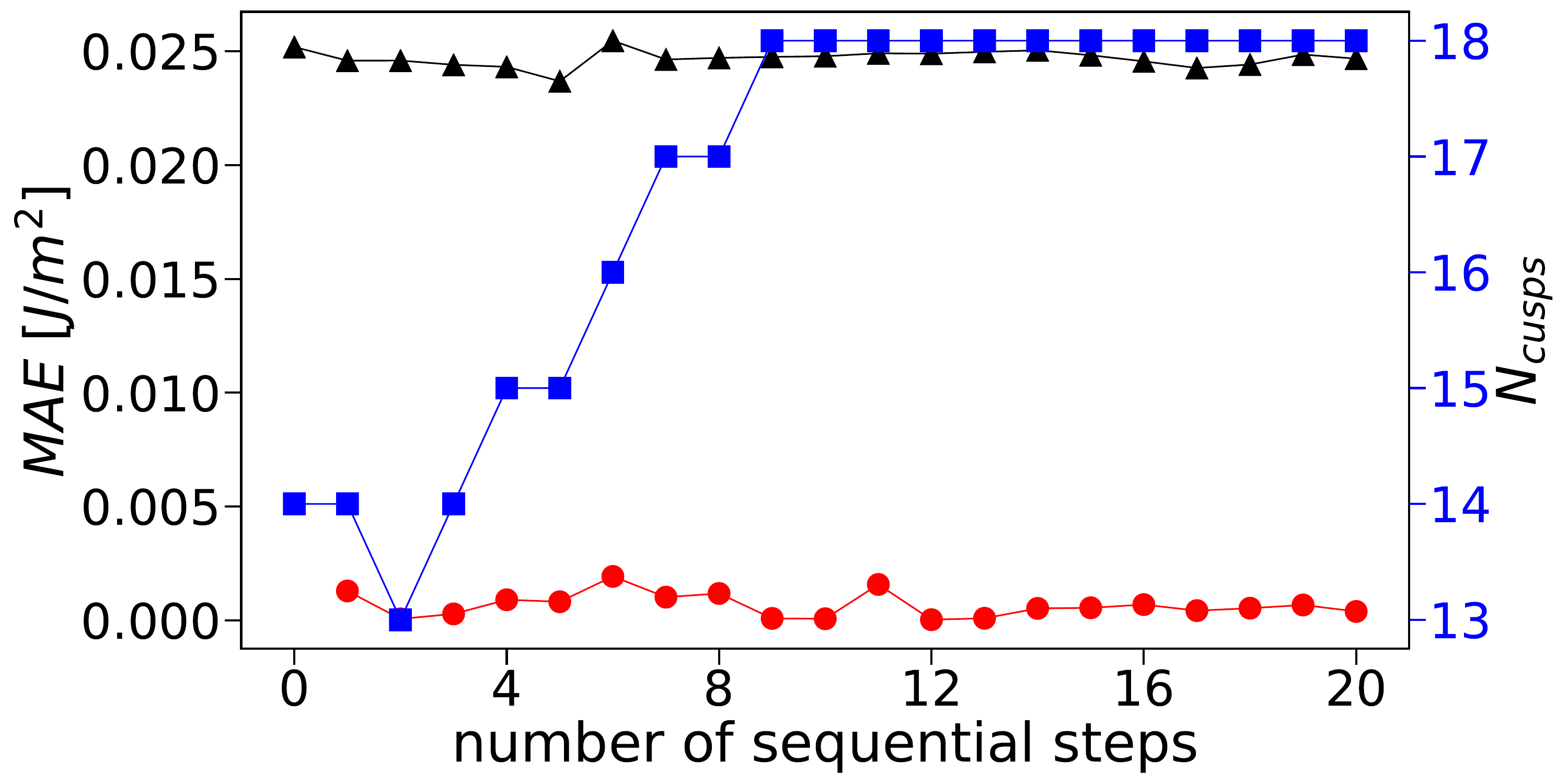} 
	\end{subfigure}
	\caption{\textcolor{black}{Mean absolute error
			$\mathrm{MAE}_{\mathrm{ref}}$ with respect to a reference database,  (left $y$-axis, $\blacktriangle$); mean absolute error
			$\mathrm{MAE}_{\mathrm{prev}}$ with respect to the previous sequential step (left $y$-axis, \textcolor{black}{\textbullet}); number of cusps, $\Ncusp$ (right $y$-axis, \textcolor{blue}{$\blacksquare$}).\\
			(a) $\Sigma 3$ subspace with $\Ninit = 50$, (b) $\Sigma 5$ subspace with $\Ninit = 20$, (c) $\Sigma 7$ subspace with $\Ninit = 28$ and (d) $[110]7.5^\circ$ subspace with $\Ninit = 40$.}}\label{fig:2D_mae}
\end{figure}

\begin{figure}[H]
	\begin{subfigure}{0.49\textwidth}
		\centering
		\caption*{(a)}
		\includegraphics[width=\textwidth]{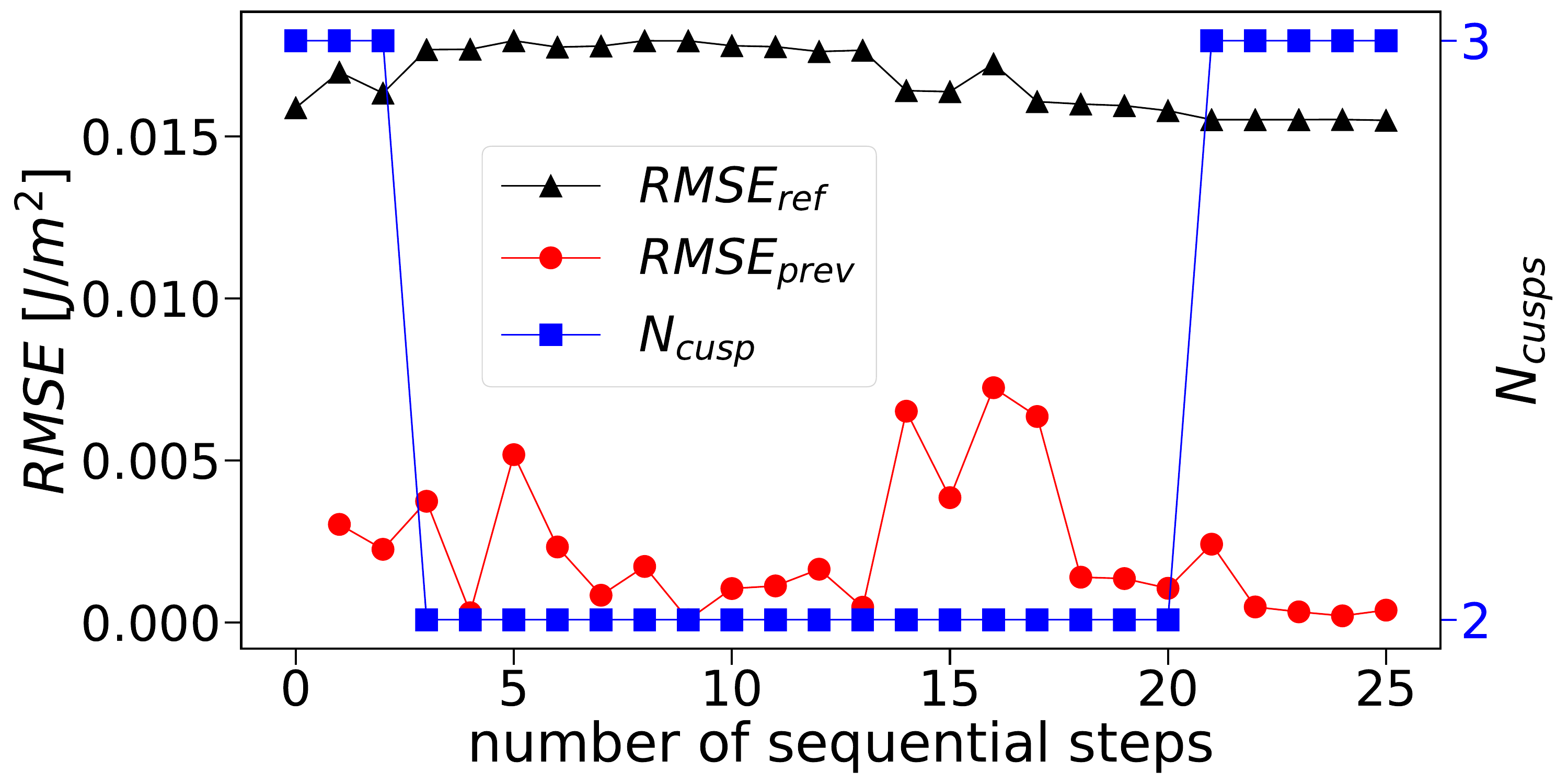}     
	\end{subfigure}
	\begin{subfigure}{0.49\textwidth}
		\centering
		\caption*{(b)}
		\includegraphics[width=\textwidth]{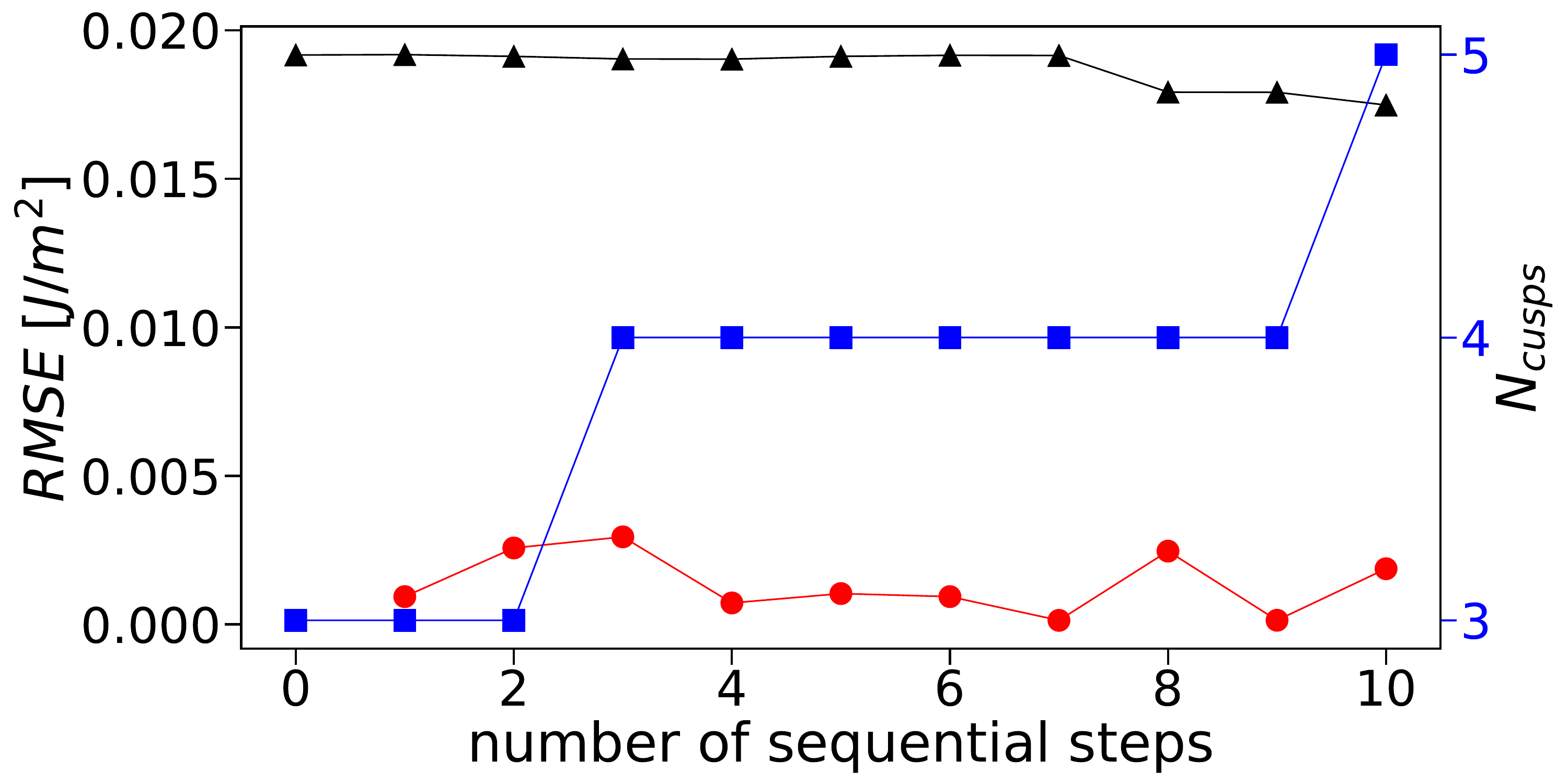}   
	\end{subfigure}
	
	\begin{subfigure}{0.49\textwidth}
		\centering
		\caption*{(c)}
		\includegraphics[width=\textwidth]{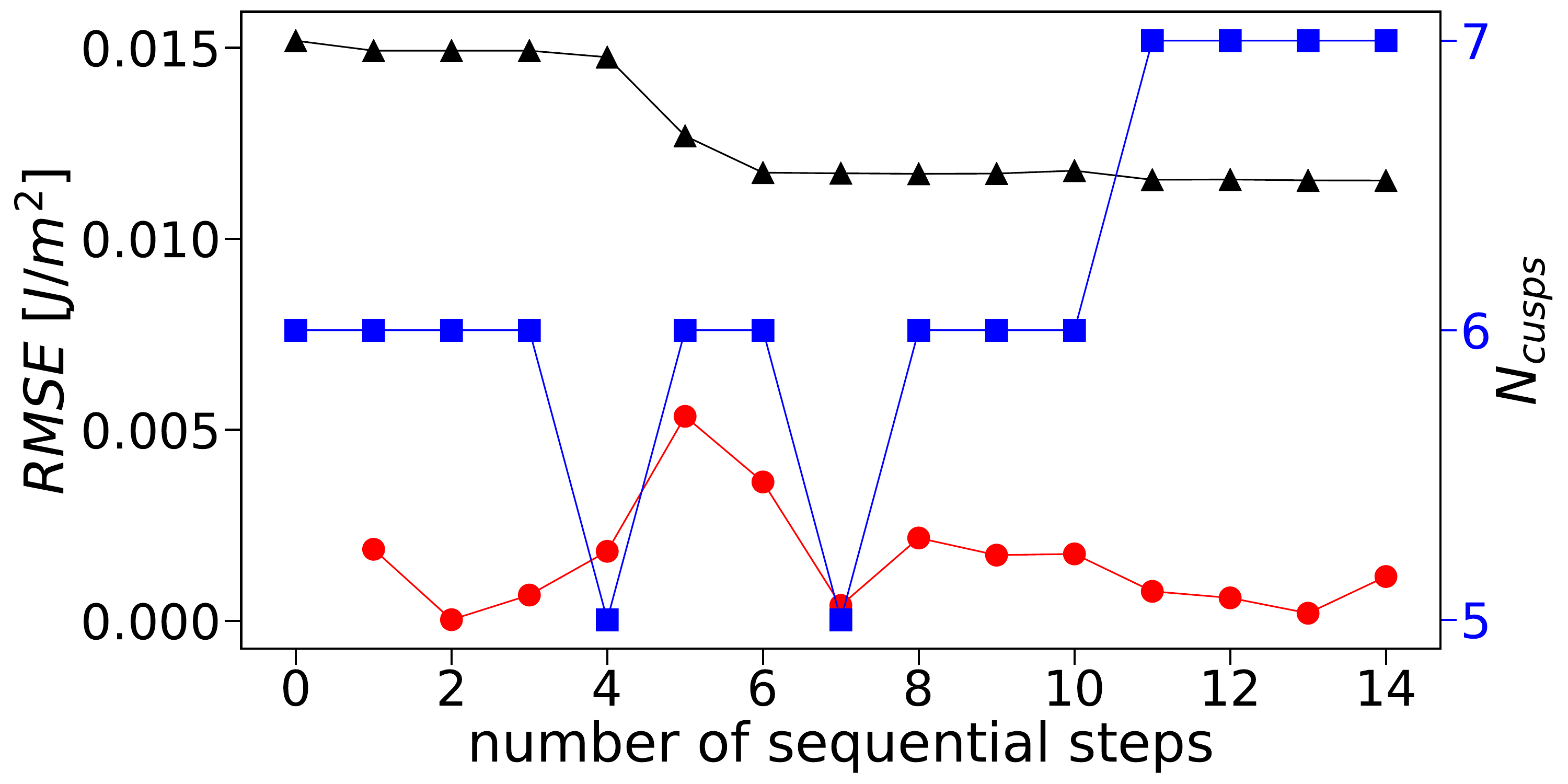}     
	\end{subfigure}
	\begin{subfigure}{0.49\textwidth}
		\centering
		\caption*{(d)}
		\includegraphics[width=\textwidth]{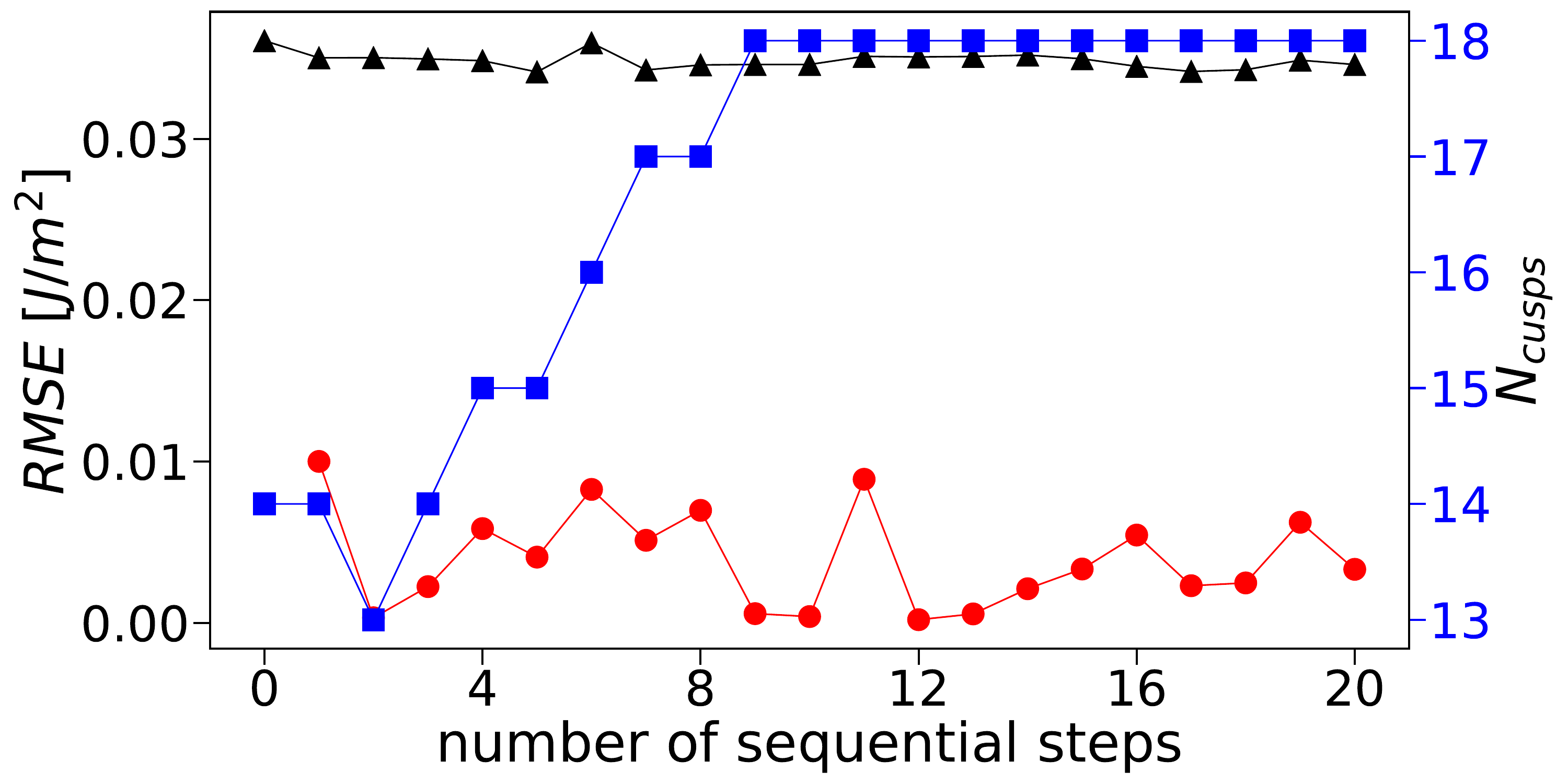} 
	\end{subfigure}
	\caption{\textcolor{black}{Rooted mean squared error
			$\mathrm{RMSE}_{\mathrm{ref}}$ with respect to a reference database,  (left $y$-axis, $\blacktriangle$); rooted mean squared error
			$\mathrm{RMSE}_{\mathrm{prev}}$ with respect to the previous sequential step (left $y$-axis, \textcolor{black}{\textbullet}); number of cusps, $\Ncusp$ (right $y$-axis, \textcolor{blue}{$\blacksquare$}).\\
			(a) $\Sigma 3$ subspace with $\Ninit = 50$, (b) $\Sigma 5$ subspace with $\Ninit = 20$, (c) $\Sigma 7$ subspace with $\Ninit = 28$ and (d) $[110]7.5^\circ$ subspace with $\Ninit = 40$.}}\label{fig:2D_rmse}
\end{figure}

\textcolor{black}{In this paper the maximum absolute error was used as the primary error measure. Since other error measures exist, we consider two alternative error measures for the stopping criterion and show how they evolve sequentially. We display the mean absolute error (MAE) in Figure~\ref{fig:STGB_mae_comparison} for the 1D STGB subspaces and in Figure~\ref{fig:2D_mae} for the 2D inclination subspaces.
	We show the root-mean-square error (RMSE) in Figure~\ref{fig:STGB_rmse_comparison} for the 1D STGB subspaces and Figure~\ref{fig:2D_rmse} for the 2D inclination subspaces. The comparison of these two error measures with the the MAE shows that MAE and RMSE are smaller, since averaged, but the overall trend for the different error measures is nearly identical.
	A significant change of the MAE is accompanied with a significant change in the estimated energy function at at least one specific location.
	This interpretation is not possible for the MAE and RMSE where a significant change of the error measure might be caused by various small changes over the overall domain.
	Moreover, significant changes at one specific location might be more difficult to detect by a global error measure.
	Therefore we decided to use the MAE in this paper.}
\subsection*{1D STGB subspaces}

\subsection*{2D inclination subspaces}

\section{Evolution of the Kriging interpolation of the $[110]7.5^\circ$ subspace}\label{app:evolution}
\begin{figure}[H]
	
	\centering
	\begin{subfigure}{0.33\textwidth}
		\centering
		\caption{}
		\includegraphics[width=\textwidth]{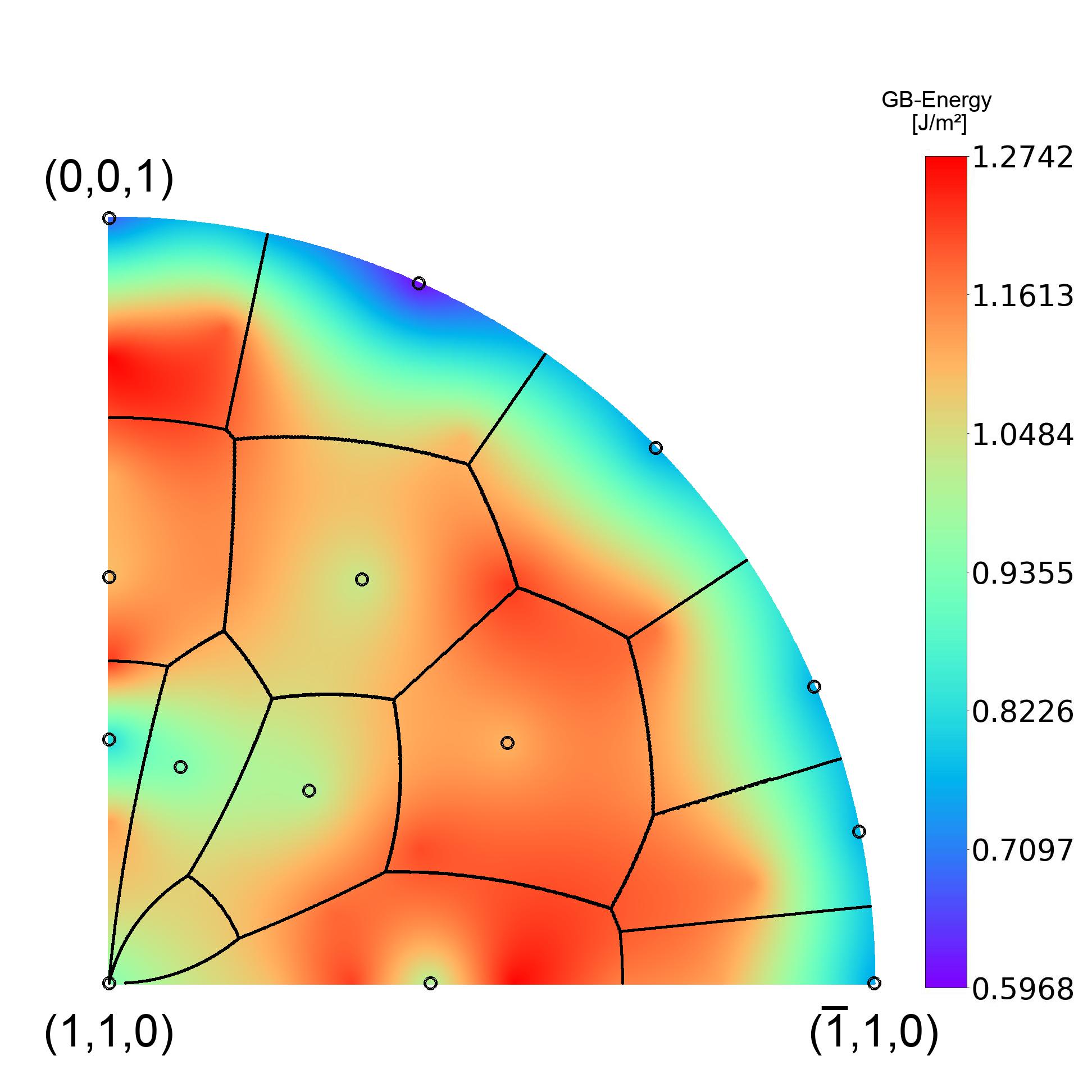}   
	\end{subfigure}
	\begin{subfigure}{0.33\textwidth}
		\centering
		\caption{}
		\includegraphics[width=\textwidth]{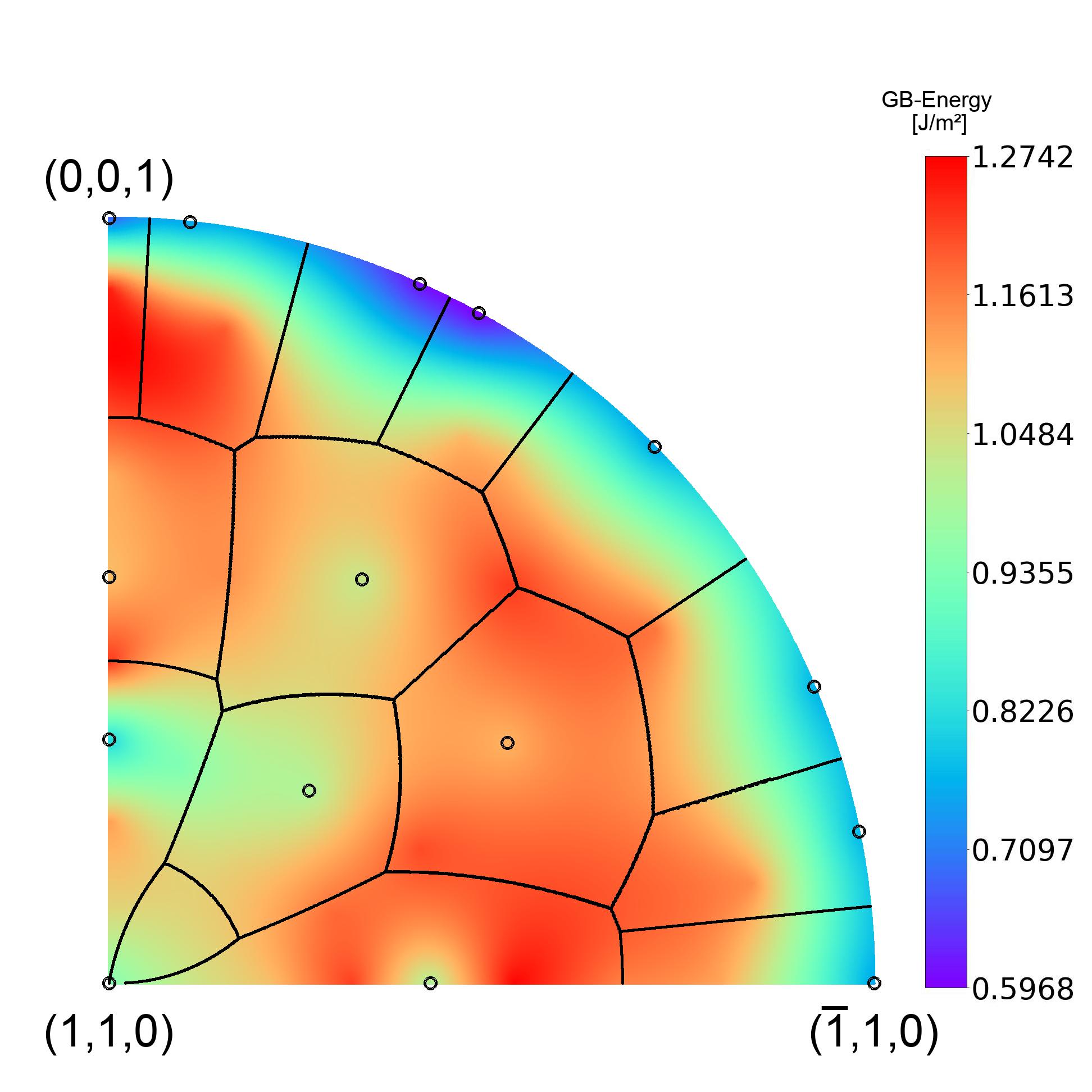}   
	\end{subfigure}
	\begin{subfigure}{0.33\textwidth}
		\centering
		\caption{}
		\includegraphics[width=\textwidth]{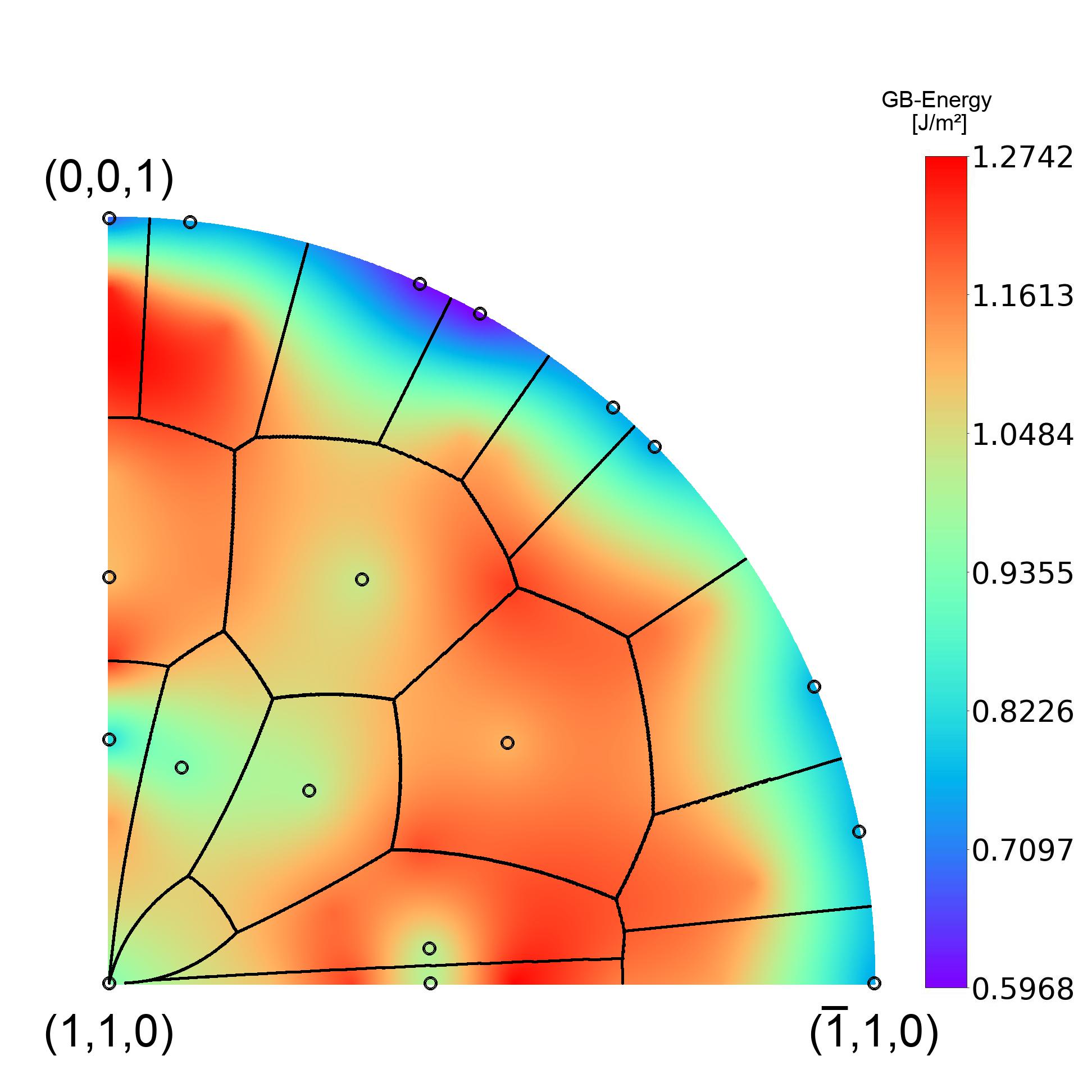}   
	\end{subfigure}
	\begin{subfigure}{0.33\textwidth}
		\centering
		\caption{}
		\includegraphics[width=\textwidth]{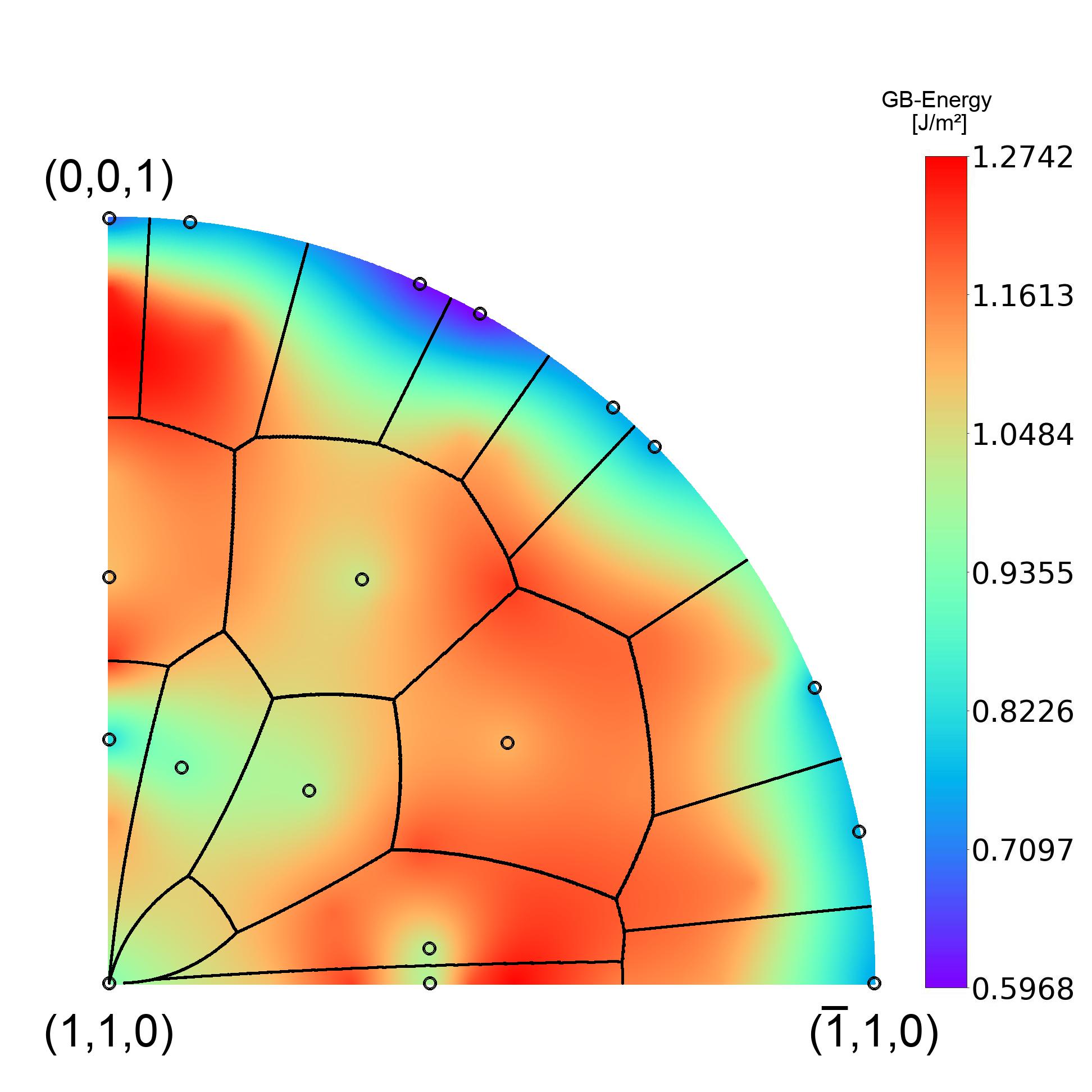}   
	\end{subfigure}
	\begin{subfigure}{0.33\textwidth}
		\centering
		\caption{}
		\includegraphics[width=\textwidth]{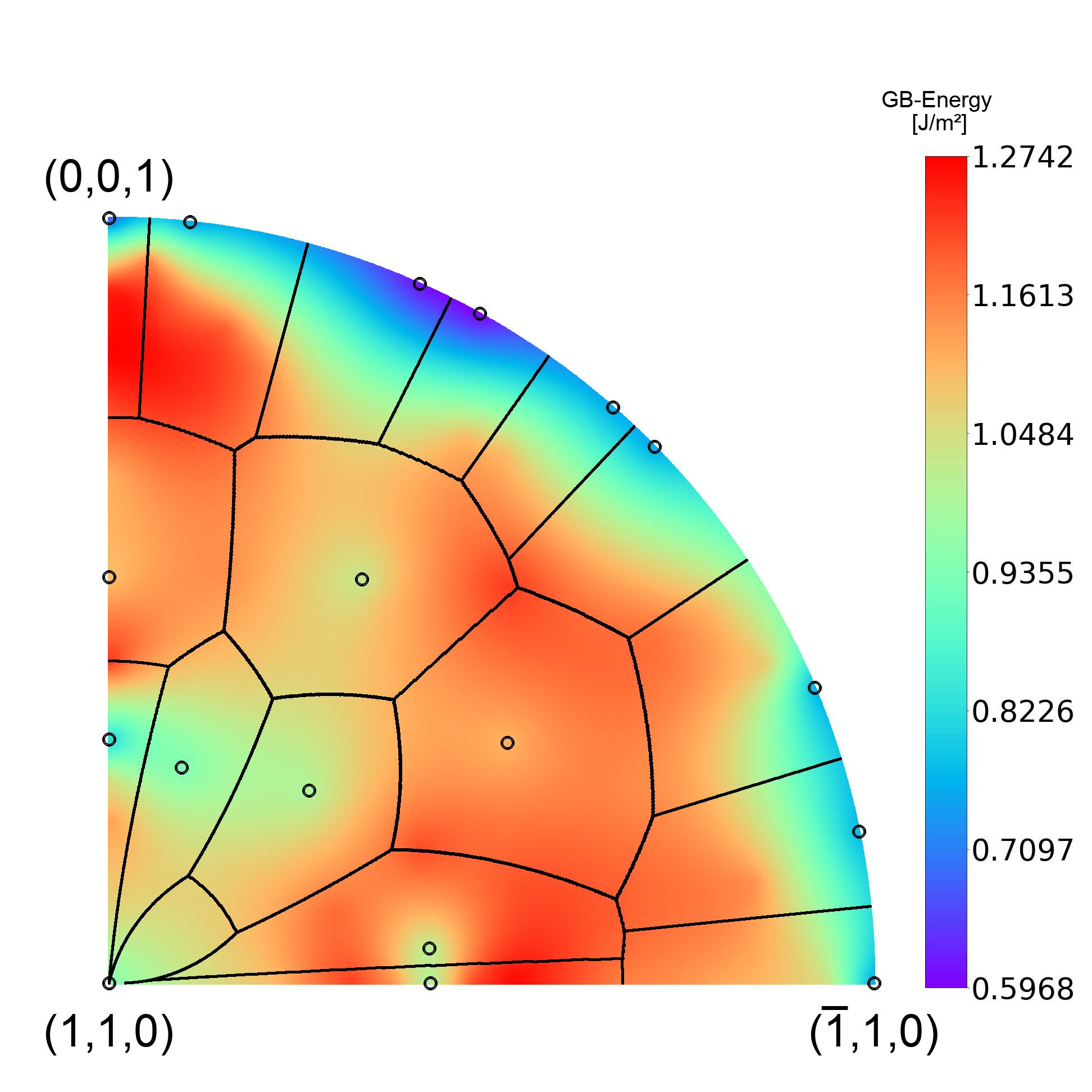}   
	\end{subfigure}
	\begin{subfigure}{0.33\textwidth}
		\centering
		\caption{}
		\includegraphics[width=\textwidth]{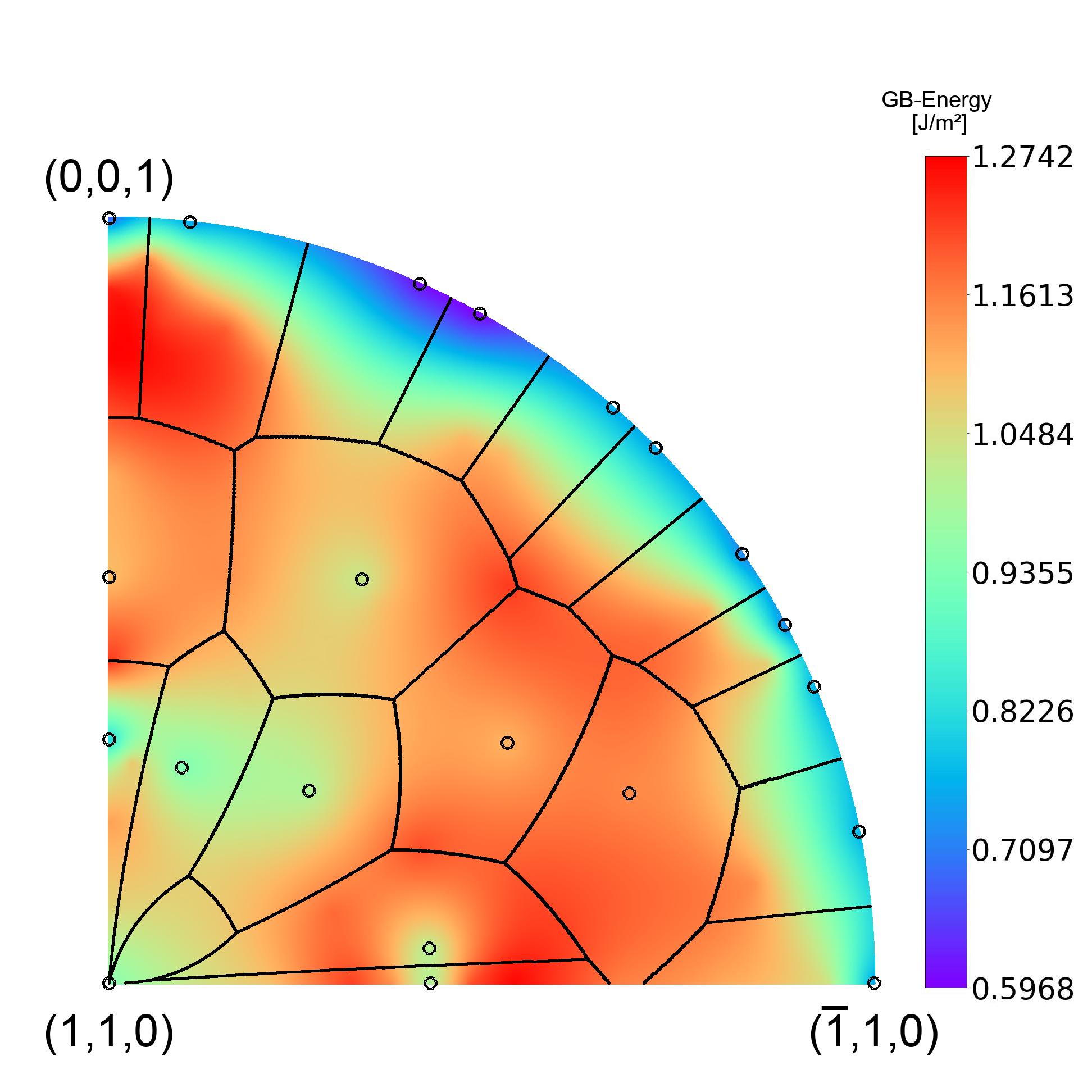}   
	\end{subfigure}
	\caption{\textcolor{black}{Kriging interpolation ($\delta = 0$) of the $[110]7.5^\circ$ subspace with $\Ninit = 40$ after (a) 0, (b) 5, (c) 10, (d) 15, (e) 20, and (f) 25 sequential iterations. The black lines indicate the boundaries of the Voronoi cells and the black circles mark the positions of the cusps.}} \label{fig_LAGB_evo}  
\end{figure}
\textcolor{black}{Figure \ref{fig_LAGB_evo} shows the evolution of the Kriging interpolation of the $[110]7.5^\circ$ subspace at different stages of the sequential sampling. It can be seen that not only new cusps are identified, but also existing cusps can disappear, e.g., between iteration $0$ and $5$. This is the case when two cusps are merged to a valley. Similarly, with further iterations a valley can also split into two distinct cusps (e.g., between iterations $5$ and $10$).
	It is also noticeable that the sequential algorithm primarily, but not exclusively, detects cusps in the tilt area of the subspace.
	This shows that the procedure tries to discover the lowest energy areas of the subspace more and more, but does not get trapped in them, because also cusps in the mixed grain boundary area are discovered by the sequential approach.} 

\end{document}